\documentclass[a4paper,10pt]{scrartcl}

\usepackage[utf8]{inputenc}
\usepackage[T1]{fontenc}
\usepackage[english]{babel}
\usepackage[dvipsnames]{xcolor}
\usepackage[labelfont={sf,bf}]{caption}
\usepackage{amssymb,amsmath,amsthm,amsfonts,bbm,stmaryrd}
\usepackage{xspace}
\usepackage{enumerate}

\usepackage{thm-restate}
\usepackage{diagbox}
\usepackage{multirow}

\usepackage{tikz}
\usetikzlibrary{arrows,automata,trees,backgrounds,decorations.pathmorphing,positioning,calc,shapes.geometric}
\tikzset{shorten >=1pt, >=stealth, auto, node distance=40, initial text=}

\definecolor{darkgray}{rgb}{0.33, 0.33, 0.33}
\definecolor{lightgray}{rgb}{0.6, 0.6, 0.6}
\definecolor{myred}{RGB}{255,15,0}
\definecolor{ao(english)}{rgb}{0.0, 0.5, 0.0} 

\definecolor{itemizecol}{named}{darkgray}
\definecolor{enumcol}{named}{darkgray}
\definecolor{desccol}{named}{darkgray}

\makeatletter
\renewcommand\subparagraph{%
 \@startsection {subparagraph}{5}{\z@ }{3.25ex \@plus 1ex
 \@minus .2ex}{-1em}{\normalsize\bfseries\sffamily }}%
\makeatother

\usepackage{hyperref}
\hypersetup{pdfauthor={Sarah Winter},pdftitle={title of document},colorlinks=true,linkcolor={myred},citecolor={ao(english)}}
\usepackage[capitalize,nameinlink]{cleveref}
\crefname{desc}{Item}{Items}
\Crefname{desc}{Item}{Items}

\crefname{desc}{Item}{Items}
\Crefname{desc}{Item}{Items}

\makeatletter
\def\th@plain{%
  \thm@headfont{\bfseries\sffamily}
  \thm@notefont{\normalfont\sffamily}
  \itshape 
}
\def\th@definition{%
  \thm@headfont{\bfseries\sffamily}
  \thm@notefont{\normalfont\sffamily}
  \normalfont 
}
\makeatother

\theoremstyle{plain}
\newtheorem{theorem}{Theorem}
\newtheorem{lemma}[theorem]{Lemma}

\theoremstyle{definition}

\newtheorem{example}[theorem]{Example}
\newtheorem{remark}[theorem]{Remark}

\theoremstyle{remark}

\usepackage{pifont}

\newcommand{\inp}{\mathbbmtt{i}}

\newcommand{\outp}{\mathbbmtt{o}}

\newcommand{\SigmaI}{\ensuremath{\Sigma_{\inp}}\xspace}
\newcommand{\SigmaO}{\ensuremath{\Sigma_{\outp}}\xspace}
\newcommand{\SigmaIO}{\ensuremath{\Sigma_{\inp\outp}}\xspace}

\newcommand{\SUM}{\ensuremath{\mathsf{Sum}}\xspace}
\newcommand{\AVG}{\ensuremath{\mathsf{Avg}}\xspace}
\newcommand{\DSUM}{\ensuremath{\mathsf{Dsum}}\xspace}

\newcommand{\Obs}{\mathcal O}
\newcommand{\obs}{\mathsf{obs}}
\newcommand{\EL}{\mathsf{EL}}
\newcommand{\plays}{\mathsf{Plays}}
\newcommand{\prefs}{\mathsf{Prefs}}
\newcommand{\last}{\mathsf{last}}
\newcommand{\Win}{\mathsf{Win}}

\newcommand{\DS}{\mathsf{DS}}
\newcommand{\MPover}{\overline{\mathsf{MP}}}
\newcommand{\MPunder}{\underline{\mathsf{MP}}}
\newcommand{\MPinf}{\mathsf{MPinf}}
\newcommand{\MPsup}{\mathsf{MPsup}}
\newcommand{\PosEn}{\mathsf{PosEn}}
\newcommand{\Thres}{\mathsf{Thres}}
\newcommand{\prefixEG}{critical prefix energy game with imperfect information\xspace}
\newcommand{\prefix}{critical prefix\xspace}
\newcommand{\prefixPosEn}{\mathsf{PrefPosEn}}
\newcommand{\crit}{\mathsf{Crit}}

\newcommand{\bestVal}{\mathsf{bestVal}}
\newcommand{\BestVal}{\mathsf{BestVal}}
\newcommand{\Approx}{\mathsf{Approx}}
\newcommand{\Safe}{\mathsf{Safe}}

\newcommand{\A}{\mathcal{A}}
\newcommand{\Adom}{\mathcal{A}_{\text{dom}}}
\newcommand{\dom}{\text{dom}}

\newcommand{\rg}{\text{rg}} 

\usepackage{fancyhdr}

\fancypagestyle{specialfooter}{%
  \fancyhf{}
  
  \fancyfoot[L]{\small\textit{This is the full version of \cite{DBLP:conf/fsttcs/FiliotLW20}.}}
  \fancyfoot[R]{\small\textit{\today}}
}

\title{Synthesis from Weighted Specifications with Partial Domains over Finite Words}
\author{Emmanuel Filiot$^a$ \and Christof Löding$^b$ \and Sarah Winter$^a$}

\date{\small $^a$Université libre de Bruxelles, Belgium \\
$^b$RWTH Aachen University, Germany}

\begin{document}

\maketitle

\thispagestyle{specialfooter}

\begin{abstract}
  In this paper, we investigate the synthesis problem of terminating
  reactive systems from quantitative specifications. Such systems
  are modeled as finite transducers whose executions are represented
  as finite words in $(\SigmaI\times \SigmaO)^*$, where $\SigmaI$,
  $\SigmaO$ are finite sets of input and output symbols,
  respectively. A weighted specification $S$ assigns a rational
  value (or $-\infty$) to words in $(\SigmaI\times \SigmaO)^*$,
  and we consider three kinds of objectives for synthesis, namely
  threshold objectives where the system's executions are required to
  be above some given threshold, best-value and approximate
  objectives where the system is required to perform as best as it
  can by providing output symbols that yield the best value and
  $\varepsilon$-best value respectively w.r.t.\ $S$.  We establish a landscape of decidability results for these three
  objectives and weighted
  specifications with partial domain over finite words given by deterministic weighted automata equipped
  with sum, discounted-sum and average measures. The resulting objectives
  are not regular in general and we develop an infinite game framework to
  solve the corresponding synthesis problems, namely the class of
  (weighted) critical prefix games.
\end{abstract}

\section{Introduction}
\label{sec:intro}

\subparagraph{Reactive synthesis.} The goal of automatic synthesis is to automatically construct programs
from specifications of correct pairs of input and output. The goal is
to liberate the developer from low-level implementation details, and
to automatically generate programs which are correct by construction. 
In the automata-based approach to synthesis~\cite{BuLa69,clarke2018handbook}, the programs to be synthesized
are finite-state reactive programs, which react continuously to
stimuli received from an environment. Such systems are not assumed to
terminate and their executions are modeled as $\omega$-words in
$(\SigmaI\SigmaO)^\omega$, alternating between input symbols in
$\SigmaI$ and
output symbols in $\SigmaO$. Specifications of such systems are then languages
$S\subseteq (\SigmaI\SigmaO)^\omega$ representing the set of
acceptable executions. The synthesis problem asks to check
whether there exists a total synchronous\footnote{$f \colon \SigmaI^\omega\rightarrow
\SigmaO^\omega$ is synchronous if it is induced by a strategy $s\colon \SigmaI^+ \to \SigmaO$ in the sense that $f(i_0i_1\dots) = s(i_0)s(i_0i_1)s(i_0i_1i_2)\dots$ for all $i_0i_1\dots \in \Sigma^\omega$} function $f \colon \SigmaI^\omega\rightarrow
\SigmaO^\omega$ such that for all input sequences $u = i_0i_1\dots$, there
exists an output sequence $v = o_0o_1\dots$ such that $f(u) = v$ and
the convolution $u\otimes v = i_0o_0i_1o_1\dots$ belongs to $S$. The
function $f$ is called a realizer of $S$. 
Automatic synthesis of non-terminating reactive systems
has first been introduced by Church~\cite{church1957applications},
and a first solution has been given by B\"uchi and Landweber
\cite{BuLa69} when the specification $S$ is $\omega$-regular. In this
setting, when a realizer exists, there is always one which can be
computed by a finite-state sequential transducer, a finite-state
automaton which alternates between reading
one input symbol and producing one output symbol. This result has sparked much further
work to make synthesis feasible in practice, see e.g.,
\cite{KupfermanPV06,FiliotJR11,BloemJPPS12}. The synthesis problem is
classically modeled as an infinite-duration game on a graph, played by two players,
alternatively picking input and output symbols. One player, representing
the system, must enforce an objective that corresponds to the
specification. Finite-memory winning strategies are in turn systems
that realize the specification. This game metaphor has triggered a lot
of research on graph games~\cite[Chapter 27]{clarke2018handbook}. 
There has also been
a recent effort to increase the quality of the automatically generated
systems by enhancing Boolean specifications with quantitative
constraints, e.g.,~\cite{bloem2009better,chatterjee2012energy,brenguier2016non,almagor2017quantitative}. This
has also triggered a lot of research on quantitative extensions of
infinite-duration games, for example mean-payoff, energy, and
discounted-sum games, see, e.g.,
\cite{ehrenfeucht1979positional,zwick1996complexity,degorre2010energy,bouyer2010timed,brazdil2010reachability,andersson2006improved,DBLP:conf/csl/HunterPR16}.

\subparagraph{Partial-domain specifications.} In the classical formulation of the synthesis problem, it
is required that a realizer $f$ meets the specification \emph{for all}
possible input sequences. In particular, if there is a single input
sequence $u$ such that $u\otimes v\not\in S$ for all output sequences
$v$, then $S$ admits no realizer. In other words, when the \emph{domain}
of $S$ is partial, then $S$ is unrealizable. Formally, the domain of $S$ is $\dom(S) = \{
u\in\SigmaI^\omega\mid \exists v\colon u\otimes v\in S\}$. As noticed recently and independently
in~\cite{AK20}, asking that the realizer meets the specification for all
input sequences is often too strong and a more realistic setting is to
make some assumptions on the environment's behaviour, namely, that the
environment plays an input sequence in the domain of the
specification. This problem is called good-enough synthesis in~\cite{AK20}
and can be formulated as follows: given a specification $S$, check whether
there exists a \emph{partial} synchronous function $f \colon \SigmaI^\omega\rightarrow
\SigmaO^\omega$ whose domain is $\dom(S)$, and such that for all input
sequence $u\in \dom(S)=\dom(f)$, $u\otimes f(u)\in S$. Decidability of
the latter problem is entailed by decidability of the classical
synthesis problem when the specification formalism used to describe
$S$ is closed under expressing the assumption that the environment
provides inputs in $\dom(S)$.  It is the case for instance when $S$ is
$\omega$-regular, because the specification $S\cup
\overline{\dom(S)}\otimes \SigmaO^\omega$ has total domain and is
effectively $\omega$-regular. \cite{AK20} investigates the more
challenging setting of $S$ being expressed by a multi-valued (in
contrast to Boolean) LTL logic. More generally, there is a series of
works on solving games under assumptions on the behaviour of the
environment~\cite{DBLP:conf/tacas/ChatterjeeH07,DBLP:conf/atva/BloemEK15,DBLP:journals/amai/KupfermanPV16,DBLP:conf/icalp/ConduracheFGR16,DBLP:journals/acta/BrenguierRS17,almagor2017quantitative}.

\subparagraph{Our setting: Partial-domain weighted specifications.} 
In this paper, motivated by the line of work on
quantitative extensions of synthesis and the latter more realistic
setting of partial-domain specifications, we investigate synthesis
problems from partial-domain weighted specifications (hereafter just
called weighted specifications). We conduct this
investigation in the setting of \emph{terminating reactive systems},
and accordingly our specifications are over finite words. 
Formally, a specification is a mapping $S\colon (\SigmaI.\SigmaO)^* \to \mathbbm Q \cup
\{-\infty\}$. The domain $\dom(S)$ of $S$ is defined as all the input
sequences $u\in\SigmaI^*$ such that $S(u\otimes v)\in \mathbbm{Q}$ for
some $v\in\SigmaO^*$. We consider three quantitative synthesis
problems, which all consists in checking whether there exists a
function $f$ computable by a finite transducer  such that $\dom(f) =
\dom(S)$ and which satisfies respectively the following conditions:
\begin{itemize}
  \item for all $u\in \mathrm{dom}(S)$ it holds that
  $S(u\otimes f(u)) \triangleright t$ for a given threshold $t\in\mathbbm Q$
  and $\triangleright \in \{>,\geq\}$, called \emph{threshold synthesis}, or
  \item 
  $S(u \otimes f(u)) = \bestVal_S(u)$, that is, the maximal value that can be
  achieved for the input $u$, i.e., $\bestVal_S(u) =
  \mathrm{sup}\{S(u\otimes v)\mid v\in\SigmaO^*\}$, called \emph{best-value synthesis}, or 
  \item $\bestVal_S(u) - S(u\otimes f(u)) \triangleleft r$ for a given threshold
$r\in\mathbbm Q$ and $\triangleleft \in \{<,\leq\}$, called \emph{approximate
synthesis}.
\end{itemize}
   
Following the game metaphor explained before, those quantitative
synthesis problems can be formulated as two-player games in which Adam (environment) and Eve (system) alternatively
pick symbols in $\SigmaI$ and $\SigmaO$ respectively. Additionally, Adam has the power to stop the game. If it does not, then
Eve wins the game. Otherwise, a finite play spells a word $u\otimes
v$. For the Boolean synthesis problem, Eve has won if either
$u\not\in\dom(S)$ where $S$ is the specification, or $u\otimes v\in
S$. Additionally, for the threshold synthesis problem, the value $S(u \otimes v)$
must be greater than the given threshold; for the best-value synthesis
problem, it must be equal to $\bestVal_S(u)$ and for approximate
synthesis it must be $r$-close to $\bestVal_S(u)$.

\subparagraph*{Contributions.}
Our main contribution is a clear picture about decidability of
threshold synthesis, best-value synthesis and approximate
synthesis for weighted specifications over
finite words defined by deterministic weighted finite automata \cite{droste2009handbook},
equipped with either sum, average or discounted-sum measure. Such
automata extend finite automata with integer weights on their
transitions, computing a value through a payoff function that combines
those integers, with sum, average, or discounted-sum. The results (presented in \cref{sec:syntesis-problems}) are summarized in \cref{table:summary}.
We also give an application of our results to the decidability of quantitative extensions of
the Church synthesis problem over infinite words, for some classes of
weighted safety specifications, which intuitively require that all
prefixes satisfy a quantitative requirement (being above a threshold,
equal to the best-value, or close to it). 

As we explain in the related works section, some of our results are
obtained via reduction to solving known quantitative games or
to the notions of $r$-regret determinization for weighted automata.
We develop new techniques to solve the strict threshold synthesis problem
for discounted-sum specifications in \textsc{NP} (\cref{thm:deterministic-threshold}), the best-value synthesis problem
for discounted-sum specifications in $\textsc{NP}\cap\textsc{coNP}$
(\cref{thm:best-value}) and approximate synthesis for average specifications (\cref{thm:approximate}), which are to the best of our knowledge new results.

Moreover, as our main tool to obtain our synthesis results, we introduce in
\cref{sec:games} a new
kind of (weighted) games called \emph{critical prefix games} tailored
to handle weighted specifications with \emph{partial} domain of
\emph{finite} words. We believe these kind of games are interesting on
their own and are described below in more detail.

\begin{table}[t!]
  \centering
  \caption{Complexity results for weighted specifications.
  Here, D stands for decidable, the suffix -c for complete, $\lambda$ for discount factor, and $n$ for a natural number.}
  \label{table:summary}
\begin{tabular}{  r | c | c | c }
    \multirow{2}{*}{\diagbox[height=1.5\line]{Problem}{Spec}} &   &  &  \\[-.5em]
      & \SUM-automata & \AVG-automata & \DSUM-automata \\ \hline
      strict threshold   & $\textsc{NP} \cap \textsc{coNP}$ & $\textsc{NP} \cap \textsc{coNP}$ & \textsc{NP} \\ \hline
      non-strict threshold  & $\textsc{NP} \cap \textsc{coNP}$ & $\textsc{NP} \cap \textsc{coNP}$ & $\textsc{NP} \cap \textsc{coNP}$ \\ \hline
      best-value               & \textsc{Ptime}~\cite{DBLP:journals/talg/AminofKL10} & \textsc{Ptime}~\cite{DBLP:journals/talg/AminofKL10} & $\textsc{NP} \cap \textsc{coNP}$ \\ \hline
      strict approximate & \textsc{EXPtime}-c~\cite{DBLP:conf/lics/FiliotJLPR17} &  D & {\footnotesize\textsc{NEXPtime}\! for \!$\lambda\!=\!1/n$} \\ \hline
      non-strict approx. & \textsc{EXPtime}-c~\cite{DBLP:conf/lics/FiliotJLPR17} & D & {\footnotesize\textsc{EXPtime}\! for \!$\lambda\!=\!1/n$} \\
      \hline
  \end{tabular}
\end{table}

\subparagraph*{Critical prefix games.} Following the classical game
metaphor of synthesis, we design weighted games into which some of our
synthesis problems can be directly encoded. Those games still have
infinite-duration, but account for the fact that specifications are on
finite words and have partial domains. In particular, the quantitative constraints must be
checked only for play prefixes that correspond to input words of the
environment which are in the domain of the specification.  
So, a critical prefix game is defined as 
a two-player turn-based weighted game with some of the vertices being declared as
critical. When the play enters a critical vertex, a quantitative
requirement must be fulfilled, otherwise Eve loses. For
instance, critical prefix \emph{threshold} games require that the
payoff value when entering a critical vertex is at least or above a certain
threshold. We show that these threshold games are all decidable 
for sum, average, and discounted-sum payoffs, see \cref{thm:threshold-games,thm:threshold-games-dsum-all}.
For solving approximate average synthesis, we use a reduction to critical prefix energy games of imperfect
information starting with fixed initial credit (the energy level must be at least zero whenever the play is in a critical vertex).
Without critical vertices
(where the energy level must be at least zero all the time) these
games are known to be decidable \cite{degorre2010energy}. We show that
adding critical vertices makes these games undecidable, in general,
see \cref{thm:prefix-energy-games-undec}.  However, a large subclass
of imperfect information critical prefix energy games, sufficient for
our synthesis problems, is shown to be decidable, see
\cref{thm:prefix-energy-reachable}.

\subparagraph*{Domain-safe weighted specifications.}
Most of our quantitative
synthesis problems reduce to two-player games. While we need games
of different natures, they all model the fact that Eve
constructs a run of the (deterministic) automaton,
given the input symbols provided by Adam so far. By
choosing outputs, Eve must make sure that this run is accepting whenever the input
word played by Adam so far is in the domain of
$S$. Otherwise Adam can stop and Eve loses. While
this condition can be encoded in the game by enriching the vertices with subsets
of states (in which Eve could have been by choosing alternative
output symbols), this would result in an exponential blow-up of the game. We instead
show that the weighted automaton can be preprocessed in polynomial-time into a
so called domain-safe automaton, in which there is no need to monitor
the input domain when playing, see \cref{thm:domain-safe}. 

\subparagraph{Related works.} 
Boolean synthesis problems for finite words have been considered
in~\cite{DBLP:conf/ijcai/ZhuTLPV17,DBLP:conf/aaai/LiRPZV19} where the specification is given as an LTL formula over
finite traces. In the quantitative setting, it has also been
considered in~\cite{DBLP:conf/concur/FiliotGR12} for weighted specifications given by
deterministic weighted automata. In these works however, it is the
role of Eve to eventually stop the game. While this makes sense for
reachability objectives and planning problems, this setting does not
accurately model a synthesis scenario where the system has no control
over the provided input sequence.
Our setting is different and needs new technical developments.

Threshold problems in quantitative infinite-duration
two-player games with discounted- and mean-payoff measures are
known to be solvable in
$\textsc{NP}\cap\textsc{coNP}$~\cite{andersson2006improved,zwick1996complexity}. Our threshold
synthesis problems all directly reduce to critical prefix threshold
games with corresponding payoff functions. The latter games, for sum
and average, are shown to reduce to mean-payoff games, so our 
$\textsc{NP}\cap\textsc{coNP}$ upper-bound follows from~\cite{zwick1996complexity}. 
For critical prefix discounted-sum games with a non-strict threshold,
we show a polynomial time reduction to infinite-duration discounted
sum games and hence our result follows from~\cite{andersson2006improved}. Such a reduction
fails for a strict threshold and we develop new techniques to solve
critical prefix discounted-sum games with strict threshold, by first
showing that memoryless strategies suffice for Eve to win, and then by
showing how to check in \textsc{PTime} whether a memoryless strategy
is winning for Eve. The latter result actually shows how to test in
\textsc{PTime} whether there exists, in a weighted graph, a path from a
source to a target vertex of discounted-sum greater or equal to some
given threshold. This result entails that the non-emptiness problem
for non-deterministic discounted-sum
max-automata\footnote{i.e., checking whether there exists a word with
  value greater or equal to some threshold, where the value is defined
by taking the max over all accepting runs.} is solvable in
\textsc{PTime} (\cref{thm:non-emptiness-dsum-automata}). To the best of our
knowledge, up to now this problem is only known to be in \textsc{PSpace} for the subcase of functional discounted-sum automata~\cite{DBLP:conf/concur/FiliotGR12,DBLP:conf/lics/BokerHO15}. 

As we show, the best-value synthesis problems correspond to 
zero-regret determinization problems for non-deterministic weighted automata, i.e.,
deciding whether there is a non-determinism resolving strategy for Eve
that guarantees the same value as the maximal value of an accepting run in the non-deterministic weighted automaton. Such a problem is in \textsc{PTime} for sum-automata~\cite{DBLP:journals/talg/AminofKL10} and the average
case easily reduces to the sum-case. For discounted-sum, zero-regret
determinization is known to be decidable in \textsc{NP} for
dsum-automata over infinite words~\cite{DBLP:conf/csl/HunterPR16}. We
improve this bound to $\textsc{NP}\cap\textsc{coNP}$ for finite words.

Finally, approximate synthesis corresponds to a problem known as
$r$-regret determinization of non-deterministic weighted automata. For
sum-automata, it is known to be \textsc{ExpTime}-complete~\cite{DBLP:conf/lics/FiliotJLPR17}. For
average-automata, there is no immediate reduction to the sum case,
because the sum value computed by an $r$-regret determinizer can be arbitrarily
faraway from the best sum, while its averaged value remains close to
the best average. Instead, we show a reduction to the new class of partial
observation critical prefix energy games. For dsum-automata
over infinite words, total domain and integral discount factor,
$r$-regret determinization is known to be
decidable~\cite{DBLP:conf/csl/HunterPR16}. Our setting does not
directly reduce to this setting, but we use similar ideas. 

\section{Preliminaries}
\label{sec:prelims}

\subparagraph{Languages and relations.}
Let $\mathbbm N$ be the set of non-negative integers.
Let $\Sigma$ be a finite alphabet. We denote by $\Sigma^*$, respectively $\Sigma^\omega$, the set of finite, respectively infinite, words over $\Sigma$, and $\Sigma^+$ the set of non-empty finite words over $\Sigma$.
The empty word is denoted by $\varepsilon$.
A \emph{language} over $\Sigma$ is a set of words over $\Sigma$.
A (binary) \emph{relation} $R$ is a subset of $\SigmaI^* \times
\SigmaO^*$, i.e., a set of pairs of words. Its domain is the set
$\mathrm{dom}(R) = \{ u \mid \exists v\colon(u,v) \in R \}$. Given a pair of words, we refer to the first (resp.\ second) component as input (resp.\ output) component, the alphabets $\SigmaI$ and $\SigmaO$ are referred to as input resp.\ output alphabet.
We let $\SigmaIO = \SigmaI \cup \SigmaO$.

\subparagraph{Automata.}
A \emph{nondeterministic finite state automaton (NFA)} is a tuple $\mathcal A = (Q,q_i,\Sigma,\Delta,F)$, where $Q$ is a finite state set, $q_i \in Q$ is the initial state, $\Sigma$ is a finite alphabet, $\Delta \subseteq Q \times \Sigma \times Q$ is a transition relation, and $F \subseteq Q$ is a set of final states.
A \emph{run} of the automaton on a word $w = a_1\dots a_n$ is a sequence $\rho = \tau_1\dots \tau_n$ of transitions such that there exist $q_0,\dots,q_n \in Q$ such that $\tau_j = (q_{j-1},a_j,q_j)$ for all $j$.
A run on $\varepsilon$ is a single state.
A run is \emph{accepting} if it begins in the initial state and ends in a final state.
The \emph{language recognized} by the automaton is defined as $L(\mathcal A) = \{ w \mid \textnormal{there is an accepting run of $\mathcal A$ on $w$}\}$.
The automaton is \emph{deterministic (a DFA)} if $\Delta$ is given as a partial function $\delta: Q \times \Sigma \to Q$.

\subparagraph{Transducers.}
A \emph{transducer} is a tuple $\mathcal T = (Q,q_i,\SigmaI,\SigmaO,\delta,F)$, where $Q$ is a finite state set, $q_i \in Q$ is the initial state, $\SigmaI$ and $\SigmaO$ are finite alphabets, $\delta\colon \bigl(Q \times \SigmaI\bigr) \to \bigl(\SigmaO \times Q\bigr)$ is a transition function, and $F \subseteq Q$ is a set of final states.
A transition is also denoted as a tuple for convenience.
A \emph{run} is either a non-empty sequence of transitions $\rho = (q_0,u_1,v_1,q_1)(q_1,u_2,v_2,q_2)\dots (q_{n-1},u_n,v_n,q_n)$ or a single state.
The \emph{input (resp.\ output)} of $\rho$ is $u = u_1\dots u_n$ (resp.\ $v = v_1\dots v_n$) if $\rho \in \Delta^+$, both are $\varepsilon$ if $\rho \in Q$.
We denote by $p \xrightarrow{u|v} q$ that there exists a run from $p$ to $q$ with input $u$ and output $v$.
A run is \emph{accepting} if it starts in the initial and ends in a final state.
The \emph{partial function recognized} by the transducer is $f_\mathcal T\colon \SigmaI^* \to \SigmaO^*$ defined as $f_\mathcal T(u) = v$ if there is an accepting run of the form $p \xrightarrow{u|v} q$.

\subparagraph{Weighted automata.}
Let $n > 0$. Given a finite sequence $\phi = j_1\dots j_n$ of integers, and a discount factor $\lambda \in \mathbbm Q$ such that $0 < \lambda < 1$, we define the following functions:
$\SUM(\phi) = \sum_{i=1}^n j_i, \quad \AVG(\phi) = \frac{\SUM(\phi)}{n}, \quad \DSUM(\phi) = \sum_{i=1}^n \lambda^i j_i$ if $\phi$ is non-empty and $\SUM(\phi) = \AVG(\phi) = \DSUM(\phi) = 0$ otherwise.
Let $V \in \{\SUM,\AVG,\DSUM\}$.
A \emph{weighted $V$-automaton (WFA)} is a tuple $\mathcal A = (Q,\Sigma,q_i,\Delta,F,\gamma)$, where $(Q,\Sigma,q_i,\Delta,F)$ is a classical \emph{deterministic} finite state automaton, and $\gamma\colon \delta \to \mathbbm Z$ is a \emph{weight function}.
Its recognized language, etc., is defined as for classical finite
state automata. %
The value $V(\rho)$ of a run $\rho = \tau_1\dots \tau_n$ is defined as $V(\gamma(\tau_1)\dots \gamma(\tau_n))$ if $\rho$ is accepting and $-\infty$ otherwise.
The value $\mathcal A(w)$ of a word $w$ is given by the total function, called the function \emph{recognized} by $\mathcal A$, $\mathcal A\colon \Sigma^* \to \mathbbm Q \cup \{-\infty\}$ defined as $w \mapsto 
V(\rho)$, where $\rho$ is the run of $\mathcal A$ on $w$, that is, the value of a word is the value of its accepting run, or $-\infty$ if there exists none.

\subparagraph{Weighted specifications.}
A \emph{weighted specification} is a total function $S \colon (\SigmaI\SigmaO)^* \to \mathbbm Q \cup \{-\infty\}$ recognized by a WFA $\mathcal A$.
Note that by our definition, $\mathcal A$ is deterministic by default.
Given $u = u_1\dots u_n \in \SigmaI^*$ and $v = v_1\dots u_n \in \SigmaO^*$, $u \otimes v$ denotes its \emph{convolution} $u_1v_1\dots u_nv_n \in (\SigmaI\SigmaO)^*$.
We usually write $S(u \otimes v)$ instead of $S(u_1v_1\dots u_nv_n)$.
The relation (or Boolean specification) of $S$, denoted by $R(S)$, is given by the set of pairs that are mapped to a rational number, i.e., $R(S) = \{ (u,v) \mid S(u \otimes v) > -\infty\}$.
We usually write $u \otimes v \in S$ instead of $(u,v) \in R(S)$.
The domain of $S$, denoted by $\mathrm{dom}(S)$, is defined as $\{ u \in \SigmaI^* \mid \exists v\in\SigmaO^*\colon u \otimes v \in S\}$.
If a weighted specification is given by some $V$-automaton, we refer to it as $V$-specification.

\subparagraph{Quantitative synthesis problems.}\label{subsec:synthdef}
The (Boolean) \emph{synthesis problem} asks, given a weighted specification $S$, whether there exists a partial function $f\colon \SigmaI^* \to \SigmaO^*$ defined by a transducer  with $\mathrm{dom}(f) = \mathrm{dom}(S)$ such that $u \otimes f(u) \in S$ for all $u \in \mathrm{dom}(f)$.

We define three quantitative synthesis problems that pose additional conditions, we only state the additions.
The \emph{threshold synthesis problem} additionally asks, given a threshold $\nu \in \mathbbm Q$, and ${\triangleright \in \{>,\geq\}}$, that $S(u \otimes f(u)) \triangleright \nu$ for all $u \in \mathrm{dom}(f)$.
The \emph{best-value synthesis problem} additionally asks that $S(u \otimes f(u)) = \bestVal_S(u)$, where $\bestVal_S(u) = \mathrm{sup}\{S(u \otimes v) \mid u \otimes v \in S\}$ for all $u \in \mathrm{dom}(f)$.
The \emph{approximate synthesis problem} additionally asks, given a threshold $\nu \in \mathbbm Q$, and $\triangleleft \in \{<,\leq\}$, that $\bestVal_S(u) - S(u \otimes f(u)) \triangleleft \nu$ for all $u \in \mathrm{dom}(f)$.

In these settings, if such a function $f$ exists, it is called \emph{$S$-realization}, a transducer that defines $f$ is called \emph{$S$-realizer}, and is said to \emph{implement an $S$-realization}.
A transducer whose implemented function $f$ only satisfies the Boolean condition is called \emph{Boolean $S$-realizer}.

\begin{example}
 Let $\SigmaI=\{a,b\}$ and $\SigmaO = \{c,d\}$, and consider the
 weighted specification $S$ defined by the following automaton $\mathcal A$.
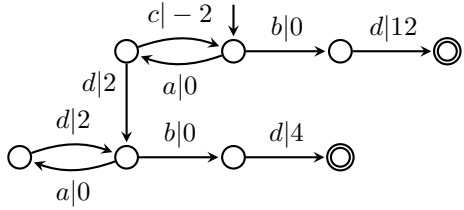
\begin{center}
  \qquad 
  \begin{tikzpicture}[thick,scale=1,every node/.style={scale=1}]
    \centering
     \tikzstyle{every state}+=[inner sep=3pt, minimum size=5pt];
     \node[state, initial above] (0) {};
     \node[state, right of=0] (1) {};
     \node[state, right of=1, accepting] (2) {};
     \node[state, left of=0] (3) {};
     \node[state, below of=3] (4) {};
     \node[state, left of=4] (5) {};
     \node[state, right of=4]  (6) {};
     \node[state, right of=6, accepting]  (7) {};
     
     \draw[->] (0) edge[bend left=20] node {$a|0$} (3);
     \draw[->] (3) edge[bend left=20] node {$c|-2$} (0);
     \draw[->] (4) edge[bend left=20] node {$a|0$} (5);
     \draw[->] (5) edge[bend left=20] node {$d|2$} (4);
     \draw[->] (0) edge node {$b|0$} (1);
     \draw[->] (1) edge node {$d|12$} (2);
     \draw[->] (3) edge node[swap,near start] {$d|2$} (4);
     \draw[->] (4) edge node {$b|0$} (6);
     \draw[->] (6) edge node {$d|4$} (7);
  \end{tikzpicture}
\end{center}
 Clearly, $S$ has a Boolean realizer (infinitely many, in fact).
 First, we view $\mathcal A$ as a \SUM-automaton.
 There exists a realizer that ensures a value of at least $6$, for example, the transducer that always outputs $d$. %
 There exists no best-value realizer.
 To see this, we look at the maximal values.
 We have $\bestVal(b) = 12$, $\bestVal(ab)=10$, and $\bestVal(a^ib) = 2i+4$ for $i > 1$.
 The maximal value for $ab$ is achieved with $cd$ and the maximal value for $aaab$ with $dddd$.
 So, the first output symbol depends on the length of the input word,
 which is unknown to a transducer when producing the first
 output symbol. %
 However, there exists an approximate realizer for the non-strict threshold $4$: the transducer that outputs $c$ solely for the first $a$.
 The difference to the maximal value is $0$ for the inputs $b$ and
 $ab$, and $4$ for all other inputs. %
 Secondly, we view $\mathcal A$ as an \AVG-automaton.
 With the same argumentation as for \SUM, it is easy to see that there exists no best-value realizer, there exists an approximate realizer for the non-strict threshold $\frac{2}{3}$:
 the transducer that outputs $c$ solely for the first $a$.
 The difference to the maximal value is $0$ for the inputs $b$ and $ab$, and $\frac{2}{i+1}$ for inputs of the form $a^ib$ for $i>1$.
 Note that the difference decreases with the input length unlike for \SUM.
\end{example}

\subparagraph{Boolean synthesis and domain-safe automata.}\label{sec:domain-safe}
The quantitative synthesis problems that we have defined, ask for
Boolean realizers that additionally satisfy a quantitative
condition. We start by showing that a weighted specification $\mathcal
A$ can be preprocessed in polynomial time such that dealing with the
Boolean part becomes very simple. Basically, we remove all parts of
$\A$ that cannot be used by a Boolean realizer. We call the result
of this preprocessing a \emph{domain-safe weighted specification}, to
be defined formally below. In \cref{subsec:threshold-synthesis}
we use domain-safe specifications.

Denote by $\dom(\A) \subseteq \SigmaI^*$ the domain of the weighted specification defined by $\A$. We
can easily obtain an NFA (with $\varepsilon$-transitions) for $\dom(\A)$
by removing the weights and turning all transitions that are labelled
by an output letter into an $\varepsilon$-transition. We call the
resulting NFA the domain automaton of $\A$, and denote it by $\Adom$.
For a state $q$ of $\A$, we denote by $L(\Adom,q)$ the language of
$\Adom$ accepted by runs starting in $q$.
An output transition $(q,a,q')$ of $\A$ is called \emph{domain-safe}
if $L(\Adom,q) = L(\Adom,q')$, i.e., it
does not restrict the language of input words that can be
accepted by $\Adom$. Otherwise, such a transition is called
\emph{domain-unsafe}. We call a weighted specification $\A$ \emph{domain-safe} if it is trim, i.e., all states are accessible and co-accessible,
and all its output transitions are domain-safe.

A transducer that produces an input/output pair
whose run in $\A$ uses a domain-unsafe transition of $\A$ cannot be a Boolean
realizer of $\A$ because it cannot complete all inputs in the
domain with an output in the relation $R(\mathcal A)$.
We now show that we can compute in polynomial time for a given
weighted specification $\A$ a sub-automaton $\A'$ of $\A$ that is
domain-safe and has the same Boolean
realizers as $\A$.
We would like to mention that there is a tight connection between
domain-safe automata and the problem of ``determinization by pruning''
(DBP) as it is studied in \cite{DBLP:journals/talg/AminofKL10}. The
following result can also be derived from the proof of
\cite[Theorem~4.1]{DBLP:journals/talg/AminofKL10}.
Furthermore, the proof of \cref{thm:domain-safe} directly yields an alternative game-based proof of the ``determinization by pruning'' problem.

\begin{restatable}{theorem}{thmdomainsafe}
\label{thm:domain-safe}
There is a polynomial time procedure that takes as input a weighted
specification $\A$, and either returns ``no realizer''
if $\A$ does not have Boolean realizers, or, otherwise,
returns a sub-automaton $\A'$ of $\A$ that is domain-safe, has the same
domain as $\A$, and has the same Boolean
realizers as $\A$.
\end{restatable}

A direct consequence of the above theorem is that the Boolean synthesis problem is decidable in polynomial time. 

\section{Critical prefix games}
\label{sec:games}

In this section we introduce the necessary definitions and notations regarding games.
Moreover, we introduce critical prefix games and establish our results for these kind of games.

\subparagraph{Games.}

A \emph{weighted game with imperfect information} is an infinite-duration two-player game played on a game arena $G = (V,v_0,A,E,\Obs,w)$, where $V$ is a finite set of vertices, $v_0 \in V$ is the initial vertex, $A$ is a finite set of actions, $E \subseteq V \times A \times V$ is a labeled transition relation, $\Obs \subseteq 2^V$ is a set of \emph{observations} that partition $V$, and $w\colon E \to \mathbbm Z$ is a weight function.
Without loss of generality, we assume that the arena has no dead ends, i.e., for all $v \in V$ there exists $a \in A$ and $v' \in V$ such that $(v,a,v') \in E$.
The unique observation containing a vertex $v$ is denoted $\obs(v)$.
A game with \emph{perfect information} is such that $\Obs = \{ \{v\} \mid v \in V\}$.
In that case we omit $\Obs$ from the tuple $G$.

Games are played in rounds in which Eve chooses an action $a \in A$, and Adam chooses an $a$-successor of the current vertex.
The first round starts in the initial vertex $v_0$.
A \emph{play} $\pi$ in $G$ is an infinite sequence $v_0a_0v_1a_1\dots$ such that $(v_i,a_i,v_{i+1}) \in E$ for all $i \in \mathbbm N$.
The prefix of $\pi$ up to $v_n$ is denoted $\pi(n)$, its last element $v_n$ is denoted by $\last(\pi(n))$.
The set of all plays resp.\ prefixes of plays in $G$ is denoted by
$\plays(G)$ resp.\ $\prefs(G)$. %
The \emph{observation sequence} of the play $\pi$ is defined as $\obs(\pi) = \obs(v_0)a_0\obs(v_1)a_1\dots$ and the finite observation sequence of the play prefix $\pi(n)$ is $\obs(\pi(n)) = \obs(v_0)a_0\dots \obs(v_n)$.
Naturally, $\obs$ extends to sets of (prefixes of) plays.

A game is defined by an arena $G$ and an \emph{objective} $\Win
\subseteq \plays(G)$ describing a set of good plays in $G$ for Eve. 
A \emph{strategy for Eve} in $G$ is a mapping $\sigma\colon \prefs(G) \to A$, it is called \emph{observation-based} if for all play prefixes $\rho,\rho' \in \prefs(G)$, if $\obs(\rho) = \obs(\rho')$, then $\sigma(\rho) = \sigma(\rho')$.
Equivalently, an observation-based strategy is a mapping $\sigma\colon \obs(\prefs(G)) \to A$.
We do not formally introduce strategies for Adam, intuitively, given a play prefix and an action $a$, a strategy of Adam selects an $a$-successor of its last vertex.
Given a strategy $\sigma$, let $\plays_\sigma(G)$ denote the set of plays compatible with $\sigma$ in $G$, and $\prefs_\sigma(G)$ denote the set of play prefixes of $\plays_\sigma(G)$.
An Eve's strategy $\sigma$ in $G$ is \emph{winning} if $\plays_\sigma(G) \subseteq \Win$.

We now define quantitative objectives. The \emph{energy level} of the play prefix $\pi(n)$ is $\EL(\pi(n)) =
\sum_{i = 1}^n w((v_{i-1},a_{i-1},v_i))$, the \emph{sum value} is
$\SUM(\pi(n)) = \sum_{i = 1}^n w((v_{i-1},a_{i-1},v_i))$, the
\emph{average value} is $\AVG(\pi(n)) = \frac{1}{n}\SUM(\pi(n))$, and
the \emph{discounted-sum value} is $\DSUM(\pi(n)) = \sum_{i = 1}^n
\lambda^i w((v_{i-1},a_{i-1},v_i))$, and we let $\DSUM(\pi) = \sum_{i = 1}^\infty \lambda^i w((v_{i-1},a_{i-1},v_i))$ (we do not explicitly mention the discount factor $\lambda$ in this notation because it is always clear from the context).
%
%

The \emph{energy objective} in $G$ is parameterized by an initial credit $c_0 \in \mathbbm N$ and is given by $\PosEn_G(c_0) = \{ \pi \in \plays(G) \mid \forall i \in \mathbbm N\colon c_0 + \EL(\pi(i)) \geq 0\}$.
It requires that the energy level of a play never drops below zero when starting with initial energy level $c_0$.
The \emph{fixed initial credit problem for imperfect information
  games} asks whether there exists an observation-based winning
strategy for Eve for the objective $\PosEn_G(c_0)$. %
%
%
The \emph{discounted-sum objective} in $G$ is parameterized by a threshold $\nu\ \in \mathbbm Q$, and $\triangleright\ \in \{>,\geq\}$.
It is given by $\DS_G^\triangleright(\nu) =  \{ \pi \in \plays(G) \mid \DSUM(\pi) \triangleright \nu\}$ and requires that the discounted-sum value of a play is greater than resp.\ at least $\nu$.
The \emph{discounted-sum game problem} asks whether there exists a winning strategy for Eve for the objective $\DS_G^\triangleright(\nu)$.


A game with perfect information is a special case of an imperfect information game.
Classically, instead of using the above model with full observation, a (weighted) perfect information game, simply called game, is defined over an arena $(V,V_{\exists},v_0,E,w)$, where the set of vertices $V$ is partitioned into $V_{\exists}$ and $V\setminus V_{\exists}$, the vertices belonging to Eve and Adam, respectively, $v_0 \in V$ is the initial vertex, $E \subseteq V \times V$ is a transition relation, and $w\colon E \to \mathbbm Z$ is a weight function.
In a play on such a game arena, Eve chooses a successor if the current vertex belongs to her, otherwise Adam chooses.
For games with perfect information the two models are equivalent and we shall use both.

\subparagraph{Critical prefix games.}

A \emph{critical prefix game} is a game, where the winning objective is parameterized by a set $C \subseteq V$ of \emph{critical vertices}, and a set of play prefixes $W \subseteq \prefs(G)$.
Its objective is defined as $\crit_{C,W}(G) = \{ \pi \in \plays(G) \mid \forall i\  \last(\pi(i)) \in C\rightarrow \pi(1)\dots\pi(i) \in W\}$.
The idea of a \prefix game is that the state of a play is only relevant whenever the play is in a critical vertex.
For convenience, in the case of \prefix games, we also refer to the set $W$ as objective.

The \emph{threshold problem for \prefix games} asks whether there exists a winning strategy for Eve for the objective $\crit_{C,W}(G)$, where $W$ is of the form $\Thres_{G}^{V\triangleright}(\nu) = \{ \varphi \in \prefs(G) \mid V(\varphi) \triangleright \nu \}$ parameterized by a threshold $\nu \in \mathbbm Q$, $\triangleright \in \{>,\geq\}$, and $V \in \{\SUM,\AVG,\DSUM\}$.

The \emph{initial credit problem for \prefix imperfect information energy games} asks whether there exists an observation-based winning strategy for Eve for the objective $\crit_{C,W}(G)$, where $W$ is of the form $\prefixPosEn_G(c_0) = \{ \varphi \in \prefs(G) \mid c_0 + \EL(\varphi) \geq 0\}$ parameterized by an initial credit $c_0 \in \mathbbm N$.

\begin{restatable}{theorem}{thmthresholdgamesall}
\label{thm:threshold-games}
 The threshold problem for \prefix games for $V \in \{\SUM,\AVG\}$ and
 a strict or non-strict threshold is decidable in $\textsc{NP} \cap
 \textsc{coNP}$.
 Moreover, positional strategies are sufficient for
 Eve to win such games. 
\end{restatable}

\begin{proof}[Proof sketch.]
 For \SUM and \AVG and a strict or non-strict threshold, the critical prefix threshold games reduce to mean-payoff games which are solvable in $\textsc{NP} \cap
 \textsc{coNP}$ \cite{zwick1996complexity}.
 Positional strategies suffice for mean-payoff games, a winning strategy in the constructed mean-payoff game directly yields a positional winning strategy in the critical prefix threshold game.
\end{proof}

\begin{restatable}{theorem}{thmthresholddsumgamesall}\label{thm:threshold-games-dsum-all}
  The threshold problem for \prefix games for $\DSUM$ and a strict resp.\ non-strict threshold is decidable in $\textsc{NP}$ resp.\ $\textsc{NP} \cap
  \textsc{coNP}$.
  Moreover, positional strategies are sufficient for
  Eve to win such games. 
 \end{restatable}

To prove the above theorem, we first show a result on weighted graphs which is interesting in itself.

\begin{restatable}{lemma}{lempathchecking}\label{lem:pathchecking}
    Given a weighted graph $G$, a source vertex $v_0\in V$, a target
    set $T\subseteq V$ and a threshold $\nu\in\mathbbm{Q}$, checking
    whether there exists a path $\pi$ from $v_0$ to some vertex $v\in
    T$ such that $\DSUM(\pi) \leq \nu$ can be done in \textsc{Ptime}.
\end{restatable}

  \cref{lem:pathchecking} can be used to show that the $\geq
  \nu$-non-emptiness problem for nondeterministic discounted-sum automata\footnote{In contrast to deterministic weighted automata, there might be serveral accepting runs on an input and the value of the word is defined as the maximal value of its accepting runs \cite{DBLP:conf/concur/FiliotGR12,haddad2015value}.}
  can be checked in \textsc{Ptime}, a result which is, to the best
  of our knowledge, new. It was known to be in \textsc{PSpace} for
  unambiguous discounted-sum
  automata~\cite{DBLP:conf/concur/FiliotGR12,DBLP:conf/lics/BokerHO15}. This
  problem asks for the existence of a word of value greater or equal
  than a given threshold $\nu$. Since the value of a word is the
  maximal value amongst its accepting runs, it suffices to check for
  the existence of a run from the initial state to an accepting
  state of discounted-sum value $\geq \nu$. By inverting the
  weights, the latter is equivalent to checking whether there exists
  a run from the initial state to an accepting
  state of discounted-sum value $\leq -\nu$. By seeing the
  (inverted) discounted-sum automaton as a weighted graph, the
  latter property can be checked in \textsc{Ptime} by
  \cref{lem:pathchecking}, thus proving the following theorem.  

\begin{theorem}\label{thm:non-emptiness-dsum-automata}
The $\geq \nu$ non-emptiness problem is decidable in
\textsc{PTime} for nondeterministic discounted-sum automata. 
\end{theorem}

We now go back to the proof of \cref{thm:threshold-games-dsum-all}.

\begin{proof}[Proof sketch of \cref{thm:threshold-games-dsum-all}]
  For \DSUM, and a non-strict threshold, the problem can be directly reduced to discounted-sum games which are solvable in $\textsc{NP} \cap
  \textsc{coNP}$ \cite{andersson2006improved}.

  For \DSUM, and a strict threshold, such a reduction fails.
  To solve the problem, we first show that positional strategies are sufficient for Eve to win in a critical prefix threshold discounted-sum game (for strict and non-strict thresholds).
    The NP-algorithm guesses a positional strategy $\sigma$ for Eve, and
    then verifies in polynomial time whether $\sigma$ is winning. Let
    $G'$ be the game restricted to Eve's  $\sigma$-edges, seen as a
    weighted graph. 
    The strategy $\sigma$ is not winning iff Adam can form a path
    in $G'$ from the initial vertex to a critical vertex that has weight
    $\le \nu$. This property can be checked in \textsc{Ptime} thanks to
    \cref{lem:pathchecking} (by taking as target set the set of
    critical vertices). 
\end{proof}

The following is shown by reduction from the halting problem for 2-counter machines.

\begin{restatable}{theorem}{thmprefixenergygamesundec}
\label{thm:prefix-energy-games-undec}
 The fixed initial credit problem for imperfect information \prefix energy games is undecidable. 
\end{restatable}

The above result contrasts the fixed initial credit problem for imperfect information energy games which is decidable \cite{degorre2010energy}.

\begin{restatable}{theorem}{thmprefixenergyreachable}
\label{thm:prefix-energy-reachable}\label{cor:fin-mem}
  The fixed initial credit problem for imperfect information \prefix
  energy games is decidable if from each vertex Adam has a strategy to
  reach a critical vertex against observation based strategies. Moreover, finite-memory strategies are sufficient for Eve
  to win.
\end{restatable}

\begin{proof}[Proof sketch.]
This problem is reduced to the fixed initial credit problem for imperfect information energy games which is decidable \cite{degorre2010energy}.
In classical energy games, Eve loses as soon as the energy goes below zero.
The idea of the reduction is that if in the critical prefix energy game the initial credit is $c_0$, then in the classical energy game we start the game with an additional buffer, i.e., with $c_0 + B$, for some computable bound $B$.
In the critical prefix energy game, if the energy level drops below $-B$ Adam can force to visit a critical vertex such that the energy level can rise by at most $B$, ensuring that a critical vertex is visited with energy level below zero.
Thus, the additional buffer $B$ suffices in the classical energy game.
\end{proof}

\section{Synthesis problems}
\label{sec:syntesis-problems}

Here, we solve the quantitative synthesis problems defined
in \cref{sec:prelims}. Recall that weighted specifications are given by weighted automata
that alternate between reading one input and one output symbol. In other words, we prove the
decidability results of \cref{table:summary}. We then show consequences of
these results to quantitative synthesis problems over infinite words.

\subparagraph{Threshold synthesis problems.}\label{subsec:threshold-synthesis}

Since weighted specifications $S$ are given by 
weighted automata, the synthesis problem naturally reduces
to a game played on the automaton. 
In order to solve threshold synthesis problems, in contrast to best-value and approximate synthesis problems, it is not necessary to compare the values of runs of the specification automaton that have the same input sequence.
Hence, it is relatively straightforward to reduce threshold synthesis problems to critical prefix threshold games.
An important point needs to be 
taken care of due to the fact the domain of $S$ might be
partial, and therefore lead Eve into the following bad situation $(\star)$:
Eve must choose her outputs in such a way that she does not go
in a state of the automaton which is non-accepting, while the input
word played by Adam so far is in the domain of $S$. Otherwise, the
pair of input and output word formed would not even be in
$S$, something which is required by the definition of synthesis
problems. So, Eve has to
monitor the domain, which is easy if the domain is total, but more involved if it is partial.
Thanks to \cref{thm:domain-safe}, this can be done in
polynomial time. More precisely, we first run the algorithm of
\cref{thm:domain-safe} which either returns that there is no Boolean realizer,
or returns a domain-safe deterministic weighted automaton $\mathcal A'$ which
has the same Boolean realizers as $S$. By
the very definition of domain-safe automata, the bad situation $(\star)$
described above cannot happen. Hence, Eve can freely play on $\mathcal A'$
without taking care of the domain constraint. Only the quantitative
constraint matters, and it has to be enforced whenever Eve is in an
accepting state of $\mathcal A'$ (this corresponds to the situation where Adam
has chosen an input word in the domain of $S$). Hence, only accepting
states of $\mathcal A'$ matter for the quantitative constraint and these are declared
as critical. To conclude, by projecting away the symbols of $\mathcal A'$
and by declaring its accepting states to be critical, we obtain a
critical prefix game. For the threshold synthesis problem, decidability follows directly from the decidability
of the threshold problem for critical prefix games
(\cref{thm:threshold-games,thm:threshold-games-dsum-all}). For \SUM- and \AVG-specifications, this can be done
in $\textsc{NP} \cap \textsc{coNP}$. We leave open whether it is
solvable in \textsc{Ptime} and show that this would also solve the long
standing open problem of whether mean-payoff games are solvable in
\textsc{Ptime}. 

\begin{restatable}{theorem}{thmdeterministicthreshold}
\label{thm:deterministic-threshold}
The threshold synthesis problem for a $V$-specification with $V \in
\{\SUM,\AVG\}$ and a strict or non-strict threshold is decidable in
$\textsc{NP} \cap \textsc{coNP}$ and \textsc{PTIME}-equivalent to
mean-payoff games. The threshold synthesis problem for a \DSUM-specification and a strict resp.\ non-strict threshold is decidable in \textsc{NP} resp.\ in $\textsc{NP} \cap \textsc{coNP}$.
\end{restatable}

\subparagraph{Synthesis and regret determinization.}
Before we prove our results about best-value and approximate synthesis, 
we highlight the tight connection between the approximate synthesis problem and the so-called regret determinization problem for nondeterministic weighted automata\footnote{In contrast to deterministic weighted automata, there might be serveral accepting runs on an input and the value of the word is defined as the maximal value of its accepting runs \cite{DBLP:conf/concur/FiliotGR12,haddad2015value}.}.
This problem has
for instance been studied in~\cite{DBLP:conf/lics/FiliotJLPR17} for
\SUM-automata and in~\cite{DBLP:conf/csl/HunterPR16} for \DSUM-automata. We formalize this connection here. 
Given $r \in \mathbbm Q$ and $\triangleleft \in \{<,\leq\}$, a
nondeterministic WFA $\mathcal A = (Q,\Sigma,q_i,\Delta,F,\gamma)$ is
called \emph{$r_\triangleleft$-regret determinizable} if there exists
a finite set of memory states $M$ and a deterministic WFA $\mathcal
A_r = (Q \times M,\Sigma,q_i^r,\Delta_r,F_r,\gamma_r)$, where
$q_i^r = (q_i,m)$ for some $m \in M$, $F_r \subseteq F \times M$,
$\bigl((q,m),a,(q,m')\bigr) \in \Delta_r$ implies that $(q,a,q') \in
\Delta$, and $\gamma_r\left(\bigl((q,m),a,(q,m')\bigr)\right) =
\gamma((q,a,q'))$ for all $m,m' \in M$, such that $L(\mathcal A) =
L(\mathcal A_r)$ and $\mathcal A(w) - \mathcal A_r(w) \triangleleft r$
for all $w \in \mathrm{dom}(L(\mathcal A))$. The \emph{regret
  determinization problem} asks, given a nondeterministic weighted
automaton $\mathcal A$, a threshold $r \in \mathbbm Q$, and
$\triangleleft \in \{<,\leq\}$, whether $\mathcal A$ is
$r_\triangleleft$-regret determinizable. 

\begin{restatable}{lemma}{lemapproxtoregret}
\label{lem:approx-to-regret}
 The approx.\ synthesis problem for weighted specifications reduces
 in linear time to the regret determinization problem for nondet.\ weighted
 automata (with the same threshold). The converse is true (in linear
 time and with the same threshold) for \SUM-automata. 
\end{restatable}

\cref{lem:approx-to-regret} is independent from any payoff function.
Regarding the converse direction, when going from the regret determinization problem to the approximate synthesis problem, a transition (for an input symbol) must be translated into two transitions (adding an output symbol).
This step can cause difficulties depending on the used payoff function, e.g., \DSUM.

\subparagraph{Best-value synthesis problems.}
\label{sec:best-value}

Best-value synthesis is equivalent to zero-regret synthesis, which is,
by \cref{lem:approx-to-regret}, equivalent to zero-regret
determinization of weighted automata. In
\cite{DBLP:conf/lics/BokerHO15}, the authors showed that if a
\SUM-automaton is zero-regret determinizable, then  no memory states
are needed, i.e., a sub-automaton suffices. We give general sufficient
conditions on weighted finite automata (which hold for \SUM-, \AVG- and \DSUM-automata) under
which the latter result can be generalized. 

Let $V\colon \mathbbm Z^* \to \mathbbm Q$ be a payoff function.
A $V$-automaton defining a $V$-specification, where $V$ is applied to runs as usual, is called \emph{$\leq$-stable} if for all runs $\rho,\rho',\rho''$ such that the end state of $\rho$ is the beginning state of $\rho'$ and $\rho''$, $w' = u \otimes v'$, and $w'' = u \otimes v''$ for some $u \in \SigmaI^*$ and $v',v'' \in \SigmaO^*$, where $w'$ and $w''$ are the words associated to $\rho'$ and $\rho''$, respectively, holds that if $V(\rho') \leq V(\rho'')$ then $V(\rho\rho') \leq V(\rho\rho'')$.

\begin{restatable}{lemma}{lemgoodproperty}
  \label{lem:good-property}
   Given a weighted specification $S$ by a $\leq$-stable weighted automaton $\mathcal A$, if there exists a transducer that implements a best-value $S$-realization, then there exists a transducer that implements a best-value $S$-realization that is defined as a sub-automaton of $\mathcal A$.
  \end{restatable}

While the above lemma can be used to obtain our decidability results for best-value synthesis, we use other techniques to obtain the complexity results stated below.

\begin{restatable}{theorem}{thmbestvalue}
\label{thm:best-value}
 The best-value synthesis problem is decidable in \textsc{Ptime} for 
\SUM-specifications and \AVG-specifications, and in
\textsc{NP}$\cap$\textsc{coNP} for \DSUM-specifications.
\end{restatable}

\begin{proof}[Proof sketch]
 For \SUM, the problem reduces to the zero-regret determinization problem for
 \SUM-automata, see \cref{lem:approx-to-regret}, aka the
 determinization by pruning problem for \SUM-automata, known to be
 decidable in \textsc{Ptime} in
 \cite{DBLP:journals/talg/AminofKL10}. For \AVG, it easily reduces to
 \SUM by interpreting the \AVG-specification as a
 \SUM-specification. For \DSUM, we show that the problem reduces in
 \textsc{Ptime} to a
 critical prefix threshold game, for non-strict threshold, which is
 solvable in \textsc{NP}$\cap$\textsc{coNP} by \cref{thm:threshold-games-dsum-all}. 
\end{proof}

Alternatively, decidability for \DSUM can be obtained by reduction to
the zero-regret determinization problem for \DSUM-automata over
infinite words which was shown to be decidable in \textsc{NP} in
\cite[Theorem 6]{DBLP:conf/csl/HunterPR16}. However, our techniques allow us to get \textsc{NP}$\cap$\textsc{coNP}.

\subparagraph{Approximate synthesis problems.}
\label{sec:approx}

We now turn to the approximate synthesis problems and show its
decidability for \SUM and \AVG. We leave the
decidability status open for \DSUM, but nevertheless show
decidability for a large class, namely when the discount factor is of
the form $\frac{1}{n}$ for $n\in\mathbbm N$. Nondeterministic
\DSUM-automata in this class have been considered
in~\cite{DBLP:journals/corr/BokerH14} and shown to be determinizable. 

\begin{restatable}{theorem}{thmapproximate}
\label{thm:approximate}
 The approximate synthesis problem is 
\begin{itemize}
\item \textsc{EXPtime}-complete for \SUM-specifications and strict or
  non-strict thresholds;
\item decidable and \textsc{EXPtime}-hard for \AVG-specifications and strict or
  non-strict thresholds;
\item in \textsc{NEXPtime} (resp. \textsc{EXPtime}) for
  \DSUM-specifications with a discount factor $\lambda$ of the form
  $\frac{1}{n}$ with $n\in \mathbbm N$ and strict (resp.\
  non-strict) thresholds. 
\end{itemize}
\end{restatable}

\begin{proof}[Proof sketch]
    For \SUM, we reduce the problem to $r$-regret determinization of
    \SUM-automata, known to be \textsc{EXPtime}-complete, using the
    back-and-forth connection given by
    \cref{lem:approx-to-regret}. 

    For an \AVG-specifications $S$, it is worth noting that even though
    $r$-approximate synthesis reduces to $r$-approximate synthesis for
    \SUM when $r=0$, interpreting $S$ as a \SUM-specification, this
    reduction is wrong for $r>0$ in general. It is because in an
    \AVG-specification, Eve can deviate more and more from the best
    sum, while the average of this difference can stay low. We instead rely on a
    reduction to critical prefix energy games of imperfect information
    and fixed initial credit (which falls into the decidable subclass
    of \cref{cor:fin-mem}). Intuitively, in this game, Adam
    constructs a run $\rho$ on a pair of words $(u,v)$ and Eve constructs a
    run $\rho'$ on some $(u,v')$. She only sees $u$ and not $\rho$. The energy
    level of such a play is set to $\SUM(\rho') + |uv|\cdot r - \SUM(\rho)$
    and must be positive whenever Adam reaches an accepting
    state. \textsc{ExpTime}-hardness is perhaps the most
    technical result of the paper, and is a non-trivial 
adaptation of reduction from countdown games used to show
   \textsc{ExpTime}-hardness of the regret determinization of
    \SUM-automata~\cite{DBLP:conf/lics/FiliotJLPR17}. 

    Finally, for \DSUM, we use that by projecting away the
    output in the \DSUM-automaton defining the specification, we obtain a
    nondeterministic weighted automaton which is determinizable
    by~\cite{DBLP:journals/corr/BokerH14}.
    This allows us to reduce
    the problem to the threshold synthesis
    problem for \DSUM, which is decidable by \cref{thm:deterministic-threshold}.
    To obtain the complexity results, we first analyze the determinization procedure.
    It yields an automaton whose states are exponential in the number of states and polynomial in the weights of the nondeterministic one.
    Its weights are polynomial in the weights of the nondeterministic one.
    For a strict threshold, the claimed complexity bound follows directly from 
    \cref{thm:deterministic-threshold}.
    For a non-strict threshold, we use that critical prefix threshold games are reduced in polynomial time to discounted-sum games.
Using value iteration \cite{zwick1996complexity} to solve discounted-sum games yields the claimed complexity bound, because it runs in polynomial time in the size of the arena, logarithmic in the absolute maximal weight of the arena, and exponential in the representation of the discount factor, i.e., polynomial in the discount factor.
\end{proof}

\subparagraph{Infinite words and Church synthesis.}\label{sec:church}

An $\omega$-specification is a subset 
$S \subseteq (\SigmaI.\SigmaO)^\omega$. 
%
%
%
The (Church) synthesis problem asks to decide whether there exists a strategy to pick a correct output sequence given longer and longer prefixes of an infinite input sequence.
Formally, an $\omega$-specification $S$ is said to be \emph{realizable} if there exists a function $\lambda\colon \SigmaI^*\rightarrow \SigmaO$ such that for all $i_1i_2\dots\in \SigmaI^\omega$, it holds that $i_1\lambda(i_1)i_2\lambda(i_1i_2)i_3\lambda(i_1i_2i_3)\dots\in S.$

Strategies of interest are those which can be represented by
a finite-state machine, and in particular a Mealy machine, that is, roughly, a transducer running on
$\omega$-words and without acceptance condition. Formally, it is a
tuple $M = (P,p_0,\delta)$ such that $P$ is a finite set of states
with initial state $p_0$, and $\delta \colon P\times \SigmaI\rightarrow
\SigmaO\times P$ is a (total) transition function. The function
$\delta$ can be extended to $\delta^* \colon P\times \SigmaI^+\rightarrow
\SigmaO\times P$ as usual. Then, $M$ defines the strategy
$\lambda_M$ such that for all $u\in \SigmaI^*$, $\lambda_M(u) =
\pi_1(\delta^*(p_0,u))$, where $\pi_1$ is the first projection. 
It is well-known that when $S$ is $\omega$-regular (given e.g.\ as a
parity automaton), it is decidable whether $S$ is
realizable~\cite{BuLa69}. Moreover, realizability implies realizability by a
Mealy machine. 




\subparagraph{Weighted safety specifications.}
 In this paper, we go beyond $\omega$-regular specifications, by considering
safety $\omega$-specifications induced by weighted specifications of finite words
defined by deterministic weighted automata. Let $W \colon (\SigmaI\SigmaO)^*\rightarrow
\mathbbm{Q}\cup \{-\infty\}$ be a weighted specification. 
For a
threshold $t\in\mathbbm{Q}$ and $\triangleright\in\{>,\geq\}$, we define the $\omega$-specification $\Thres^{\triangleright t}(W) = \{ i_1o_1\dots \in
(\SigmaI.\SigmaO)^\omega\mid \forall k\geq 0, i_1\dots i_k\in\dom(W)\rightarrow W(i_1o_1\dots i_ko_k)\triangleright t\}$.
In words, an $\omega$-word $w$ is
in $\Thres^{\triangleright t}(W)$ iff for all finite prefixes
$u=i_1o_1\dots i_ko_k$ of $w$, either $i_1\dots i_k\not\in\dom(W)$ or
$W(u)\triangleright t$. So, the quantitative condition is checked only for
prefixes whose input belongs to $\dom(W)$. The $\omega$-specification $\Thres^{\triangleright t}(W)$
is a safety specification~\footnote{A language of $\omega$-words $S$
  is a safety language if any $\omega$-word $w$
      whose finite prefixes $u$ are such that $uv_u\in S$ for some
      $\omega$-word $v_u$, belongs to
      $S$~\cite{ChangMP93}.}. More generally, any set  $S\subseteq (\SigmaI.\SigmaO)^*$ induces a
safety $\omega$-specification $\Safe(S)= \{ i_1o_1\dots \in
(\SigmaI.\SigmaO)^\omega\mid \forall k\geq 0, i_1\dots
i_k\in\dom(S)\rightarrow i_1o_1\dots i_ko_k\in S\}$.

For example, we have the equality $\Thres^{\triangleright t}(W) =
\Safe(\{ u\in(\SigmaI.\SigmaO)^* \mid W(u)\triangleright t\})$. Likewise, we define best-value and approximate safety
$\omega$-specifications. Formally, given a finite word $i_1\dots
i_k\in\SigmaI^*$ and $\triangleleft\in\{<,\leq\}$, we let $\BestVal(W) = \Approx^{\leq}(W,0)$ where for all $r\in
\mathbbm{Q}_{\geq 0}$ we have $\Approx^{\triangleleft}(W,r) =  \Safe(\{ u= i_1o_1\dots i_ko_k\mid
\bestVal_W(i_1\dots i_k) - W(u)\triangleleft r\}$.
Note that the three notions of safety $\omega$-specifications we have defined are not
necessarily $\omega$-regular, even if $W$ is given by a deterministic weighted
automaton. Nevertheless, an immediate consequence of the results we
have obtained previously on finite words is that

\begin{restatable}{theorem}{thmchurch}
\label{thm:church}
    The synthesis problem for an $\omega$-specification $O\subseteq
    (\SigmaI.\SigmaO)^\omega$ is decidable when $O$ is given by a
    deterministic $V$-automaton defining a weighted $V$-specification
    of finite words $W$ s.t. $ O\in \{\Thres^{> t}(W),\Thres^{\geq t}(W), \BestVal(W),
    \Approx^{<}(W,r),
    \Approx^{\leq}(W,r)\}$    and $V = \SUM$, $V = \AVG$ or $V = \DSUM$ with discount factor
    $1/n$ for $n\in\mathbbm{N}$. Moreover, if $O$ is realizable, it is
    realizable by a Mealy machine.
\end{restatable}


\section{Future work}
\label{sec:summary}

In this paper, weighted
specifications are defined by deterministic weighted
automata. Nondeterministic, even unambiguous, weighted automata, are strictly more
expressive than their deterministic variant in general, and in
particular for $\SUM$, $\AVG$ and $\DSUM$. An interesting direction is
to revisit our quantitative synthesis problems for specifications
defined by nondeterministic weighted automata. Using
similar ideas as the undecidability of critical prefix energy games of
imperfect information, it can be shown that threshold synthesis
becomes undecidable for unambiguous sum- and
avg-specifications. The problem is open for best-value and
approximate synthesis, and we plan to investigate it.

Two other directions seem
interesting as future work, both in the setting of infinite
words. First, natural measures in this setting are discounted-sum and
mean-payoff. While the threshold synthesis problems directly reduce to known
results and best-value/approximate  synthesis for dsum has
been studied in~\cite{DBLP:conf/csl/HunterPR16}, 
nothing is known to the best of our knowledge about
best-value/approximate synthesis for mean-payoff. We expect the
techniques to be different because such a measure is
prefix-independent, unlike our measures in the setting of finite words. As a second
direction, we have seen how our results apply to synthesis on infinite
words through weighted safety conditions. An
interesting direction is to consider such weighted requirements in conjunction
with $\omega$-regular conditions such as parity, in the line
of~\cite{DBLP:journals/tcs/ChatterjeeD12} that combines energy and
parity objectives in games.



\bibliographystyle{plain}
\bibliography{bib}

\appendix

\section{Proofs of \cref{sec:prelims} (Domain-Safe automata)}

\subsection{Proof of \cref{lem:domain-safe}}

Intuitively, the following lemma implies that 
if the specification $\A$ is domain-safe, then for an input $u$ it
suffices to follow some run in $\A$ in order to produce an output $v$
such that $(u,v) \in R(\A)$.

\begin{restatable}{lemma}{lemdomainsafe}
  \label{lem:domain-safe}
  If $\A$ is a domain-safe weighted specification, then each run of $\A$
  on $u \otimes v$ with $u \in \dom(\A)$ ends in a final state.
  \end{restatable}


\begin{proof}
Let $(x,y)$ be a pair of input/output words of equal length, such that
$\A$ has a run on $x \otimes y$ ending in some state $q$. By induction on
$|x|$ one can easily show that $L(\Adom,q) = x^{-1}\dom(\A) = \{ x' \mid xx' \in \dom(\A)\}$: For
$(x,y) = (\varepsilon, \varepsilon)$ the claim is obvious because
$L(\Adom,q_0) = \dom(\A)$. If $(x,y) = (x'a,y'b)$, let the run of $\A$
on $x' \otimes y'$ end in $q'$. By induction, $L(\Adom,q') =
(x')^{-1}\dom(\A)$. With $q'' = \delta(q',a)$, we obtain $L(\Adom,q'')
= a^{-1}L(\Adom,q') = a^{-1}((x')^{-1}\dom(\A)) = (x'a)^{-1}\dom(\A)$. The last transition of the run on $x'a \otimes y'b$ is $(q'',b,q)$. Since $\A$ is domain-safe, $L(\Adom,q) = L(\Adom,q'') = (x'a)^{-1}\dom(\A)$. This finishes the induction.

For $x \in \dom(\A)$, we obtain $\varepsilon \in x^{-1}\dom(\A) =
L(\Adom,q)$. Thus, $q$ has to be a final state.
\end{proof}

\subsection{Proof of \cref{thm:domain-safe}}

\thmdomainsafe*

\begin{proof}
  Consider the following game between Adam and Eve that tracks
  two runs of $\A$ (formalized below). The input word of the two runs
  is the same, and chosen by Adam. For the first run, the output
  sequence is played by Eve, and for the second run, the output
  sequence is played by Adam, who makes in each round his choice after
  the choice of Eve. Adam wins if his run is accepting while Eve's run
  is not accepting.

Since $\mathcal A$ alternatively reads input and output symbols,
we assume that the set of states is partitioned into a set
of input states $Q_\inp$ from which only input symbols are read, and
set of output states $Q_\outp$. Note that $q_i\in Q_\inp$.

  Formally, the vertex set of Adam is $(Q_\inp \times Q_\inp) \cup
  (Q_\inp \times Q_\outp)$, and the vertex set of Eve is $Q_\outp \times
  Q_\outp$. The initial vertex is $(q_0,q_0) \in Q_\inp \times Q_\inp$. A vertex $(q,q')$ has
  the following outgoing edges:
  \begin{itemize}
  \item If $(q,q') \in Q_\inp \times Q_\inp$, there are edges to
    $(\delta(q,a),\delta(q',a)) \in Q_\outp \times Q_\outp$ for each
    $a \in \SigmaI$ (Adam chooses next input letter).
  \item If $(q,q') \in Q_\outp \times Q_\outp$, then there is an edge to
    $(\delta(q,a),q') \in Q_\inp \times Q_\outp$ for each letter $a \in \SigmaO$ (Eve chooses her next output letter).
  \item If $(q,q') \in Q_\inp \times Q_\outp$, then there is an edge to
    $(q,\delta(q',a)) \in Q_\inp \times Q_\inp$ for each letter $a \in \SigmaO$ (Adam chooses his next output letter).
  \end{itemize}

  We only consider vertices that are reachable from the initial vertex
  $(q_0,q_0)$ by these edges.
  
  Adam wins if the play is in a vertex from $(Q\setminus F) \times F$ (since $F$ is
  a subset of $Q_\inp$, this means that such a vertex is from $Q_\inp
  \times Q_\inp$).
  
  If $(q_0,q_0)$ is not in the winning region of Eve, then $\A$ has no
  Boolean realizers. To see this, take a
  transducer $\mathcal{T}$. This transducer
  induces a strategy $\sigma$ for Eve in which she always plays the
  next output as determined by the transducer (independent of the
  outputs chosen by Adam). Since $(q_0,q_0)$ is not in the winning
  region of the game, Adam has a strategy against $\sigma$ to reach a
  configuration $(q,q')$ such that $q' \in F$ and $q \notin F$. The
  unique play resulting from the two strategies corresponds to an
  input word $u$ chosen by Adam, an output word $v$ chosen by Eve
  (according to the transducer), and an output word $v'$
  chosen by Adam. The run on $u \otimes v$ ends in $q \notin F$, and the run
  on $u \otimes v'$ ends in $q' \in F$. Thus, $u$ is in the domain of $\A$
  but the transducer produces an output $v$ such that $(u,v) \notin
  R(\A)$. Hence, $\mathcal{T}$ is not a Boolean realizer of $\A$.

  If $(q_0,q_0)$ is in the winning region of Eve, then obtain $\A'$ as
  follows: Remove all states $q$ from $\A$ such that $(q,q)$ is not in
  the winning region of Eve. Among the remaining states, remove those
  transitions $(p,a,q)$ from $\A$ such that $(q,p)$ exists in the game and is not in the
  winning region of Eve. Finally, make the resulting automaton trim.

  We first show that all transitions $(p,a,q)$ that are domain-unsafe
  in $\A$ are removed.  In $\Adom$, the transition $(p,a,q)$
  corresponds to an $\varepsilon$-transition $(p,\varepsilon,q)$. This
  means that $L(\Adom,q) \subseteq L(\Adom,p)$. Since $(p,a,q)$ is
  domain-unsafe, there is a word $u \in L(\Adom,p) \setminus
  L(\Adom,q)$. Consider the vertex $(q,p)$ in the game. If Adam plays
  the input sequence $u$ from this vertex, Eve cannot reach an
  accepting state since $u \notin L(\Adom,q)$. But Adam can make his
  moves corresponding to an accepting run of $L(\Adom)$ starting in
  $p$. Hence, $(q,p)$ is not in the winning region of Eve, and the
  transition $(p,a,q)$ is removed in the construction of $\A'$.

  We now show that the states remaining in $\A'$ have the same domain
  as in $\A$. This implies that $\A'$ has the same domain as $\A$,
  and that $\A'$ is domain-safe because any transition that is domain-unsafe in $\A'$ would then also be domain-unsafe in $\A$.
  
  Let $p$ be a state of $\A'$. Clearly, $L(\Adom',p) \subseteq
  L(\Adom,p)$ because $\A'$ is a subautomaton of $\A$. We show by
  induction of the length of input words $u$ that $u \in L(\Adom,p)$
  implies that $u \in L(\Adom',p)$, implying that $L(\Adom',p) =
  L(\Adom,p)$.
   We distinguish the cases of $p \in Q_\inp$ and $p \in Q_\outp$.
  
  For $u = \varepsilon$ and $p \in Q_\inp$, the claim is clear because
  from input states $p$ there are no $\varepsilon$-transitions, and $p$
  is final in $\A$ iff it is final in $\A'$. Assume the inductive
  claim is true for words of length at most $n \ge 0$ at input states,
  and consider the case $p \in Q_\outp$. Since $p$ is in $\A'$,
  $(p,p)$ is in the winning region of Eve, and thus there must be an
  edge from $(p,p)$ to some $(q,p)$ that is also in the winning region
  of Eve. The move from $(p,p)$ to $(q,p)$ corresponds to some output
  transition $(p,a,q)$. Since Adam can move from $(q,p)$ to $(q,q)$ by
  just imitating Eve's last move, we obtain that also $(q,q)$ is in
  the winning region of Eve. Hence, the transition $(p,a,q)$ also
  exists in $\A'$. The state $q$ is an input state and thus accepts
  the same input words of length at most $n$ in $\Adom$ and $\Adom'$
  by our assumption. We have seen earlier that all domain-unsafe
  transitions of $\A$ are removed in $\A'$. Thus, $L(\Adom,p) =
  L(\Adom,q)$. Now let $u$ be an input word of length at most $n$ with
  $u \in L(\Adom,p)$. Then $u \in L(\Adom,q)$, and by assumption also
  $u \in L(\Adom',q)$. In $\Adom'$ there is an $\varepsilon$-transition
  $(p,\varepsilon,q)$, and thus $u \in L(\Adom',p)$, as we wanted to
  show.

  Now consider the inductive step for an input state $p$ and for $u =
  au' \in L(\Adom,p)$ of length $n+1$. There is a unique transition
  $(p,a,q)$ in $\Adom$, and thus $u' \in L(\Adom,q)$. Since $p$ exists
  in $\A'$, the vertex $(p,p)$ is in the winning region of Eve. Since
  Adam can move from $(p,p)$ to $(q,q)$, also $(q,q)$ must be in the
  winning region of Eve. Hence, the transition $(p,a,q)$ exists in
  $\A'$. By induction, $u' \in L(\Adom',q)$, and thus also $u = au'
  \in L(\Adom,p)$.
  
  It remains to show that $\A'$ has the same 
  Boolean realizers as $\A$. Clearly, since $R(\A') \subseteq
  R(\A)$ and the two relations have the same domain, every realizer
  of $\A'$ is also a realizer of $\A$. Let $\mathcal{T}$ be a Boolean realizer of $\A$. Eve can use
  $\mathcal{T}$ as a winning strategy in the game by playing the
  output sequences generated by $\mathcal{T}$. This ensures that for
  every input $u \in \dom(\A)$ that is played by Adam, she answers with an output
  sequence $v$ such that $u \otimes v$ is accepted by $\A$. Hence, the run
  that she plays in the game ends in an accepting state of $\A$ for
  each such $u \in \dom(\A)$. Since she never leaves her winning region, all
  transitions used by $\A$ in the run on $u \otimes v$ also exist in
  $\A'$. Thus $u \otimes v$ is also accepted by $\A'$, and therefore
  $\mathcal{T}$ is also a realizer of $\A'$.

  Since the winning region of Eve can be computed in polynomial time
  (because safety games are solvable in polynomial time),
  we obtain a polynomial time procedure as claimed in the statement of
  the theorem.
\end{proof}


\section{Proofs of \cref{sec:games} (Critical prefix games)}

\subsection{Proof of \cref{thm:threshold-games}}

Before we prove the result, we formally introduce mean-payoff games.

The \emph{mean-payoff values} of the play $\pi$ are $\MPover(\pi) = \mathrm{lim\ sup}_{n \to \infty} \frac{1}{n} \SUM(\pi(n))$ and $\MPunder(\pi) = \mathrm{lim\ inf}_{n \to \infty} \frac{1}{n} \SUM(\pi(n))$.
A \emph{mean-payoff objective} in $G$ is parameterized by a threshold $\nu\ \in \mathbbm Q$, and $\triangleright\ \in \{>,\geq\}$.
They are given by $\MPsup_G^\triangleright(\nu) =  \{ \pi \in \plays(G) \mid \MPover(\pi) \triangleright \nu\}$ and $\MPinf_G^\triangleright(\nu) =  \{ \pi \in \plays(G) \mid \MPunder(\pi) \triangleright \nu\}$ and require that the mean-payoff value of a play is at least resp.\ greater then $\nu$.
The \emph{mean-payoff game problem} asks whether there exists a winning strategy for Eve for the objective $\MPsup_G^\triangleright(\nu)$ or $\MPinf_G^\triangleright(\nu)$.
Note that the objectives are equivalent for games with perfect
information \cite{ehrenfeucht1979positional}. %

\thmthresholdgamesall*


\begin{proof}
 First, we show that for \SUM the threshold problems reduce to mean-payoff games.
 We begin with the non-strict case.
 Let $G$ be the arena of a \prefix threshold game, let $G'$ denote the arena that is obtained from $G$ by removing all vertices from where Adam can not force a visit to a critical vertex against Eve's strategies.
 It is a reachability game property, hence this step can be done in polynomial time.
 If the initial vertex is removed, then clearly Eve has a winning strategy by always avoiding a critical vertex.
 Otherwise, we keep only the connected component to which the initial vertex belongs.
 Since the original game arena is assumed to be deadlock-free, this results in a graph with no deadlocks, because a deadlock would imply that every previously possible transition leads to a vertex from where Eve can avoid critical vertices.
 Furthermore, from every critical vertex add an edge to the initial vertex with weight $-\nu$, where $\nu$ denotes the threshold.
 We assume that all critical vertices belong to Adam, if not, for a critical vertex $v$ that belongs to Eve we add a non-critical copy $v'$ of this vertex that belongs to Eve, the original vertex $v$ then belongs to Adam.
 From the original vertex $v$ Adam can either go with weight zero to its copy $v'$ or with weight $-\nu$ to the initial vertex.
 In her copy, Eve can make the moves that were previously possible from the original vertex $v$.

 We prove that Eve has a winning strategy for the objective $\Thres_G^{\SUM\geq}(\nu)$ in the \prefix threshold game if, and only if, she has a winning strategy for the objective $\MPsup_{G'}^\geq(0)$ in the mean-payoff game.
 Assume that Eve has a winning strategy $\sigma$ in $G$.
 Consider that $\sigma$ guarantees that critical vertices are avoided. 
 Then $G'$ is empty. There is nothing to show.
 Consider that $\sigma$ can not avoid seeing critical vertices, $G'$ is not empty and contains the initial vertex.
 We define a strategy $\sigma'$ in $G'$ that behaves like $\sigma$ if possible and arbitrary otherwise and show that it is winning.
 We explain that $\plays_{\sigma}(G') \subseteq \plays_{\sigma}(G)$.
 Assume that it is not the case, then in a play according to $\sigma'$ Eve's next move from some vertex $v$ in $G'$ is not equal to Eve's next move according to $\sigma$ from $v$ in $G$.
 This is only the case if the target chosen in $G$ does not exist in $G'$, but this means that from $v$ Eve can avoid critical vertices, thus $v$ is not in $G'$ which is a contradiction.
 Playing according to $\sigma$ guarantees that every path from the initial vertex to a critical vertex has value $\geq \nu$.
 This means that every cycle that uses an edge from a critical vertex to the initial vertex has value $\geq 0$.
 Thus, all plays according to $\sigma'$ that infinitely often see such edges have value $\geq 0$.
 Consider plays according to $\sigma'$ that only finitely often see such edges.
 We argue that these plays also have value $\geq 0$.
 Pick such a play $\pi$, towards a contradiction, we assume that $\MPover(\pi) < 0$.
 Let $i$ be the index of last time that an edge from a critical vertex to the initial vertex was used, and let $\pi'$ such that $\pi = \pi(i)\pi'$.
 We have $\MPover(\pi') < 0$, meaning that we can pick some $j$ such that $\SUM(\pi'(j))$ is lower than some arbitrary negative value.
 The prefix $\pi'(j)$ is a valid play prefix according to $\sigma$ in $G$ (we assume that moves from a critical vertex to its copy are removed).
 Since $\SUM(\pi'(j))$ is sufficiently low and from $\last(\pi'(j))$ Adam can force the play into a critical vertex, the sum reached in the critical vertex is less than the threshold $\nu$.
 This contradicts that $\sigma$ is winning in $G$.
 
 For the other direction, assume that $\sigma'$ is a winning strategy for Eve in $G'$, we can assume $\sigma'$ to be positional \cite{ehrenfeucht1979positional}.
 Remove from $G'$ all edges belonging to Eve that are not chosen by $\sigma'$, let $G'_{\sigma'}$ denote the result.
 Since $\sigma'$ guarantees that the value of each play according to the strategy is $\geq 0$, we know that $G'_{\sigma'}$ does not contain a negative cycle.
 This implies that the sum of every path from the initial vertex to a critical vertex is $\geq \nu$.
 The strategy $\sigma'$ is easily translated into a positional strategy $\sigma$ in $G$ which is clearly winning.
 Regarding the strict case, given a play $\pi$, it is easy to see that $\SUM(\pi(i)) > x$ for some $x$ implies that $\SUM(\pi(i)) \geq x-1$, because the weights in $G$ are from $\mathbbm Z$.
 Thus, deciding whether a winning strategy for the objective $\Thres_G^{\SUM>}(\nu)$ exists is equivalent to deciding whether a winning strategy for the objective $\Thres_G^{\SUM\geq}(\nu-1)$ exists. 
 Since mean-payoff games can be solved in $\textsc{NP} \cap \textsc{coNP}$ \cite{zwick1996complexity} the complexity bound follows.

 Secondly, we show that for \AVG the threshold problems reduce to threshold problems for \prefix games for \SUM.
 Let $G$ be the arena of a \prefix threshold game, and let $G'$ be the arena obtained from $G$ by subtracting $\nu$ from every weight.
 Given a play prefix $\varphi$, we have that $\AVG(\varphi) \triangleright \nu \Leftrightarrow \SUM(\varphi) - \nu|\varphi| \triangleright 0$.
 Thus, given a strategy $\sigma$ for Eve, for all $\varphi \in \prefs_\sigma(G) = \prefs_\sigma(G')$ holds $\varphi \in \Thres_G^{\AVG\triangleright}(\nu) \Leftrightarrow \varphi \in \Thres_{G'}^{\SUM\triangleright}(0)$.
\end{proof}

\subsection{Proof of \cref{thm:threshold-games-dsum-all}}

\thmthresholddsumgamesall*

We split the proof of the above theorem into several theorems.
First, \cref{thm:threshold-games-dsum-strict} shows the decidability of discounted-sum critical prefix threshold games for a strict threshold.
Secondly, \cref{thm:threshold-games-dsum-positional} shows that positional strategies suffice for discounted-sum critical prefix threshold games (for strict and non-strict thresholds).
Lastly, \cref{thm:threshold-games-dsum-strict} shows the decidability of discounted-sum critical prefix threshold games for a strict threshold using \cref{thm:threshold-games-dsum-positional}. 

\subsection{Proof of \cref{thm:threshold-games-dsum-all}: Discounted-sum with a non-strict threshold}

\begin{theorem}\label{thm:threshold-games-dsum}
  The threshold problem for \prefix games for $\DSUM$ and a non-strict
  threshold is decidable in $\textsc{NP} \cap \textsc{coNP}$. 
 \end{theorem}

\begin{proof}
 We show that for \DSUM the non-strict threshold problem reduces to discounted-sum games.
 Let $G$ be the arena of a \prefix threshold game, we remove all vertices from where Adam can not force a visit to a critical vertex against all strategies of Eve.
 If the initial vertex was removed, Eve has a winning strategy in the \prefix threshold game that avoids critical vertices.
 Let $G'$ denote the resulting game arena, to make the arena deadlock-free, we add a new sink vertex $q_\bot$ that has a self-loop with weight zero.
 From each deadlock an edge with weight zero leads to $q_\bot$.

 In the reduction we want to achieve that every time a critical vertex is entered, Adam can decide whether he wants to stop the game.
 This should be achieved by going to the sink vertex $q_\bot$ with weight zero.
 However, in the original game, a critical vertex can belong to Eve, so for these vertices we have to give Adam the choice without introducing a copy vertex as intermediate vertex as done, e.g., in the proof of \cref{thm:threshold-games}, to not change the discounted-sum value of corresponding plays by introducing longer paths.

 Towards this, for each critical vertex $v$ and edge $e$ with source $v$ that belongs to Eve, we add a vertex $(v,e)$ that belongs to Adam which has two outgoing edges, one that has weight zero to the sink $q_\bot$, and one that has weight $w(e)$ and the same target as $e$.
 For each critical vertex that belongs to Adam, we add an edge with weight zero to the sink $q_\bot$.
 
 Let $v$ be a critical vertex that belongs to Eve, and $e$ be an edge with target $v$ and source $u$.
 If $u$ belongs to Adam, we add an edge with weight $w(e)$ from $u$ to $q_\bot$.
 If $u$ belongs to Eve, for each edge $e'$ with source $v$, we add an edge with weight $w(e)$ to $(v,e')$.
 Then, the original edge $e$ is removed.
 
 We show that Eve has a winning strategy in the critical prefix threshold game for the objective $\Thres_{G}^{\DSUM,\geq}(\nu)$ if, and only if, she has a winning strategy in the discounted-sum game for the objective $\DS_{G'}^{\geq}(\nu)$.

 Let $\sigma'$ be a winning strategy for Eve in $G'$, we transform $\sigma'$ into a strategy $\sigma$ for Eve in $G$.
 Moves that go from a vertex $u$ to a vertex $(v,e)$ are transformed into a move from $u$ to $v$, in $v$ the next edge that is chosen is $e$.
 It is easy to see that $\sigma$ is a winning strategy for Eve in $G$.

 For the other direction, let $\sigma$ be a winning strategy for Eve in $G$, we transform $\sigma$ into a strategy $\sigma'$ in $G'$.
 According to $\sigma$, if Eves moves from $u$ to a critical vertex $v$ that belongs to Eve and subsequently takes an edge $e$, then $\sigma'$ defines a move from $u$ to $(v,e)$.
 We prove that $\sigma'$ is winning in $G'$.
 Assume the contrary, let $\alpha'$ be a play according to $\sigma'$ that Eve loses.
 We distinguish whether $\alpha'$ contains a move to $q_\bot$.

 First, assume that $\alpha'$ contains such a move, let $\pi'$ denote the play prefix such that $\alpha' = \pi'e_\bot^\omega$, where $e_\bot$ is the self-loop in $q_\bot$.
 The prefix $\pi'$ is translated back into a play prefix $\pi$ in $G$, $\pi$ is play prefix according to $\sigma$.
 Since $\DSUM(\alpha') = \DSUM(\pi') = \DSUM(\pi)$, and $\DSUM(\alpha) < \nu$, we directly obtain that $\sigma$ is not a winning strategy for Eve in $G'$, which is a contradiction.

 Secondly, assume that $\alpha'$ does not contain a move to the sink.
 The play $\alpha'$ is translated back into a play $\alpha$ in $G$, $\alpha$ is a play according to $\sigma$.
 Let $w_{\mathrm{max}}$ denote the largest weight.
 Since $\DSUM(\alpha') = \DSUM(\alpha)$, and $\DSUM(\alpha') < \nu$, we can pick an $i$ such that $\DSUM(\alpha(i)) < \nu$ and $\DSUM(\alpha(i)) + \Sigma_{j = i}^{i+n} \lambda^j w_{\mathrm{max}} < \nu$, where $n$ is the number of vertices in $G$.
 Now consider a play that begins with $\alpha(i)$ in $G$, and then Adam forces the play into a critical vertex in at most $n$ steps.
 This play is a play according to $\sigma$ but is not winning for Eve, which is a contradiction.

 Since discounted-sum games can be solved in $\textsc{NP} \cap \textsc{coNP}$ \cite{andersson2006improved} the complexity bound follows.
\end{proof}

The following remark shows that the above reduction fails for strict thresholds.

\label{remark-dsum-strict-threshold}

\begin{remark}
 Consider the following critical prefix threshold game $G$ with discounted-sum measure and discount factor $\lambda = \frac{1}{2}$ and objective $\Thres_{G}^{\DSUM,>}(1)$, all vertices belong to Adam, $v_0$ is the initial and $v_1$ is a critical vertex.
 
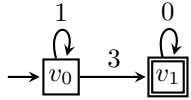
\begin{center}
  \begin{tikzpicture}[thick]
    \tikzset{square/.style={regular polygon,regular polygon sides=4,inner sep=0}}
    \tikzstyle{every state}+=[inner sep=3pt, minimum size=5pt,square];
    \node[state, initial] (0) {$v_0$};
    \node[state, right of=0, accepting] (1) {$v_1$};

    \draw[->] (0) edge[loop above] node {$1$} ();
    \draw[->] (0) edge node {$3$} (1);
    \draw[->] (1) edge[loop above] node {$0$} ();
  
   \end{tikzpicture}
 \end{center}
 Eve wins this game.
 If Adam stays in $v_0$, the discounted-sum value converges to $1$, if Adam proceeds to the critical vertex $v_1$, the value is (and stays) higher than $1$, i.e., Eve has won; the critical vertex is either not seen or always with a value $>1$.

 Now, consider the obtained discounted-sum game $G'$ with objective $\DS_{G'}^>(1)$ as in the proof of \cref{thm:threshold-games-dsum}.
The value of the initial vertex $c_0$ is $1$ (Adam stays in $v_0$), thus, our reduction would wrongly yield that Eve can not win the critical prefix game.

There seems to be no reduction to a discounted-sum game (using the same or another threshold) that correctly handles strategies of Adam which avoid critical vertices.

In the above correctness proof for the presented reduction regarding $\geq$-thresholds, we have shown that if the value of the initial vertex in the constructed discounted-sum game is $\geq \nu$, then Eve has a winning strategy the critical prefix game.
Alternatively, we could have shown that if the value is $< \nu$, then Adam has a winning strategy in the critical prefix game.
We argue intuitively why in these cases, the situation as shown in the example for strict thresholds, is not problematic.
Assume the value is $<\nu$ and is achieved by Adam never visiting a critical vertex.
This implies that Adam can stay long enough away from critical vertices to ensure that the desired threshold becomes unreachable.
Since in the constructed discounted-sum game, Adam can force to see a critical vertex, there exists also a strategy of Adam that reaches a critical vertex with a value $<\nu$.
\end{remark}

\subsection{Proof of \cref{thm:threshold-games-dsum-all}: Positional strategies for discounted-sum}

\begin{theorem}\label{thm:threshold-games-dsum-positional}
  If Eve has a winning strategy in a \prefix games for $\DSUM$ and a strict or non-strict threshold, then she also has a positional winning strategy.
 \end{theorem}

\begin{proof}
  Let $\sigma$ be a winning strategy for Eve in the game. For each vertex $v$ of Eve, we define a positional choice by comparing the $\sigma$-play prefixes that reach $v$. For each such $\sigma$-play prefix ending in $v$, we measure how ``difficult'' it is for Eve to win with this given prefix of the play. Intuitively, the positional strategy $\sigma_p$ picks a move that is chosen by $\sigma$ in the most difficult situations. We then show that this positional strategy $\sigma_p$ is at least as good as $\sigma$.

  We first explain how to compare play prefixes. Consider a play prefix $\pi$ ending in vertex $v$. Then $\sigma$ ensures that for each continuation $\pi\pi_2$ of $\pi$ that ends in a critical vertex, the threshold condition $\DSUM(\pi\pi_2) \triangleright \nu$ is satisfied (for $\triangleright \in \{>,\geq\}$, depending on whether we consider a strict or non-strict threshold).
  Using the equality $\DSUM(\pi\pi_2) = \DSUM(\pi) + \lambda^{|\pi|}\DSUM(\pi_2)$, we can rewrite $\DSUM(\pi\pi_2) \triangleright \nu$  into a property that $\pi_2$ must satisfy: $\DSUM(\pi_2) \triangleright (\nu - \DSUM(\pi))\frac{1}{\lambda^{|\pi|}}$.
  Therefore, the larger the value on the right-hand side of this second inequation is, the harder it is for a strategy of Eve to ensure the threshold condition. We call it the relative gap for $\pi$ and $\nu$. Since $\nu$ is fixed, we just denote it by $\rg(\pi)$. 
  \[
\rg(\pi) := (\nu - \DSUM(\pi))\frac{1}{\lambda^{|\pi|}}.
  \]
  Let us note two simple but important properties of this value, which are routine to check using the definition of $\rg(.)$:
  \begin{enumerate}[(A)]
  \item $\DSUM(\pi) \triangleright \nu$ iff $0 \triangleright \rg(\pi)$. \\
    Proof of (A): Clearly, ($\rg(\pi) = 0$ iff $\DSUM(\pi) = \nu$), and ($\rg(\pi) < 0$ iff $\DSUM(\pi) > \nu$)

  \item Let $\pi'$ be a play prefix ending in $v'$, and assume there is an edge from $v'$ to $v$ with weight $j$. Let $\pi$ be the play prefix $\pi'$ extended by this edge, denoted by $\pi'v'\xrightarrow{j}v$. Then $\rg(\pi) = \frac{1}{\lambda}\rg(\pi') - j$.\\
    Proof of (B): $\rg(\pi) = (\nu - \DSUM(\pi))\frac{1}{\lambda^{|\pi|}} = (\nu - (\DSUM(\pi') + \lambda^{|\pi'|}\lambda
    j))\frac{1}{\lambda^{|\pi'|+1}} =\frac{1}{\lambda}\rg(\pi') - j$.
  \end{enumerate}
  We now explain how to define the positional strategy $\sigma_p$.
  Denote the $\sigma$-play prefixes ending in $v$ by $\prefs(\sigma,v)$. If $v$ belongs to Eve, we define the move $\sigma_p(v)$ of the positional strategy to be a move that satisfies
  \[
\forall \pi \in \prefs(\sigma,v) \exists \pi' \in \prefs(\sigma,v):\; \rg(\pi') \ge \rg(\pi) \mbox{ and } \sigma(\pi') = \sigma_p(v). \hfill (*)
  \]
Such a move always exists. Either, there is a ``most difficult'' $\pi'
\in \prefs(\sigma,v)$ with $\rg(\pi') = \sup\{\rg(\pi)\mid \pi \in
\prefs(\sigma,v)\}$. Then we can choose $\sigma_p(v) = \sigma(\pi')$, which 
satisfies $(*)$. Now consider the case that there is no such most
difficult $\pi' \in \prefs(\sigma,v)$, and let $x := \sup\{\rg(\pi)\mid \pi
\in \prefs(\sigma,v)\}$. Consider a sequence $\pi_1,\pi_2, \ldots$ of play prefixes from $\prefs(\sigma,v)$ whose $\rg$-values converge to $x$, that is, $\lim_{i \rightarrow \infty} \rg(\pi_i) = x$.
Since there are only finitely many moves from vertex $v$, there must be one move that is chosen by $\sigma$ for infinitely many of the $\pi_i$. We choose such a move for $\sigma_p(v)$, which then satisfies property $(*)$.

It remains to show that $\sigma_p$ is a winning strategy. 
Let $\pi_p \in \prefs(\sigma_p,v)$ be a $\sigma_p$-play prefix ending in $v$. We show the following claim: there is a $\sigma$-play prefix $\pi \in \prefs(\sigma,v)$ with $\rg(\pi) \ge \rg(\pi_p)$. When we have shown this claim, we can conclude that $\sigma_p$ is a winning strategy as follows. If $v$ is a critical vertex, then $\pi$ satisfies the threshold condition. Using property (A) from above, we obtain that $\pi_p$ also satisfies the threshold condition.

To finish the proof, we show the claim by induction on the length of $\pi_p$. For length $0$ the claim is clear because then $v$ has to be the starting vertex, and the empty play is also in $\prefs(\sigma,v)$.
Now assume that $\pi_p' \in \prefs(\sigma_p,v')$ and $\pi_p \in \prefs(\sigma_p,v)$ is obtained from $\pi_p'$ by one move from $v$ to $v'$ with weight $j$. We now define a play prefix $\pi' \in \prefs(\sigma,v')$ with $\rg(\pi') \ge \rg(\pi_p')$ such that $\pi:=\pi'v'\xrightarrow{j}v \in \prefs(\sigma,v)$. With property (B) from above we then obtain
\[
\rg(\pi) = \frac{1}{\lambda}\rg(\pi') - j \ge \frac{1}{\lambda}\rg(\pi'_p) - j =  \rg(\pi_p)
\]
as in the claim.

For defining $\pi'$, first note that by induction, there is $\hat{\pi} \in \prefs(\sigma,v')$ with $\rg(\hat{\pi}) \ge \rg(\pi_p')$.

If $v'$ is a vertex of Adam, we let $\pi' := \hat{\pi}$ because extending $\hat{\pi}$ by the edge from $v'$ to $v$ yields a play prefix from $\prefs(\sigma,v)$.

If $v'$ is a vertex of Eve, then by $(*)$ from the definition of $\sigma_p$, there is some $\pi' \in \prefs(\sigma,v')$ with $\rg(\pi') \ge \rg(\hat{\pi})$ and $\sigma(\pi') = \sigma_p(v')$. Hence
$\pi := \pi'v'\xrightarrow{j}v \in \prefs(\sigma,v)$.  
\end{proof}

\subsection{Proof of \cref{thm:threshold-games-dsum-all}: Discounted-sum with a strict threshold}

%

\newcommand{\mrg}{\mathrm{mrg}}

\begin{theorem}\label{thm:threshold-games-dsum-strict}
  The threshold problem for \prefix games for $\DSUM$ and a strict threshold is decidable in $\textsc{NP}$.
 \end{theorem}

To prove this theorem, we first show a result on weighted graphs $G =
(V,E,\gamma:E\rightarrow \mathbbm{Z})$ which is interesting in itself.


\lempathchecking*

\begin{proof}
We remove from $G$ all the vertices from which there is no path to a
target vertex. Let $G'$ be the resulting subgraph of $G$. If $G'$ is
empty, then we can return that there is no path $\pi$ as in the
statement of the lemma. Otherwise, we make use
  of the notion of relative gap from the proof of \cref{thm:threshold-games-dsum-positional}, which is defined
for a finite path $\pi$ by  
  \[
\rg(\pi) := (\nu - \DSUM(\pi))\frac{1}{\lambda^{|\pi|}}.
\]
In the proof of \cref{thm:threshold-games-dsum-positional} we
have already used the  properties (A) and (B) below. We add two
properties (C) and (D) about the relative gaps for the concatenation of paths. In all the statements below $\pi,\pi_1,\pi_2,\pi_3$ are finite paths.  
  \begin{enumerate}[(A)]
  \item $\DSUM(\pi) > \nu$ iff $0 > \rg(\pi)$.
  \item Let $\pi'$ be a path ending in $v'$, and assume there is an edge from $v'$ to $v$ with weight $j = \gamma(v',v)$. Let $\pi$ be  $\pi'$ extended by this edge, denoted by $\pi'v'\xrightarrow{j}v$. Then $\rg(\pi) = \frac{1}{\lambda}\rg(\pi') - j$.
  \item $\rg(\pi_1\pi_2) = \frac{1}{\lambda^{|\pi_2|}}(\rg(\pi_1) - \DSUM(\pi_2))$. \\
    Proof of (C): $\rg(\pi_1\pi_2) = (\nu - \DSUM(\pi_1\pi_2))\frac{1}{\lambda^{|\pi_1\pi_2|}} = (\nu - \DSUM(\pi_1) - \lambda^{|\pi_1|}\DSUM(\pi_2)))\frac{1}{\lambda^{|\pi_1\pi_2|}}$\\
    $=\frac{1}{\lambda^{|\pi_2|}}(\rg(\pi_1) - \DSUM(\pi_2))$.
  \item if $\rg(\pi_1) \geq \rg(\pi_1\pi_2)$ and $\pi_2$ is a loop, then 
    $\rg(\pi_1\pi_3) \geq \rg(\pi_1\pi_2\pi_3)$. \\
    Proof of (D): by $(C)$, $\rg(\pi_1\pi_3)=
    \frac{1}{\lambda^{|\pi_3|}}(\rg(\pi_1) - \DSUM(\pi_3))$ and
    $\rg(\pi_1\pi_2\pi_3) =
    \frac{1}{\lambda^{|\pi_3|}}(\rg(\pi_1\pi_2) - \DSUM(\pi_3))$. 
    So, $\rg(\pi_1\pi_3) - \rg(\pi_1\pi_2\pi_3) =
    \frac{1}{\lambda^{|\pi_3|}}(\rg(\pi_1)-\rg(\pi_1\pi_2))\geq 0$. 
  \end{enumerate}
From (A) we obtain that there exists a path $\pi$ from $v_0$ to $T$ such that
$\DSUM(\pi)\leq \nu$ iff there exists a path $\pi$ from $v_0$ to $T$
such that  $\rg(\pi) \ge 0$.

Our algorithm computes for each vertex $v$ the maximal relative gap
$\mrg_i(v)$ of paths of lengths at most $i$ ending at $v$ (and starting at the initial vertex $v_0$). For $i = 0$, we obtain that
\[
\mrg_0(v) =
\begin{cases}
  \nu & \mbox{if $v = v_0$} \\
  - \infty & \mbox{otherwise}
\end{cases}
\]
where $\mrg_i(v) = -\infty$ indicates that there is no path of length at most $i$ to $v$.
We use (B) to compute $\mrg_i(v)$ from the $\mrg_{i-1}$ values by
\[
\mrg_i(v) = \max\left(\{\frac{1}{\lambda}\mrg_{i-1}(v') - \gamma(v',v) \mid (v',v) \in E'\} \cup \{\mrg_{i-1}(v)\}\right).
\]
Let $n>0$ be the number of vertices of $G'$. The algorithm computes the values $\mrg_i(v)$ for all $i \le n$, which can be done in polynomial time, using the above formula. By definition of $\mrg_i(v)$, we obtain that $\mrg_i(v) \le \mrg_{i+1}(v)$.


We claim that there is no path $\pi$ from $v_0$ to $T$ such that
$\DSUM(\pi)\leq \nu$ ($\star$) iff
\begin{enumerate}[(i)]
\item $\mrg_{n}(v) < 0$ for all $v\in T$, and
\item $\mrg_{n-1}(v) = \mrg_{n}(v)$ for all vertices $v$ of $G'$.
\end{enumerate}
Clearly, if (i) does not hold, there is a path to a target vertex
$v$ that has relative gap $\ge 0$, and hence $(\star)$ does not hold.
So assume now that (i) is satisfied.

If (ii) is satisfied, then $\mrg_j(v) = \mrg_n(v)$ for all $j \ge n$
because the values do not change anymore once a fixpoint is
reached. We can conclude that there is no path of any length to a
target vertex with relative gap $\ge 0$.

It remains to show that $(\star)$ does not hold if (ii) is not satisfied.
So let $v$ be such that $\mrg_{n-1}(v) < \mrg_{n}(v)$. Any shortest
path $\pi$ ending in $v$ such that $\rg(\pi) = \mrg_{n}(v)$ has length $n$ necessarily, otherwise 
we would have $\mrg_{n-1}(v) = \mrg_{n}(v)$. By the choice of
$n$, there is a repetition of vertices $v_j = v_k$ in $\pi$ with
$j<k$. So we can write $\pi = \pi_1\pi_2\pi_3$ for a loop $\pi_2$,
that is, $\pi_1 = v_0 \cdots v_j$ and $\pi_2 = v_{j+1} \cdots
v_k$. Let $u = v_j$.

We have $\rg(\pi_1)  <  \rg(\pi_1\pi_2)$, otherwise
by $(D)$, we would obtain $\rg(\pi_1\pi_3) \geq  \rg(\pi_1\pi_2\pi_3)
= \mrg_n(v)$ which contradicts that $\pi = \pi_1\pi_2\pi_3$ is a
shortest path witnessing $\mrg_n(v)$.

Now, define $z := \rg(\pi_1\pi_2) - \rg(\pi_1) > 0$.
We show that pumping the loop $\pi_2$ gives paths to $u$ with arbitrarily large relative gaps. We prove that each additional loop increases the relative gap by at least $z$, and hence
\[
\rg(\pi_1\pi_2^\ell) \ge \ell \cdot z + \rg(\pi_1).
\]
So we formally show that for $\ell \ge 1$:
\[
\rg(\pi_1\pi_2^\ell) - \rg(\pi_1\pi_2^{\ell-1}) \ge z.
\]
For $\ell =1$ this is satisfied by definition of $z$. For $\ell > 1$ we obtain with (C):
\[
\begin{array}{cl}
  \rg(\pi_1\pi_2^\ell)  -\rg(\pi_1\pi_2^{\ell-1}) & = \frac{1}{\lambda^{|\pi_2|}}(\rg(\pi_1\pi_2^{\ell-1})-\DSUM(\pi_2)) - \frac{1}{\lambda^{|\pi_2|}}(\rg(\pi_1\pi_2^{\ell-2})-\DSUM(\pi_2)) \\
  &= \frac{1}{\lambda^{|\pi_2|}}((\rg(\pi_1\pi_2^{\ell-1}) - \rg(\pi_1\pi_2^{\ell-2}))\\
 \mbox{(by induction)} &\ge \frac{1}{\lambda^{|\pi_2|}} z \ge z.
\end{array}
\]
Now let $\pi_4$ be a path from $u$ to a target vertex. Then we can choose $\ell$ such that $\ell \cdot z + \rg(\pi_1) \ge \DSUM(\pi_4)$ and obtain:
\[
\rg(\pi_1\pi_2^\ell\pi_4) = \frac{1}{\lambda^{|\pi_4|}}(\rg(\pi_1\pi_2^{\ell})-\DSUM(\pi_4)) \ge \frac{1}{\lambda^{|\pi_4|}}(\ell \cdot z + \rg(\pi_1) -\DSUM(\pi_4)) \ge 0,
\]
and therefore $(\star)$ does not hold, as we wanted to show.
\end{proof}

\begin{proof}[Proof of \cref{thm:threshold-games-dsum-strict}]
By \cref{thm:threshold-games-dsum-positional}, Eve has a positional winning strategy if she has one at all.
  The NP-algorithm guesses a positional strategy $\sigma$ for Eve, and
  then verifies in polynomial time whether $\sigma$ is winning. Let
  $G'$ be the game restricted to Eve's  $\sigma$-edges, seen as a
  weighted graph. 
  The strategy $\sigma$ is not winning iff Adam can form a path
  in $G'$ from the initial vertex to a critical vertex that has weight
  $\le \nu$. This property can be checked in \textsc{Ptime} thanks to
  \cref{lem:pathchecking} (by taking as target set the set of
  critical vertices). 
\end{proof}

\subsection{Proof of \cref{thm:prefix-energy-games-undec}}

\thmprefixenergygamesundec*

\begin{proof}
  We give a reduction that constructs from a given 2-counter machine $M$ a game $G_M$ such that Eve has an observation-based winning strategy in $G_M$ iff $M$ does not halt.
  We assume that in $M$ no decrement operation is applied to a counter with value $0$ (before each decrement we can test if the counter is $0$).
  
  The idea for the reduction is as follows. The initial credit is $0$.
  The game uses four copies of the machine that cannot be distinguished by Eve. At the beginning, Adam picks one of these copies.
  The vertices in each copy correspond to the instruction numbers of the machine. Let $c_1,c_2$ be the counters of the machine.
  We also use $c_1,c_2$ to denote the values of the counters. In the four different copies, the energy level is supposed to correspond to $c_1$, $-c_1$, $c_2$, and $-c_2$.

  At vertices corresponding to increment or decrement operations, there is no choice for the players. There is a unique transition moving to the next instruction.
  The weight of this edge is chosen to track the desired energy level. For an increment on $c_1$, it is $1$, $-1$, $0$, $0$ in the four copies, respectively.
  Similarly for the other increment and decrement instructions.

  At vertices corresponding to an instruction of the form ``IF($c_i = 0$) THEN GOTO $m$'', Eve can choose between two actions $0$ and $>0$,
  where $0$ means that Eve claims the counter is $0$, and $>0$ means that Eve claims that the counter is not $0$.
  In every copy, Adam can choose to execute the instruction corresponding to Eve's claim. That is, if Eve claims $0$, then
  go to vertex $m$, and if Eve claims $>0$, then go to the next instruction (all these edges have weight $0$).

  Then we also add actions for Adam to ensure that Eve loses if she makes a wrong claim.
  If Eve claims $0$, then in the copy for $-c_i$, Adam can move to a critical vertex with an edge of weight $0$.
  So if Eve's claim was wrong and Adam has chosen the copy for $-c_i$ at the beginning, then she loses the game.
  If Eve claims $>0$, then Adam can move to a critical vertex from the copy for $c_i$ with an edge of weight $-1$.
  Again, if Eve's claim was wrong and Adam has chosen the copy for $c_i$ at the beginning, then she loses the game.

  From this description, it should be clear that Eve can only win with an observation based strategy if she simulates the machine correctly (always makes correct claims on the counter values).
  Now, in all copies, we make the vertices corresponding to the STOP instruction of the machine critical, and add a self loop for Adam with negative weight on them.
  So if such a vertex is reached, Eve loses.

  In summary, if the machine halts, Eve has no observation based winning strategy.
  If the machine does not halt, then Eve can go on with the correct simulation forever. 
\end{proof}

\subsection{Proof of \cref{thm:prefix-energy-reachable}}

\thmprefixenergyreachable*

\begin{proof}
We show that the problem reduces to the fixed initial credit problem for imperfect information games which is decidable \cite{degorre2010energy}.

 Let $G$ be the arena of a \prefixEG with set of vertices $V$. First
 note that by assumption, from each vertex of the game, Adam has a
 strategy to reach a critical vertex, and since it is a reachability
 objective, it is well-known that he can do so in at most $|V|$
 steps (against any strategy, and against observation based strategies in
 particular).

 Let $B = |V|w_{\mathrm max}$, where $w_{\mathrm max}$ is the maximal weight in $G$.
 We construct a \prefix energy game with imperfect information.
 Let $G'$ denote it's game arena obtained from $G$ as follows.
 For each critical vertex $v$ and action $a$, add a transition with
 label $a$ to a new sink vertex $v_B$ with weight $-B$.
 Every outgoing transition from $v_B$ leads back to it with weight zero.
 The observation of $v_B$ is $\{v_B\}$.

 We show that Eve has an observation-based winning strategy in the \prefix energy game $G$ with imperfect information with initial credit $c_0$ if, and only if, Eve has an observation-based winning strategy in the energy game $G'$ with imperfect information with initial credit $c_0 + B$.

 Let $\sigma$ be an observation-based strategy for Eve in $G$.
 The strategy $\sigma$ naturally defines a strategy $\sigma'$ in $G'$,
 which consists of playing as $\sigma$ as long as the play is not in
 the sink $v_B$. Towards a contradiction, assume that $\sigma$ is winning in $G$, but $\sigma'$ is not winning in $G'$.
 Then there exists a (finite) prefix $\pi$ of a play in $G'$
 compatible with $\sigma'$ such that $c_0 + B + \EL_{G'}(\pi) <
 0$. 
 Note that $\last(\pi)$ is either some vertex $v$ from $G$ or the new sink $v_B$.
 In the first case, $\pi$ is also a valid play prefix according to $\sigma$ in $G$, and $\EL_{G'}(\pi) = \EL_{G}(\pi)$.
 Adam can force the play from $v$ in at most $|V|$ steps into a
 critical vertex, by assumption and since $v\neq v_B$, which rises the energy level by at most $B$.
 Thus, there exists a finite play continuation $\alpha$ according to $\sigma$ such that $\pi\alpha \in C$ and $c_0 + \EL_{G}(\pi\alpha) < 0$, because $c_0 + \EL_{G}(\pi) < -B$.
 This contradicts that $\sigma$ is winning in $G$.
 In the second case, let $\pi'$ be the shortest prefix of $\pi$ such
 that $\pi'$ does not contain the sink state, i.e., $\pi = \pi' a v_B$
 for some action $a$, and $\pi'$ is compatible with $\sigma$ in
 $G$. By definition of $G'$, $\last(\pi')$ is critical, and since
 the loop on the sink has weight $0$, we have
 $\EL_{G'}(\pi) = \EL_{G'}(\pi') - B = \EL_{G}(\pi') - B$. Therefore,
 $c_0 + \EL_{G}(\pi') = c_0+\EL_{G'}(\pi) + B < 0$, which contradicts
 that $\sigma$ is winning in $G$.

 For the other direction, let $\sigma'$ be an observation based
 strategy for Eve in $G'$. Consider the restriction $\sigma$ of
 $\sigma'$ to the play prefixes in $G$. It is an observation based strategy of $G$. Towards a contradiction, assume that $\sigma'$ is winning in $G'$, but $\sigma$ is not winning in $G$.
 Then there exists a play prefix $\pi$ in $G$ compatible with $\sigma$
 such that $\last(\pi) \in C$ and $c_0 + \EL_{G}(\pi) < 0$. The prefix
 $\pi$ is also compatible with $\sigma'$ in $G'$. In $G'$, from
 $\last(\pi)$, whatever action Eve picks, Adam can force the play to
 the sink, with weight $-B$. In particular, the prefix
 $\rho = \pi.\sigma'(\pi).v_B$ is compatible with $\sigma'$ and satisfies 
$\EL_{G'}(\rho) = \EL_{G'}(\pi) - B = \EL_G(\pi) - B$. Therefore, $c_0
+ B + \EL_{G'}(\rho) = c_0  + \EL_G(\pi) < 0$, which contradicts that
$\sigma'$ is winning in $G'$. 


By \cite{degorre2010energy}, finite-memory strategies are sufficient
for Eve to win energy games of imperfect information and fixed initial
credit. Our translation of winning strategies in $G'$ to winning
strategies in $G$ preserves finite-memoryness, entailing the second
statement of the theorem. 
\end{proof}


\section{Proofs of \cref{sec:syntesis-problems} (Synthesis
  problems)}

\subsection{Threshold synthesis}

\subsubsection{Proof of \cref{thm:deterministic-threshold}}

\thmdeterministicthreshold*

\begin{proof}
 Let $\mathcal A$ be the $V$-automaton with $V \in \{\SUM,\AVG,\DSUM\}$ that defines the weighted specification, $\nu$ be the threshold, and $\triangleright \in \{\geq, >\}$.
 Let $\mathcal A'$ be a domain safe variant of $\mathcal A$ which can be obtained in polynomial time, see \cref{thm:domain-safe}.
 We interpret $\mathcal A'$ as a weighted game arena, to make the arena deadlock free, we add a new sink state $q_\bot$ that has a self-loop with weight zero and from each deadlock the sink state is entered with weight zero.
 Let the critical vertices correspond to the final states of $\mathcal A'$.

 It is easy to see that the threshold synthesis problem with the given parameters reduces to deciding whether Eve has a winning strategy in the \prefix game played on $\mathcal A'$ with objective $\Thres_{\mathcal A'}^{V\triangleright}(\nu)$.
 The decidability and complexity results follow from \cref{thm:threshold-games,thm:threshold-games-dsum,thm:threshold-games-dsum-strict}.

 It is left to show that the threshold synthesis problems for \SUM and \AVG are \textsc{PTime}-equivalent to mean-payoff games.
 The direction from threshold synthesis to mean-payoff games follows directly from the reduction to critical prefix games which are reduced to mean-payoff games (see proof of \cref{thm:threshold-games}).

 For the other direction, consider a mean-payoff game $G$ and the non-strict threshold zero, i.e., the objective is $\MPsup_G^\geq(0)$.
 Wlog., we assume that moves of Adam and Eve alternate in $G$, this can be easily achieved by introducing extra vertices and edges with weight zero if necessary.
 Let $m$ resp.\ $n$ be the maximal out-degree of a vertex that belongs to Adam resp.\ Eve in $G$.
 We define $\SigmaI = \{a_1,\dots,a_m,\bot\}$ and $\SigmaO=\{b_1,\dots,b_n\}$.

 First, we reduce the problem to the synthesis problem for \SUM and the non-strict threshold zero.
 We construct a weighted finite automaton $\mathcal A = (V \cup \{q_i,q,q_\bot,q_f\},\SigmaIO,q_i,\Delta,\{q_f\},\gamma)$ defining a \SUM-specification $S$ from $G$ as follows.
 The arena of $G$ is used as the transition graph of $\mathcal A$, the labels of a transition is $a_i$ resp.\ $b_i$ if its source belongs to Adam resp.\ Eve in $G$ and the transition corresponds to the $i$th outgoing edge, the weights are as in $G$.
 Such an edge always exists as $G$ is deadlock-free.
 
 Mean-payoff games do not have a notion that corresponds to a domain, so we need to make sure that for the specification that we construct realizability does not depend on the domain.
 To achieve this, in the automaton we add a transition $(v,a_i,v')$ with weight $\gamma\bigl((v,a_1,v')\bigr)$ for all $i \in \{2,\dots,m\}$ such that $i \geq \mathrm{outdeg}(v)$ and $v$ belongs to Adam in $G$.
 
 From the new initial state $q_i$ there is a transition with label $a_1$ and weight zero to the new state $q$, from $q$ there is a transition with label $b_1$ and weight $N$ to $v_i$, where $N$ is the sum of all absolute values of weights that appear in $G$ and $v_i$ is the initial vertex of $G$.

 From every state that belongs to Adam in $G$, there is a transition with label $\bot$ to the new state $q_\bot$ with weight zero, and from $q_\bot$ a transition with label $b_1$ to the new (and only) final state $q_f$ with weight zero.

 We show that there exists an $S$-realizer that ensures a value $\geq 0$ for each pair if, and only if, Eve has a winning strategy in the mean-payoff game $G$ for the objective $\MPsup_G^\geq(0)$.

 Assume Adam has a winning strategy in $G$, then there exists also a positional one \cite{ehrenfeucht1979positional}, say $\tau$.
 Thus, all cycles seen in plays according to $\tau$ are negative.
 Towards a contradiction, assume that there exists a transducer that is an $S$-realizer that ensures a value of at least zero.
 Consider an input sequence that is long enough that corresponds to a play according to $\tau$.
 Eventually, enough negative cycles have been seen to make the current value of the run according to $\mathcal A$ negative.
 Then, if the next input is $\bot$, the transducer can not ensure a value of at least zero which is a contradiction.

 Assume Eve has a winning strategy in $G$, then there exists also a positional one, say $\sigma$.
 Thus, all cycles seen in plays according to $\sigma$ are non-negative.
 Let $\mathcal T_\sigma$ be transducer that produces output according to $\sigma$.
 Note that for inputs that do not directly correspond to plays in $G$ according to $\sigma$, the transducer works as if the input has been $a_1$, then $\sigma$ can be applied.
 We argue that $\mathcal T_\sigma$ is an $S$-realizer that ensures a value of at least zero.
 After the first output symbol, the value of the current run according to $\mathcal A$ is $N$ and the state is $v_i$.
 For every continuation according to the outputs that $\mathcal T_\sigma$ produces holds that the final state $q_f$ is reached with value at least zero.
 Assume otherwise, then there exists a path inside of $V$ according to $\sigma$ such that its sum is smaller than $-N$.
 Since the sum of all absolute values of weights in $G$ is $N$, this implies that there has been a cycle according to $\sigma$ with a negative value, which is a contradiction.

 It is left to show the reduction for \SUM with a strict threshold, and for \AVG with a strict and a non-strict threshold.

 Regarding \SUM with a strict threshold, it is easy to see that the above mean-payoff game with non-strict threshold zero can also be reduced to the same \SUM-specification $S$ defined by $\mathcal A$ and asking to ensure a value $>-1$ instead of $\geq 0$, because the weights in $\mathcal A$ are integers.

 Regarding \AVG with a non-strict threshold, it suffices to interpret the weighted specification defined by $\mathcal A$ as an \AVG-specification and ask to ensure a value $\geq 0$.

 Regarding \AVG and a strict threshold, we interpret the specification defined by $\mathcal A$ as an \AVG-specification and ask to ensure a value $> \frac{-1}{\ell}$, where $\ell$ is chosen as follows.
 In case that Adam has a winning strategy in the mean-payoff game, he has a positional winning strategy, and can enforce a negative cycle.
 To show that there is no realizer that ensures threshold $> \frac{-1}{\ell}$, the number $\ell$ should reflect the number of input and output symbols that it takes (at most) to go to a negative cycle, take it a sufficient number of times in order to bring the sum of the current run sufficiently into the negative, and end the computation.

 Therefore, consider all simple cycles with a negative sum in $G$, and for each of these cycles $c_1,\dots,c_i$ compute $\ell_1,\dots,\ell_i$, where $\ell_j = |c_j|\cdot x_j$ is the smallest multiple of the length of $c_j$ such that $|x_j\cdot\SUM(c_j)| > 2N$; recall $N$ is the sum of all absolute values of weights that appear in $G$.
 Let $\ell_\mathrm{max}$ be the maximal value of the $\ell_j$s.

 Then, let $\ell = \ell_\mathrm{max} + |V| + 4$.
 In the automaton, $2$ computation steps are needed to reach $v_i$, the initial vertex of the game, and $2$ steps are needed to go from a vertex in $V$ to the final state $q_f$.
 At most $|V|$ computation steps are needed to arrive at a negative cycle.
 At most $\ell_\mathrm{max}$ computation steps (inside some cycle) are needed to bring the value of the run below zero, because the sum up to reaching the cycle is at most $2N$.
 Thus, if Adam has a winning strategy, for any realizer, there exists a computation of length at most $\ell$ according to this realizer that arrives at $q_f$ with a negative sum.
 This negative sum is $\le -1$, thus the value of the run is $\leq\frac{-1}{\ell}$, i.e., the desired value of $>\frac{-1}{\ell}$ can not be ensured.
 
 If Eve has a winning strategy, we argued above that there is a realizer that ensures a sum of at least zero for each run according to this realizer, thus the average value is also at least zero, which makes it $>\frac{-1}{\ell}$.
\end{proof}

\subsection{Synthesis and regret determinization}

\subsubsection{Proof of \cref{lem:approx-to-regret}}

\lemapproxtoregret*

The proof of the above lemma is split into the two lemmas below, \cref{lem:approx-2-regret,lem:regret-to-approx}.

\begin{lemma}\label{lem:approx-2-regret}
  The approximate synthesis problem for weighted specifications reduces in linear time to the regret determinization problem for nondeterministic weighted automata (with the same threshold).
\end{lemma}

\begin{proof}
 We start with the direction from left to right.
 Let $S$ be a weighted specification defined by a WFA $\mathcal A = (Q,\SigmaIO,q_i,\Delta,F,\gamma)$.
 We construct a nondeterministic WFA $\mathcal A'$ from $\mathcal A$ as follows,
 we pick an output symbol, say $b \in \SigmaO$, and replace every output label with $b$.
 The language of $\mathcal A'$ is $\{ a_1b\dots a_nb \mid a_1\dots a_n \in \mathrm{dom}(S)\}$.
 
 Given a threshold $r \in \mathbbm Q$, and $\triangleleft \in \{<,\leq\}$, we show that there exists a transducer $\mathcal T$ that implements an $r_\triangleleft$-approximate $S$-realization if, and only if, $\mathcal A'$ is $r_\triangleleft$-determinizable.

 Let $\mathcal T$ be a transducer that implements an $r_\triangleleft$-approximate $S$-realization.
 The state space $Q_\mathcal T$ of $\mathcal T$ and $Q_\mathcal T \times \SigmaI$ serves as the set of memory states to construct an $r_\triangleleft$-regret deterministic variant of $\mathcal A'$.
 If $\mathcal T$ has a transition $s \xrightarrow{\sigma|\sigma'} t$, then $\mathcal A'_r$ has transitions $\bigl((p,s),\sigma,(q,(s,\sigma))\bigr)$ and $\bigl((q,(s,\sigma)),b,(q,t)\bigr)$ if $(p,\sigma,q) \in \Delta$ and $(q,\sigma',r) \in \Delta$.
 It is easy to see that $\mathcal A'_r$ is an $r_\triangleleft$-regret deterministic variant of $\mathcal A'$.

 For the other direction, let $\mathcal A'_r$ be an $r_\triangleleft$-regret deterministic variant of $\mathcal A'$.
 We construct a transducer $\mathcal T$ from $\mathcal A'_r$ with the following transitions.
 If $\mathcal A'_r$ has a transitions $\bigl((p,m_1),\sigma,(q,m_2)\bigr)$ and $\bigl((q,m_2),b,(r,m_3)\bigr)$, then $\mathcal T$ has a transition $(p,m_1) \xrightarrow{\sigma|\sigma'}(r,m_3)$, where $\sigma' \in \SigmaO$ is an output symbol such that $(q,\sigma',r) \in \Delta$.
 If there are more than one output symbols that lead from $q$ to $r$, chose one that yields the maximal transition weight.
 It is easy to see that $\mathcal T$ implements an $r_\triangleleft$-approximate $S$-realizer.
\end{proof}

The converse of this result, for \SUM-automata, is shown in
\cref{lem:regret-to-approx} below. 

\begin{lemma}\label{lem:regret-to-approx}
 The regret determinization problem for nondeterministic \SUM-automata
 reduces in linear time to the approximate synthesis problem for
 \SUM-specifications (with the same threshold). 
\end{lemma}

\begin{proof}
 Let $\mathcal A = (Q,\Sigma,q_i,\Delta,F,\gamma)$ be a nondeterministic \SUM-automaton.
 We construct a (deterministic) \SUM-automaton $\mathcal B = (Q',\Sigma',q_i,\Delta',F,\gamma')$ with $Q' = Q \cup \{ q_a \mid q \in Q, a \in \Sigma\}$, $\Sigma' = \SigmaI \cup \SigmaO$, where $\SigmaI = \Sigma$ and $\SigmaO = \{ \tau \mid \tau \in \Delta \}$, $\Delta' = \{ (p,a,p_a) \mid p \in Q \text{ and } a \in \SigmaI\} \cup \{ (p_a,\tau,q) \mid (p,a,q) = \tau\}$, $\gamma'\bigl( (p_a,\tau,q) \bigr) = \gamma(\tau)$ and $\gamma'\bigl( (p,a,p_a) \bigr) = 0$.
 Let $S$ denote the \SUM-specification defined by $\mathcal B$.

 Given a threshold $r \in \mathbbm Q$, and $\triangleleft \in \{<,\leq\}$, it is easy to show that there exists a transducer $\mathcal T$ that implements an $r_\triangleleft$-approximate $S$-realization if, and only if, $\mathcal A$ is $r_\triangleleft$-determinizable.
\end{proof}

\subsection{Best-value synthesis}

\subsubsection{Proof of \cref{lem:good-property}}

\lemgoodproperty*

\begin{proof}
Let $V\colon \mathbbm Z^* \to \mathbbm Q$ be a payoff function, and let $\mathcal A$ be a $\leq$-stable $V$-automaton defining a specification $S$.
 Let $\mathcal T$ be a trim transducer that implements a best-value $S$-realization.
 We build the transducer $\mathcal A \times \mathcal T$ and show that if $\mathcal A \times \mathcal T$ contains two states of the form $(q,m_1)$ and $(q,m_2)$ then it suffices to keep only one of them, e.g.,\ $(q,m_1)$, and replace every target $(q,m_2)$ with $(q,m_1)$.
 Repeating this construction eventually leads to a transducer that is a sub-automaton of $\mathcal A$.
 
 We show the correctness of this construction.
 Let $\mathcal B$ denote the transducer $\mathcal A \times \mathcal T$, and consider two states $(q,m_1)$ and $(q,m_2)$.
 Let $f_\mathcal B((q,m_i),u) = S_q\bigl((u \otimes \mathcal B_{(q,m_i)}(u))\bigr)$, where $S_q$ is the $V$-specification defined by $\mathcal A_q$, that is, $\mathcal A$ with initial state $q$, and $\mathcal B_{(q,m_i)}$ denotes the $\mathcal B$ with initial state $(q,m_i)$ for $i=1,2$.

 First, we prove that for all $u \in \mathrm{dom}(R(\mathcal A_q))$ holds that $f_\mathcal B((q,m_1),u) = f_\mathcal B((q,m_2),u)$.
 Let ($\ast$) denote this property.

 Towards a contradiction, assume that there is a $u$ such that $f_\mathcal B((q,m_1),u) < f_\mathcal B((q,m_2),u)$, let $\rho'$ resp.\ $\rho''$ denote the corresponding run of $\mathcal B_{(q,m_1)}$ resp.\ $\mathcal B_{(q,m_2)}$ with output $v'$ resp.\ $v''$.
 Pick a word $x \otimes y \in (\SigmaI\SigmaO)^*$ such that $\mathcal B\colon(q_0,m_0) \xrightarrow{x|y} (q,m_2)$, let $\rho$ be the run of $\mathcal A$ on $x \otimes y$, $\rho$ ends in $q$.
 Since $\mathcal A$ is $\leq$-stable we obtain that $V(\rho\rho'') < V(\rho\rho')$.
 We have that $\mathcal B(xu) = yv''$, $S(xu \otimes yv'') = V(\rho\rho'')$, and $S(xu \otimes yv') = V(\rho\rho')$, this contradicts that $\mathcal B$ implements a best-value $S$-realizer.

 Secondly, we prove that we can safely remove $(q,m_2)$ from $\mathcal B$.
 Let $\mathcal B'$ denote the transducer where $(q,m_2)$ is removed and the transitions pointing to $(q,m_2)$ now point to $(q,m_1)$.
 For this, we show that for all $u \in \mathrm{dom}(R(\mathcal A_q))$ it holds that $f_\mathcal B((q,m_1),u) = f_{\mathcal B'}((q,m_1),u)$, where $f_{\mathcal B'}((q,m_1),u) = S_q(u \otimes \mathcal B'_{(q,m_1)}(u))$.
 Towards a contradiction, pick a shortest $u\in\mathrm{dom}(R(\mathcal A_q))$ such that $f_{\mathcal B}((q,m_1),u) > f_{\mathcal B'}((q,m_1),u)$.
 Let $\rho$ resp.\ $\rho'$ be the run on the unique $u \otimes v$ resp.\ $u \otimes v'$ of $\mathcal B$ resp.\ $\mathcal B'$ such that $V(\rho) = f_{\mathcal B}((q,m_1),u)
 $ and $V(\rho') = f_{\mathcal B'}((q,m_1),u)$.
 Since $V(\rho)> V(\rho')$, $\rho$ must visit $(q,m_2)$, thus 
 $\rho$ has a factorization $\rho_1\rho_2$ such that the corresponding run of $\mathcal B$ is of the form
 \[\mathcal B\colon \underbrace{(q,m_1) \xrightarrow{u_1|v_1}}_{\rho_1} \underbrace{(q,m_2) \xrightarrow{u_2|v_2} (p,m)}_{\rho_2},\]
 where $u = u_1u_2$, $v = v_1v_2$ and $(p,m)$ is a final state of $\mathcal B$, and $\rho_1$ does not visit $(q,m_2)$. By construction of $\mathcal B'$, the run $\rho'$ is of the form $\rho_1\rho_2'$ as follows:
 \[\mathcal B'\colon \underbrace{(q,m_1) \xrightarrow{u_1|v_1}}_{\rho_1} \underbrace{(q,m_1) \xrightarrow{u_2|v_2'} (p',m')}_{\rho_2'},\]
 where $v' = v_1v_2'$, and $(p',m')$ is a final state of $\mathcal B'$.
 The property ($\ast$) implies that $f_{\mathcal B}((q,m_1),u_2) =
 f_{\mathcal B}((q,m_2),u_2) = V(\rho_2)$.
 Since $\mathcal A$ is $\leq$-stable, it follows that $V(\rho_2) > V(\rho_2')$.
 Hence, $f_{\mathcal B}((q,m_1),u_2) > f_{\mathcal B'}((q,m_1),u_2)$, which contradicts the assumption that $u$ is a shortest counter-example. 

 Finally, we show that $\mathcal{B'}$ is also a best-value
 $S$-realizer. Let $u\in \mathrm{dom}(S)$ and let
 $f_{\mathcal{B}}(u)$ resp. $f_{\mathcal{B'}}(u)$ be equal to
 $S(u \otimes \mathcal{B}(u))$ resp. $S(u \otimes \mathcal{B'}(u))$. We show that 
$f_{\mathcal{B}}(u) = f_{\mathcal{B'}}(u)$. Let $\rho$ be the run of
$\mathcal{A}$ on $(u,\mathcal{B}(u))$, and $\rho'$ be the run of 
$\mathcal{A}$ on $(u,\mathcal{B'}(u))$. If $\rho$ does not visit
$(q,m_2)$ then $\rho = \rho'$ and the claim follows. If $\rho$ visits
$(q,m_2)$, then $\rho$ and $\rho'$ can be decomposed as:
 \[\mathcal B\colon \underbrace{(q_0,m_0) \xrightarrow{u_1|v_1}}_{\rho_1} \underbrace{(q,m_2) \xrightarrow{u_2|v_2} (p,m)}_{\rho_2}\qquad\mathcal B'\colon \underbrace{(q_0,m_0) \xrightarrow{u_1|v_1}}_{\rho_1} \underbrace{(q,m_1) \xrightarrow{u_2|v_2'} (p',m')}_{\rho_2'}.\]

By ($\ast$) we have $f_\mathcal B((q,m_1),u_2) = f_\mathcal
B((q,m_2),u_2)$ and by the second property we have shown, we have
$f_\mathcal B((q,m_1),u_2) = f_{\mathcal B'}((q,m_1),u_2)$. In other
words, we get $V(\rho_2) = V(\rho'_2)$, from which we conclude that
$V(\rho_1\rho_2) = V(\rho_1\rho'_2)$ (because $\mathcal{A}$ is
$V$-stable), i.e. $f_{\mathcal{B}}(u) = f_{\mathcal{B'}}(u)$.
This concludes the proof, as $\mathcal{B}$ is a best-value $S$-realizer.
\end{proof}

As a consequence of the latter lemma, the fact that
$V$-specifications are $\leq$-stable and have decidable inclusion
problem~\cite{DBLP:conf/concur/FiliotGR12} for $V\in\{\SUM,\AVG,\DSUM\}$, we obtain decidability
of the best-value synthesis problem for those measures. While this
technique is general to obtain decidability results, it does not yield
optimal complexities, and in the following theorems, we use other
techniques to obtain better bounds.

\subsubsection{Proof of \cref{thm:best-value} (continued)}

\thmbestvalue*

It remains to prove \cref{thm:best-value} for \AVG- and
\DSUM-specifications. This is done below in two separate theorems,
\cref{thm:best-value-avg} and
\cref{thm:best-value-dsum}.

\subsubsection{Proof of \cref{thm:best-value}: Average}

\begin{theorem}\label{thm:best-value-avg}
 The best-value synthesis problem for an \AVG-specification is decidable in \textsc{Ptime}.
\end{theorem}

\begin{proof}
 The problem reduces to the best-value synthesis problem for \SUM-specifications.

 Let $\mathcal A$ be a \AVG-automaton defining an \AVG-specification $S$, and let $\mathcal A_\SUM$ be the same automaton but interpreted as a \SUM-automaton defining the \SUM-specification $S_\SUM$.
 We have that $R(S) = R(S_\SUM)$.

 Let $\mathcal T$ be a such that $\mathrm{dom}(T) = \mathrm{dom}(S)$ and $u \otimes \mathcal T(u) \in S$ for all $u \in \mathrm{dom}(S)$.
 It is easy to see that $S(u \otimes \mathcal T(u)) = \bestVal_S(u)$ if, and only if, $S_\SUM(u,\mathcal T(u)) = \bestVal_{S_\SUM}(u)$ for all $u \in \mathrm{dom}(S)$, because $\bestVal_{S_\SUM}(u) = \bestVal_{S}(u)|uu|$.
\end{proof}

\subsubsection{Proof of \cref{thm:best-value}: Discounted-sum}

\begin{theorem}\label{thm:best-value-dsum}
 The best-value synthesis problem for a \DSUM-specification is decidable in \textsc{NP}$\cap$\textsc{coNP}.
\end{theorem}

\begin{proof}

\newcommand{\T}{\mathcal{T}}
\newcommand{\dsafe}{\Delta_\outp^s}

Let $\mathcal A = (Q,\Sigma,q_0,\Delta,F,\gamma)$ be a $\DSUM$-automaton (with discount factor $\lambda$) defining a weighted specification $S$, for which we want to solve the best-value synthesis problem. We first give a property that characterizes best-value realizers of $S$, and then use a game to check this property.
  
For a state $q \in Q$ we write $\bestVal_q$ for the function $\bestVal_{S_q}$, where $S_q$ is the weighted specification defined by $\A$ with initial state $q$. Let $\T$ be a transducer. Then $\T$ is a best-value realizer of $S$ if, and only if, the following property is satisfied:
\begin{itemize}
\item[$(*)$] Assume an input $u \in \Sigma_\inp^*$, $\T$ has produced
  some output $v$, such that $\A$ reaches state $q$ on
  $u \otimes v$, and the corresponding transition sequence of $\A$ has
  discounted-sum $x$. Then for each $u' \in \Sigma_\inp^*$ such that $uu'\in \mathrm{dom}(S)$:
  \[
  x +\lambda^{2|u|} \bestVal_q(u') = \bestVal_S(uu').
  \]
\end{itemize}
We argue that $(*)$ indeed characterizes the best-value realizers of $S$. Clearly, always $x +\lambda^{2|u|} \bestVal_q(u') \le \bestVal_S(uu')$ because $x +\lambda^{2|u|} \bestVal_q(u')$ is a possible value for $uu'$. So if $(*)$ is not satisfied, then there is a $u'$ such that $uu' \in \mathrm{dom}(S)$, and $x +\lambda^{2|u|} \bestVal_q(u') < \bestVal_S(uu')$. But then $\T$ cannot produce the best value for input $uu'$.
If $(*)$ is satisfied, then for $u \in \mathrm{dom}(S)$, we obtain that $\T$ produces weight $x = \bestVal(u)$ by choosing $u' = \varepsilon$.
This finishes the argument that $(*)$ characterizes the best-value realizers of $S$.

We now show how to reduce the problem of checking $(*)$ to critical prefix $\DSUM$ non-strict threshold games in polynomial time, which yields the claimed complexity by \cref{thm:threshold-games-dsum}. 
The idea for the game is the same as for the game used in the proof of \cref{thm:domain-safe} for the construction of domain-safe automata: Adam and Eve simultaneously build runs of $\A$, both starting in the initial state. The weights are chosen such that they give the difference of the weights of the two runs (run of Eve $-$ run of Adam). Eve's goal is to ensure that the difference is $\ge 0$ for inputs that are in the domain.

In the Boolean case (\cref{thm:domain-safe}), the output transitions where chosen in two moves, first by Eve, then by Adam. For $\DSUM$ this is problematic because for building the weight difference of the two runs, we have to ensure to use the right discount factors. Therefore, we modify the game a bit to take this into account as follows:
\begin{itemize}
\item In a first part of the game, there is only one run. Adam chooses inputs, and Eve chooses output transitions. The weights of these moves are 0. This corresponds to the part where Adam's run is the same as Eve's run (and thus the difference is $0$).
\item Whenever Eve has chosen an output transition, Adam can either follow this output transition and pick the next input, staying in the first part of the game, or choose a different output, branching off another run and going to the second part of the game (the idea is that Adam chooses this move the first time Eve would reach a state violating property $(*)$).
\item In the second part, the game traces two runs. Adam always chooses the input and his next output at the same time. Then Eve chooses her output. The weights are the differences between the transitions in the two runs. Since the move to the second part of the game shifts the number of moves in the game by one w.r.t.\ the transitions in the run, the weights have to be corrected, dividing them by $\lambda$.
\end{itemize}
Since a best-value realizer of $S$ is also a Boolean realizer of $S$, we also need to take care of the Boolean condition (the run of Eve is accepting if the run of Adam is accepting). This could be done by first solving a safety condition on the game that we build. However, since we have already solved the Boolean case in \cref{thm:domain-safe}, we can use those results.

We now turn to the formal definition of the game.
Let $\A^s$ with transitions $\Delta^s \subseteq \Delta$ be a domain-safe subautomaton of $\A$ according to \cref{thm:domain-safe}.  Let $Q_\inp$, $Q_\outp$ be the partition of the state set of $\A$ into input and output states, and let $\Delta_\inp = \Delta \cap (Q_\inp \times \Sigma_\inp \times Q_\outp)$ and $\Delta_\outp = \Delta \cap (Q_\outp \times \Sigma_\outp \times Q_\inp)$ be the transitions starting from an input and output states, respectively. The transitions that Eve can use are the ones in $\dsafe = \Delta_\outp \cap \Delta^s$.
\begin{itemize}
\item The vertices of Adam are $\{q_0\} \cup \dsafe \cup (Q_\inp \times Q_\inp)$.
\item The vertices of Eve are $Q_\outp \cup (Q_\outp \times Q_\outp \times \Sigma_\outp)$.
\item The critical vertices are $Q \times F$.
\item The edges are
  \begin{itemize}
  \item $q_0 \rightarrow p$ with weight $0$ for all $(q_0,a,p) \in \Delta_\inp$ (Adam's very first input)
  \item For $p \in Q_\outp$: \\ $p \rightarrow (p,b,q)$ with weight $0$ for all $(p,b,q) \in \dsafe$ (Eve's output in the first part of the game)
  \item For $(p,b,q) \in \dsafe$: \\ $(p,b,q) \rightarrow p'$ with weight $0$ for all $(q,a,p') \in \Delta_\inp$ (Adam follows Eve's output and chooses the next input in the first part of the game)
  \item For $(p,b,q) \in \dsafe$: \\ $(p,b,q) \rightarrow (q,q')$ with weight $\frac{\gamma(p,b',q') - \gamma(p,b,q)}{\lambda}$ for each $(p,b',q') \in \Delta_\outp$ (Adam chooses a different output than Eve, moving to the second part of the game; the length of the game is now one more than the length of the runs, so the weights are corrected by $\frac{1}{\lambda}$)
  \item For $(q,q') \in (Q_\inp \times Q_\inp)$: \\
    $(q,q') \rightarrow (p,p',b')$ with weight $\frac{\gamma(q',a,p') - \gamma(q,a,p)}{\lambda}$ for each $a \in \Sigma_\inp$, $b' \in \Sigma_\outp$, and $(q,a,p),(q',a,p') \in \Delta_\inp$ (Adam chooses next input and his next output in the second part of the game)
  \item For $(p,p',b') \in Q_\outp \times Q_\outp \times \Sigma_\outp$: \\
    $(p,p',b') \rightarrow (q,q')$ with weight $\frac{\gamma(p',b',q') - \gamma(p,b,q)}{\lambda}$ for each $b \in \Sigma_\outp$ with $(p,b,q) \in \dsafe$, and $(p',b',q') \in \Delta_\outp$ (Eve chooses her next output in the second part of the game).
  \end{itemize}
  We claim that Eve has a winning strategy in this critical prefix $\DSUM$ game with threshold $\ge 0$ iff there is a best-value realizer of $S$.

  We view the first part of the game as Adam playing the same outputs as Eve. With this view, any play ending in a vertex of Adam corresponds to an input word $u$, two output words $v,v'$ of the same length as $u$, a run of $\A^s$ on $u \otimes v$, and a run of $\A$ on $u \otimes v'$. The weight of such a play is the difference of the weights of these two runs, which can easily be shown by induction on the length of the play. 

  If there is a best-value realizer of $S$, then it only uses transition of $\A^s$ (see \cref{thm:domain-safe}). Eve can play her moves according to this realizer, ensuring that her run produces the best value for each input in the domain of $S$. Hence, the weight of plays ending in $Q \times F$ is $\ge 0$.

  Now assume that Eve has a winning strategy $\sigma$ in the game,
  which can be assumed to be positional by
  \cref{thm:threshold-games-dsum-positional}.

 On the first part of the game, this strategy corresponds to a transducer $\T_\sigma$. We show that $\T_\sigma$ satisfies property $(*)$. Assume that $\T_\sigma$ violates $(*)$, and pick a shortest word $u$ for which $(*)$ is not satisfied. Let $v$ be the output produced by $\T_\sigma$, and $x$, $q$ as in $(*)$.

  If $u = v = \varepsilon$, then $x=0$ and $q=q_0$, and thus the equation from $(*)$ is satisfied. Hence, $u = a_1 \cdots a_n$  and $v = b_1 \cdots b_n$ with $n \ge 1$. Consider the corresponding play $\pi$ in the first part of the game, which is of this form (where the $a_i$ are not part of the game but are added for readability):
  \[
\pi = q_0 \xrightarrow{a_1} p_1 \rightarrow (p_1,b_1,q_1)  \xrightarrow{a_2} p_2 \rightarrow \cdots \rightarrow q_{n-1} \xrightarrow{a_n} p_n \rightarrow (p_n,b_n,q_n)
 \] 
 with $q = q_n$. Let $u'= a_{n+1} \cdots a_{n+m}$ be a witness that $(*)$ is violated, i.e., such that 
  \[
  x +\lambda^{2|u|} \bestVal_q(u') < \bestVal_S(uu').
  \]
  We can conclude that for all $b_{n+1} \cdots b_{n+m}$:
  \[
\A(a_1 b_1 \cdots a_{n+m} b_{n+m}) < \bestVal_S(uu'). \hfill (1)
  \]
 Since $u$ was chosen as shortest word, property $(*)$ is satisfied for $w = a_1 \cdots a_{n-1}$, $q_{n-1}$, and the weight $y = \A(a_1b_1 \cdots a_{n-1}b_{n-1})$, i.e.,
  \[
  y +\lambda^{2|w|} \bestVal_{q_{n-1}}(a_nu') = \bestVal_S(wa_nu') = \bestVal_S(uu').
  \]
  So there are $b_n' b_{n+1}' \cdots b_{n+m}'$ such that
  \[
\A(a_1 b_1 \cdots a_{n-1} b_{n-1} a_n  b_n' a_{n+1} b_{n+1}' \cdots a_{n+m} b_{n+m}') = \bestVal_S(uu'). \hfill (2)
  \]
  This means that Adam can continue the play $\pi$ by first playing $b_n'$, taking the play into a configuration $(q_n,q_n')$ in the second part of the game, and then playing $a_{n+1} b_{n+1}' \cdots a_{n+m} b_{n+m}'$, reaching a critical vertex. For all responses of Eve (corresponding to some outputs $b_{n+1} \cdots b_{n+m}$), the value reached at the end of the play is $<0$ because of $(1)$ and $(2)$. 
\end{itemize}

\end{proof}

\subsection{Approximate synthesis}

\thmapproximate*

\phantom{blub}

We split the proof into several theorems.
First, we prove the theorem for sum, see \cref{thm:approximate-sum}.
Then, we prove the theorem for average, see \cref{thm:approximate-avg} for decidability and \cref{thm:approx-avg-hardness} for hardness.
Finally, we prove the theorem for discounted-sum, see \cref{thm:approximate-dsum}.

\subsubsection{Proof of \cref{thm:approximate}: Sum}

\begin{theorem}\label{thm:approximate-sum}
 The approximate synthesis problem for a \SUM-specification and a strict or non-strict threshold is decidable.
 The problem is \textsc{EXPtime}-complete.
\end{theorem}

\begin{proof}
 The approximate synthesis problem for a \SUM-specification reduces to the regret determinization problem for nondeterministic \SUM-automata, see \cref{lem:approx-to-regret}.
 The latter problem was shown to be \textsc{EXPtime}-complete in \cite{DBLP:conf/lics/FiliotJLPR17}, so we obtain \textsc{EXPtime}-membership.
 We obtain \textsc{EXPtime}-hardness by \cref{lem:regret-to-approx}. 
\end{proof}

\subsubsection{Proof of \cref{thm:approximate}: Average}

\begin{theorem}\label{thm:approximate-avg}
 The approximate synthesis problem for an \AVG-specification and a strict or non-strict threshold is decidable.
\end{theorem}

\begin{proof}
Let $\mathcal A = (Q,\Sigma, q_i, \Delta, F,\gamma)$ be a complete \AVG-automaton defining an \AVG-specification $S$.
We assume that $S$ has a non-empty domain, otherwise we can directly
conclude that the specification is realizable. Since $\mathcal A$ alternatively reads input and output symbols,
we assume that the set of states is partitioned into a set
of input states $Q_\inp$ from which only input symbols are read, and
set of output states $Q_\outp$. Note that $q_i\in Q_\inp$.

Given a non-strict threshold $r$, the goal is to decide whether there exists a transducer $\mathcal T$ such that $\mathrm{dom}(\mathcal
 T) = \mathrm{dom}(S)$ and $S(u \otimes \mathcal T(u)) \geq \bestVal_S(u)-r$ for all $u\in \mathrm{dom}(S)$.
 The latter is equivalent to $S(u \otimes \mathcal T(u))|u\mathcal T(u)| - \bestVal_S(u)|u\mathcal T(u)| + |u\mathcal T(u)|r \geq 0$.
 Let $\rho$ be the unique run of $\mathcal A$ on $u \otimes v$, let $v'$ be such that $S(u \otimes v') = \bestVal_S(u)$, and let $\rho'$ be the unique run of $\mathcal A$ on $u \otimes v'$.
 We can express the above equation as $\SUM(\rho) - \SUM(\rho') + |\rho|r \geq 0$.
 With this formulation in mind, we are ready to present a decision method.

 We reduce the problem to deciding whether Eve has an
 observation-based winning strategy in a critical prefix energy game with
 imperfect information with  initial credit zero, in which for
 all states of the game, Adam has a strategy to enforce a visit to a
 critical state. Such games are decidable according to
 \cref{thm:prefix-energy-reachable}. Intuitively in the game, Adam constructs a run $\rho$
 of $\mathcal A$ but Eve only sees the input word $u$ of $\rho$ and not the
 output word nor its states, using imperfect information. Eve also
 constructs a run $\rho'$ of $\mathcal A$ on input $u$. The weights are defined
 so as to maintain the invariant property that the energy level is
 $\SUM(\rho') - \SUM(\rho) + |\rho|r$. 
 Moreover, the construction makes sure that
 Adam has always a strategy to reach an accepting state, i.e., that
 the run $\rho$, if non-accepting, can be continued into an accepting run. To do so,
 Adam's actions are restricted to those which maintain this
 invariant.

 Let us now describe the construction formally. Consider $Q^{trim}\subseteq Q$
the subset of states which are both reachable
from the initial state and from which an accepting state can be
reached. Let $Q^{trim}_\inp = Q^{trim}\cap Q_\inp$ and $Q^{trim}_\outp = Q^{trim}\cap Q_\outp$.
Note that $q_i\in Q^{trim}_\inp$ since we assume that $\mathrm{dom}(S)\neq
\varnothing$, and $F\subseteq Q^{trim}_\inp$, since accepted words end
with an output symbol.


 
 The game arena $G$ has the set of vertices $(Q_\inp \times Q_\inp^{trim})\ \cup \ (Q_\outp \times Q^{trim}_\outp \times \SigmaI)\ \cup \ \{q_\bot\}$, and the set of available actions are $\SigmaO \cup \{\mathit{choose}\}$.
 The arena contains the following edges
 \begin{itemize}
    \item $(p,q)$ to $(\delta(p,a),\delta(q,a),a)$ with action
      $\mathit{choose}$ and weight $\gamma(\delta(p,a)) -
      \gamma(\delta(q,a)) + r$ for all $p,q \in Q$ and $a \in
      \SigmaI$ such that $\delta(q,a)\in Q^{trim}_\outp$. \quad \emph{This edge indicates that Adam chooses the input symbol $a$.}
    \item $(p,q,a)$ to $(\delta(p,b),\delta(q,b'))$ with action $b$
      and weight $\gamma(\delta(p,b)) - \gamma(\delta(q,b')) + r$ for
      all $p,q \in Q$, $a \in \SigmaI$ and $b,b' \in \SigmaO$ such
      that $\delta(q,b')\in Q^{trim}_\inp$.\quad \emph{This edge indicates that Eve chooses the output symbol $b$ and Adam chooses $b'$ instead.}
    \item $(p,q)$ to $q_\bot$ with action $\mathit{choose}$ and weight zero for all $p \in Q\setminus F$ and $q \in F$ \quad \emph{This edge indicates that Adam has completed a valid input word, but Eve was not able to produce a valid output word.}
    \item $q_\bot$ to $q_\bot$ with action $\mathit{choose}$ and weight $-1$ \quad \emph{This is a sink state used to bring the energy level below zero.}
      \item $(p,q)$ to $(p,q)$ with action $\mathit{choose}$ and
        weight $0$. \quad \emph{This is to make the arena deadlock-free.}
 \end{itemize}
Note that the arena is deadlock free thanks to the loop on states
$(p,q)$ and for states of the form $(p,q,a)$, there always exist
$b,b'$ such that $\delta(p,b)$ is defined (because $\mathcal A$ is
complete) and $\delta(p,b')\in Q_\inp^{trim}$ since $q\in
Q_\outp^{trim}$ and $F\subseteq Q_\inp^{trim}$.

The observation of a vertex of the form $(p,q)$ is $p$, of the form
$(p,q,a)$ is $(p,a)$ and of $q_\bot$ is $q_\bot$.
 The set of critical vertices contains all vertices of the form
 $(p,q)$ with $q \in F$ and $q_\bot$. The initial vertex is
 $(q_i,q_i)$.

 First, let us show that Adam, from any vertex of the game, has a
 strategy to reach a critical vertex, an assumption required to apply
 \cref{cor:fin-mem}. Let consider a non-critical vertex of the form
 $(p,q)$ or of the form $(p,q,a)$ such that $q\not\in
 F$. Since $q\in Q^{trim}$, Adam can, in the game, construct a run
 (correspond to the second components of the vertices) to an accepting
 state of $\mathcal A$, leading in the game to a critical vertex.

Second, we show that Eve has an observation-based winning strategy in the
 constructed \prefix energy game with imperfect information with
 initial credit zero, if, and only if, there exists a transducer $\mathcal T$ such that $\mathrm{dom}(\mathcal
 T) = \mathrm{dom}(S)$ and for all $u\in \mathrm{dom}(S)$,
$S(u \otimes \mathcal T(u)) \geq \bestVal_S(u)-r$.

 Assume that Eve has an observation-based winning strategy with initial credit zero.
 Let $\sigma$ be a finite-memory observation-based winning strategy, as is guaranteed to exist by \cref{cor:fin-mem}.
 The strategy can be translated into a 
 transducer $\mathcal T_\sigma$. Indeed, a finite-memory
 observation-based strategy is of type $M\times
 O\rightarrow (\mathit{choose}\cup \SigmaO)\times M$, where $M$ is a
 finite set of memory states and $\mathcal O$ is the set of observations of the
 game. Note that there, an observation $O$ is either a pair $(p,a)$, a
 single state $p$ or $q_\bot$. Since $\sigma$ is winning and $q_\bot$
 is losing, $q_\bot$ is never reached. Initially, the transducer is in
 state $(q_i,m_0)$ where $m_0$ is the initial memory state of
 $\sigma$. The states of $\mathcal T_\sigma$ are pairs $(p,m)$. Let $a\in
 \Sigma_\inp$ be some input symbol and assume that Eve gets
 observation $(p' = \delta(p,a),a)$. From this observation, we get the pair
 $\sigma(m,(p',a)) = (b,m')$ meaning that she plays $b$ and
 the memory is updated to $m'$. Then, she can only observes that she is in
 $p'\in Q_\inp$ and her only possible action is $\mathit{choose}$. 
 Let $m''$ such that $\sigma(m',p') = (\mathit{choose},m'')$. 
 From state $(p,m)$ and input $a$, $\mathcal T_\sigma$ outputs
 $b$ and moves to state $(\delta(p',b),m')$. The accepting states of
 $\mathcal T_\sigma$ are pairs $(p,m)$ such that $p\in F$.

 We show that $\mathcal T_\sigma$ implements an $r$-approximate $S$-realization.
 From the construction of the game graph it is clear that $R(\mathcal
 T_\sigma) \subseteq R(S)$, since any accepting run of $\mathcal T_\sigma$ can
 be translated back to an accepting run of $\mathcal A$. We prove that
 their domains are equivalent. Towards a contradiction assume otherwise.
 Then there exists some input word $u = u_1\dots u_n \in \mathrm{dom}(S)$ such that $u \notin \mathrm{dom}(\mathcal T_\sigma)$.
 We pick some $v = v_1\dots v_n$ such that $u \otimes v \in S$.
 Consider the play $\pi$ according to $\sigma$ where Adam chooses the
 edges that correspond to input $u$ and output $v$, eventually the
 play reaches the vertex $(p,q) = (\delta(u_1v'_1 \dots
 u_nv'_n),\delta(u_1v_1 \dots u_nv_n))$. Since $u \otimes v \in S$ we have
$q \in F$ and since $u\not\in \mathrm{dom}(T_\sigma)$, $p \notin F$,
because by construction of $\mathcal T_\sigma$, the projection of $\pi$ on its
first component corresponds to a run of $\mathcal T_\sigma$. Then, 
Adam can move to $q_\bot$ and bring the energy level below zero.
 Since $q_\bot$ is a critical vertex, Eve loses. This contradicts that
 $\sigma$ is winning.

 It is left to show that for all $u \in \mathrm{dom}(S)$ holds that
 $S(u \otimes \mathcal T_\sigma(u)) \geq \bestVal_S(u) - r$.
 Towards a contradiction assume $S(u \otimes \mathcal T_\sigma(u)) < \bestVal_S(u) - r$.
 Let $v$ be an output word such that $u \otimes v \in S$ and $S(u \otimes v) =
 \bestVal_S(u)$. 
 Let $\rho$ and $\rho'$ be the unique runs of $\mathcal A$ on $u \otimes v$
 and $u \otimes \mathcal{T}_\sigma(u)$, respectively.
 Now, we consider the play $\pi$ in $G$, where Eve plays according to $\sigma$ and Adam makes his choices such that they correspond to the input word $u$ and the run $\rho$.
 Eve does not observe $\rho$.
 Eve choices of actions spell the run $\rho'$.
 By construction of the game graph, the play ends up in a critical vertex, say after $i$ steps, because $u \in \mathrm{dom}(S)$.
 Furthermore, the energy level of the play prefix is $\EL(\pi(i)) = \SUM(\rho') - \SUM(\rho) + r|\rho|$.
 Since $S(u \otimes \mathcal T_\sigma(u)) = \frac{\SUM(\rho')}{|\rho|}$ and $S(u \otimes v) = \frac{\SUM(\rho)}{|\rho|} = \bestVal_S(u)$, it follows from $S(u \otimes \mathcal T_\sigma(u)) < \bestVal_S(u) - r$ that $\SUM(\rho') + r|\rho| < \SUM(\rho)$.
 Thus, $\EL(\pi(i)) < 0$, which contradicts that $\sigma$ is an observation-based winning strategy with initial credit zero.

 For the other direction, assume that $\mathcal T$ is an $S$-realizer which non-strictly $r$-approximates $S$. 
 The transducer $\mathcal T$ can be directly translated into an
 observation-based finite-state strategy $\sigma_\mathcal T$ for Eve
 as follows. Let $M$ be the states of $\mathcal T$ with initial state $m_0$. The states of
 $\sigma_\mathcal T$ are $M\times\{\inp,\outp\}$, with initial state
 $(m_0,\inp)$. In state $(m,\inp)$ and for an observation $(p,a)$, 
$\sigma_\mathcal T$ picks action $b$ such that there exists a
transition $(m,a,b,m')$ in $T$, and the strategy moves to state $m'$. 
From a state $(m,\outp)$ and for an observation $p$, Eve can only pick
action $\mathit{choose}$ and the strategy moves to state $(m,\inp)$.

We show that $\sigma_\mathcal T$ is winning with initial credit zero.
 Clearly, plays according to $\sigma_\mathcal T$ never reach the critical vertex $q_\bot$.
 We prove that in the energy level in every other critical vertex that is reached is not below zero.
 Consider some play $\pi$ according to $\sigma_\mathcal T$ and assume $\last(\pi(i))\in C$ for some $i$ and assume $\pi(i)$ describes some input word $u$.
 Eve has made choices such that she picks the run $\rho'$ of $\mathcal A$ on the pair $(u,\mathcal T(u))$.
 Let $\rho$ denote the alternative run that Adam has picked which corresponds to some pair $(u,v)$.
 We have that $\frac{\SUM(\rho)}{|\rho|} \leq \bestVal_S(u)$.
 Since $\mathcal T$ is an $r$-approximate $S$-realizer, we know that $\frac{\SUM(\rho')}{|\rho'|} + r \geq \bestVal_S(u)$.
 Thus, $\EL(\pi(i)) = \SUM(\rho') - \SUM(\rho) + r|\rho| \geq 0$,
 since $|\rho| = |\rho'|$.

 Given a strict threshold $r$, we show that the problem reduces to deciding if Eve has an observation based winning strategy in a slightly different \prefix energy game with imperfect information with initial credit zero.
 For all $u,v$ holds that $S(u \otimes v) > \bestVal_S(u)-r$ implies that $S(u \otimes v)|uv| \geq \bestVal_S(u)|uv|-r|uv|+1 \Leftrightarrow S(u \otimes v)|uv| - \bestVal_S(u)|uv| + r|uv| - 1 \geq 0$, because $\mathcal A$ has weights in $\mathbbm Z$.
 Thus, it easy to see that it suffices to subtract one from the weights of the edges outgoing of the initial state.
\end{proof}

\subsubsection{Proof of \cref{thm:approximate}: Average (lower bound)}

\begin{theorem}\label{thm:approx-avg-hardness}
 The approximate synthesis problem for an \AVG-specification and a strict or non-strict threshold is \textsc{EXPtime}-hard.
\end{theorem}

\begin{proof}
 We provide a reduction from countdown games.
 A countdown game is a weighted game with weights over $\mathbbm N\setminus \{0\}$ with perfect information in which the actions of a transition are equal to their weight.
 In this setting an action is called duration and the sum value of a play prefix is called counter value.
 The countdown objective is parameterized by a bound $N \in \mathbbm N$ and is given by $\mathsf{Count}_G(N) = \{ \pi \in \plays(G) \mid \exists i\in \mathbbm N: \SUM(\pi(i)) = N\}$.
 Deciding whether Eve has a winning strategy for the objective $\mathsf{Count}_G(N)$, where $N$ and all durations are given in binary, is \textsc{EXPtime}-complete \cite{jurdzinski2007model}.

 Note that Adam has clearly won a play if it is currently in a vertex $s$ and the counter value is $c < N$ such that for all $\tau = (s,d,s') \in T$ holds that $c+d > N$.
 Let $n = \lfloor \mathrm{log}_2 N \rfloor + 2$, let $D \subseteq \mathbbm N$ be the set of durations used in the counter game.
 
 We first consider the strict case.
 We design an \AVG-automaton $\mathcal A$ defining a weighted specification $S$ such that there exists a $\mathcal T$ that implements a strictly $1$-approximate $S$-realization if, and only if, Eve has a winning strategy in the countdown game.
 Recall, strictly $1$-approximate means that $S(u \otimes \mathcal T(u)) > \bestVal_S(u) - 1$ for all $u\in\mathrm{dom}(S)$.

 \begin{figure}[th!]
   \centering
   \begin{tikzpicture}[thick]
      \tikzstyle{every state}+=[inner sep=1pt, minimum size=7pt];
      \node[state, initial above]    (in) {$q_{in}$};
      \node[state, below of=in]              (q) {};
      \node[state, below left of=q]        (qn) {$q_n$};
      \node[state, below right of=q,xshift=2cm]        (qy) {$q_y$};
      \node[state, below of=qn]              (qn2) {};
      \node[state, below of=qy]              (qy2) {};

      \node[state, below of=qy2]              (qy3) {};
      \node[state, right of=qy3]              (vin) {$v_{in}$};
      \node[text=gray, below of=vin]                     (vin2) {Game part};

      \node[state, left of=qn,xshift=-.25cm]   (q1sq)    {$q_{1,\square}$};
      \node[state, above of=q1sq, accepting]   (fin1) {$q_{fin}$};

      \node[state, right of=qy,xshift=.25cm]   (q0sq)    {$q_{0,\square}$};
      \node[state, above of=q0sq, accepting]   (fin0) {$q_{fin}$};

      \node[state, below of=qn2,xshift=-2.8cm]    (nx0) {$\overline{x_0}$};
      \node[state, below of=qn2,xshift=-1.4cm]    (nx1) {$\overline{x_1}$};
      \node[state, below of=qn2,xshift=0cm]    (nx2) {$\overline{x_2}$};
      \node[below of=qn2,xshift=1cm]         (ndots) {$\dots$};
      \node[state, below of=qn2,xshift=2cm]    (nxn) {$\overline{x_n}$};

      \node[state, below of=nx0]    (x0) {${x_0}$};
      \node[state, below of=nx1]    (x1) {${x_1}$};
      \node[state, below of=nx2]    (x2) {${x_2}$};
      \node[below of=ndots]         (dots) {$\dots$};
      \node[state, below of=nxn]    (xn) {${x_n}$};

      \node[state, below of=x0,yshift=-1.5cm,xshift=-1.75cm]    (qx0) {$\hat{q_1}$};
      \node[state, below of=x1,yshift=-1.5cm,xshift=-1.75cm]    (qx1) {$\hat{q_1}$};
      \node[state, below of=x2,yshift=-1.5cm,xshift=-1.75cm]    (qx2) {$\hat{q_1}$};
      \node[state, below of=xn,yshift=-1.5cm,xshift=-1.75cm]    (qxn) {$\hat{q_1}$};


      \node[state,below of=xn, yshift=-1.5cm]    (m) {$\hat{q_1}$};

      \node[text=gray, below of=x1,yshift=.75cm]    (counter) {Counter part};

      \node[state, left of=nx0,yshift=-.75cm]    (q1) {$q_1$};
      \node[state, right of=vin,xshift=0.5cm]    (qtri) {$q_\blacktriangleright$};
      \node[state, above of=qtri]    (q0) {$q_0$};
      \node[state, right of=q0,yshift=0.5cm]    (q00) {$\hat{q_0}$};
      \node[state, left of=q1,yshift=0.5cm]    (q11) {$\hat{q_1}$};

      \draw[->] (in)  edge[] node {$\inp|1$}   (q);
      \draw[->] (q)  edge[] node {$n|1$}   (qn);
      \draw[->] (q)  edge[] node {$y|\!-\!1$}   (qy);

      \draw[->] (qn)   edge[] node {$\square|\!-\!3$}   (q1sq);
      \draw[->] (qy)   edge[] node {$\square|4$}   (q0sq);
      \draw[->] (q1sq) edge[] node {$\square|1$}   (fin1);
      \draw[->] (q0sq) edge[] node {$\square|0$}   (fin0);

      \draw[->] (qn) edge[] node {$\inp|1$} (qn2);
      \draw[->] (qy) edge[swap] node {$\inp|0$} (qy2);

      \draw[->] (qy2) edge[swap] node {$\outp|0$} (qy3);
      \draw[->] (qy3) edge[swap] node {$\inp|0$} (vin);

      \draw[->] (qn2) edge[swap,] node {$0|1$} (nx0);
      \draw[->] (qn2) edge[] node {\hspace{-.2cm}$1|1$} (nx1); 
      \draw[->] (qn2) edge[] node {$2|1$} (nx2); 
      \draw[->] (qn2) edge[] node {$n|1$} (nxn);

      \draw[->] (q0) edge[bend left=20] node {$\inp|0$} (q00);
      \draw[->] (q00) edge[bend left=20] node {$\outp|0$} (q0);
      \draw[->] (q1) edge[bend left=20] node {$\inp|1$} (q11);
      \draw[->] (q11) edge[bend left=20,near start] node {$\outp|1$} (q1);

      \draw [draw=gray] ($(nx0) - (0.5,-1.0)$) rectangle ($(xn) + (0.5,-1.0)$);

      \draw [draw=gray] ($(nx0) - (0.3,-0.3)$) rectangle ($(x0) + (0.3,-0.3)$);
      \draw [draw=gray] ($(nx1) - (0.3,-0.3)$) rectangle ($(x1) + (0.3,-0.3)$);
      \draw [draw=gray] ($(nx2) - (0.3,-0.3)$) rectangle ($(x2) + (0.3,-0.3)$);
      \draw [draw=gray] ($(nxn) - (0.3,-0.3)$) rectangle ($(xn) + (0.3,-0.3)$);

      \draw[->,draw=blue,text=blue] (nx0) edge[bend left=10] node[swap,near end,xshift=0.1cm] {$\blacktriangleright|0$} (qx0);
      \draw[->] (x0) edge[bend left=10] node[near end,xshift=-0.1cm] {$\blacktriangleright|1$} (qx0);

      \draw[->] (nx1) edge[bend left=10] node[swap,near end,xshift=0.1cm] {$\blacktriangleright|1$} (qx1);
      \draw[->,draw=blue,text=blue] (x1) edge[bend left=10] node[near end,xshift=-0.1cm] {$\blacktriangleright|0$} (qx1);

      \draw[->] (nx2) edge[bend left=10] node[swap,near end,xshift=0.1cm] {$\blacktriangleright|1$} (qx2);
      \draw[->,draw=blue,text=blue] (x2) edge[bend left=10] node[near end,xshift=-0.1cm] {$\blacktriangleright|0$} (qx2);

      \draw[->] (nxn) edge[bend left=10] node[swap,near end,xshift=0.1cm] {$\blacktriangleright|1$} (qxn);


      \draw[->] (q1) edge[bend left=40] node {$\square|1$} (q1sq);

      \draw[->] (xn) edge[swap,near end] node {$\inp|1$} (m);

      \draw[dashed,->] ($(xn)+ (0.3,-0.7)$) node[swap,yshift=-1.5cm,xshift=2cm] {$\textit{counting error }\!|1$} to[bend left=80] (m);

      \draw[dashed,->,draw=blue,text=blue] ($(nx0)+ (0.2,0.7)$) node[swap,yshift=+.75cm] {$\square|0$} to[] (q1sq);

      \draw [draw=gray] ($(vin) - (1,-0.5)$) rectangle ($(vin2) + (1,-0.5)$);

      \draw[dashed,->] ($(vin2)+ (0.8,0.5)$) node[swap,xshift=0.7cm] {$\blacktriangleright|0$} to[bend right=20] (qtri);
      \draw[dashed,->] ($(vin)+ (0.2,0.4)$) node[swap,yshift=1cm,xshift=0.5cm] {$\square|0$} to (q0sq);
      \draw[->] (qtri) edge[swap] node {$\blacktriangleright|0$} (q0);
      \draw[->] (q0) edge[swap] node {$\square|0$} (q0sq);

      \draw[dashed,->,draw=blue,text=blue] ($(vin2)+ (0.1,-0.4)$) node[swap,xshift=0.7cm,yshift=-0.75cm] {$\textit{input error }\!|1$} to[bend right=120] (q00);

   \end{tikzpicture}
  \caption{Overview of the automaton.
  The initial part is designed to force a strictly $1$-approximate realization to be in the game part (if it exists).
  The counter part is always entered with average one, and all transitions in the counter part have weight one.
  Each bit of the counter has its own gadget.
  The counter part is exited with weight one when either the $n$th bit is flipped to one, or a counting error occurred, or after seeing the input $\blacktriangleright$ where the counter bits do not correspond to the bits of $N$, $N$ is the counter value to be reached in the counter game.
  This figure shows the example where $N$ corresponds to the little-endian bit sequence $011$.
  The counter part is exited with weight zero when either the end input $\square$ is seen, or after seeing the input $\blacktriangleright$ where the counter bits correspond to the bits of $N$.
  These transitions are shown in blue.
  The game part is always entered with average zero. All transitions in the game part have weight zero. The game part is exited with weight zero when either the end input $\square$ is seen, or the output $\blacktriangleright$ is seen. 
  The next input then has to be $\blacktriangleright$.
  The output $\blacktriangleright$ is intended to be taken when the counter value is $N$. 
  The game part is exited with weight one if the input has an error, meaning input does not simulate the game play correctly.
  This transition is shown in blue.
  }
  \label{fig:automaton-overview}
 \end{figure}
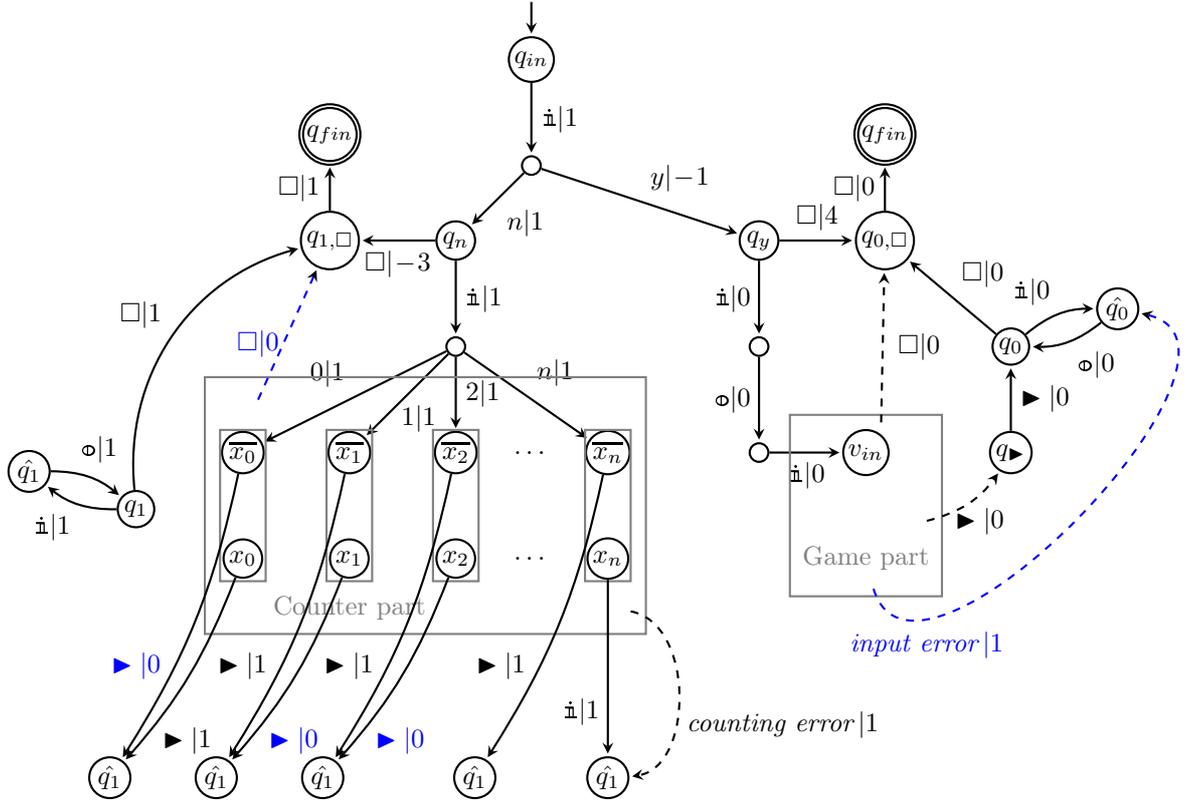

 The idea behind the automaton is as follows, see \cref{fig:automaton-overview} for an overview.
 The automaton has three parts, the initial part, the game part and the counter part.

 The initial part is used to force a strictly $1$-approximate $S$-realization to be in the game part, if there exists one.
 This is done as follows.
 Consider the input $\sigma\square \in \mathrm{dom}(S)$, $\sigma$ is any input symbol.
 The possible output combinations are $n\square$ and $y\square$, with $S(\sigma\square \otimes n\square) = 0$ and $S(\sigma\square \otimes y\square) = 1$.
 Since the difference is one, the first output must be $y$ in a sequential transducer in order to be a strictly $1$-approximate $S$-realizer.
 
 The remainder of the automaton is split into two parts, one part -- the game part -- is used to simulate the moves of Eve and Adam in the counter game, the other part -- the counter part -- is used to represent the binary encoding of the current sum of durations.
 It suffices to use an $n$-bit counter, because if the $n$th bit is one, the counter represents a value of at least $2^n > N$.
 If the counter has reached this value without seeing $N$, it is no longer possible to reach the counter value $N$ required for Eve to win the game.
 In the counter part each bit is represented by its own gadget.

 We give a brief overview of the weights in these parts and their functions.
 The game part is always reached with current value zero, the counter part with value one.
 The value of a run while in the game part remains zero, and while in the counter part remains one.
 In the end, given a transducer $\mathcal T$, for a given input sequence $u \in \mathrm{dom}(S)$, we compare the run that was induced by the produced output sequence $\mathcal T(u)$ to all other runs that could be induced with the same input sequence $u$ and different output sequences $v$ such that $u \otimes v \in S$.
 For $\mathcal T$ to be a strictly $1$-approximate $S$-realizer, we require that the difference between $S(u \otimes \mathcal T(u))$ and $S(u \otimes v)$ is less than one for all $v$ such that $u \otimes v \in S$.
 The goal is that runs can exit the counter part (triggered by the special output $\blacktriangleright$ in the game part, explained in detail further below) with a value slightly less than one when the counter value represents $N$ (the input sequence can continue after the exit).
 Also, if the counter represents a value less or equal to $N$ and the input sequence ends, then the counter part is exited with a value slightly less than one.
 In \cref{fig:automaton-overview} these transitions correspond to the transitions exiting the counter part which are shown in blue.
 Runs exit the counter part with value exactly one if either the $n$th bit has flipped to one, which indicates that the value $N$ has been surpassed, or if an error in updating the counter has been observed.
 Moreover, if exiting the counter part was triggered by the special output $\blacktriangleright$ in the game part, then it should not be beneficial if the counter value is not exactly $N$.
 Thus, in these cases the counter part is exited maintaining value one.
 The game part is always exited with value zero, unless the exit was caused by an error in the input sequence, then the value is above zero.
 The input errors are symbolized by the blue transition exiting the game part in \cref{fig:automaton-overview}.
 These properties ensure that the difference between compared runs in these two parts will always be less than one if, and only if, Eve has a winning strategy in the counter game that is copied by a transducer.


 We are ready to go into more detail.
 First, we give the alphabet explicitly, the other components are given implicitly by describing the structure of the automaton.

 The alphabet $\Sigma$ is the union of the following sets: the set $\{0,\dots,n\}$, the number $i$ is used as output to lead the run into the gadget for the $i$th bit of the counter; the set of vertices $V$, vertices are used as input to choose the successor of a transition; the set of durations $D$, durations are used as output to choose a duration; $\{ b_i \mid 0 \leq i \leq n\}$ and $\{ c_i \mid 1 \leq i \leq n\}$ are used as input, each $b_i$ is used to indicate that the $i$th bit is one, each $c_i$ is used to indicate that the $i$th carry bit in an addition is one; $\{y,n\}$, for yes and no, are used as output to indicate whether the next input should set a carry bit or not; $\{\diamond\}$, $\diamond$ is used as input for carry bits that remain zero; finally, $\{\blacktriangleright, \square\}$, these two symbols are used as both input and output, $\blacktriangleright$ to force the runs into special states and $\square$ to end input and output.
 Let $\SigmaI$ and $\SigmaO$ denote the subset that is used as input and output, respectively.

 We use the symbol $\inp$ resp.\ $\outp$ to denote any input resp.\ output symbol that is not $\blacktriangleright$ or $\square$.
 Furthermore, when depicting (parts of) the automaton, we allow to label transitions with two symbols, one input and one output symbol, e.g., $(p,b_1\outp,q)$ is used a shorthand for $(p,b_1,r)$ and $(r,\outp,q)$ that uses an intermediate state which is not shown to make the depictions more readable.

 The automaton has special states, $q_0$, $\hat{q_0}$, $q_{0,\square}$, $q_1$, $\hat{q_1}$ $q_{1,\square}$, $q_{\mathit fin}$, and $q_\blacktriangleright$, see \cref{fig:automaton-overview}.
 The state $q_{fin}$ is a deadlock and the only final state of the automaton.
 From $q_i$, reading one input (except $\square$) and one output symbol leads back to $q_i$ via $\hat{q_i}$ with sum $i+i$ for $i = 0,1$. 
 From $q_i$, reading $\square$ leads to $q_{i,\square}$ with weight $i$, and from $q_{i,\square}$, reading $\square$ leads to $q_{\mathit fin}$ with weight $i$ for $i = 0,1$.
 The purpose of $q_\blacktriangleright$ is explained further below in the paragraphs that describe the game part. 

 We describe the counter part. 
 Each transition with source and target in this part has weight one.
 For each bit of the counter there is one gadget, see \cref{fig:counter}.
 The counter part is designed to update the counter based on the durations seen in the game part of the automaton, the addition of bits and carry bits is done in the order $b_0c_1b_1\dots b_nc_n$, see the description of the game part.
 The outputs play no role in the bit gadgets, the targets of transitions are determined by the input.
 We explain the gadget for the $i$th bit.
 The design is such that a run that is in state $\overline x_i$ resp.\ $x_i$ indicates that the $i$th bit of the counter is currently zero resp.\ one.
 In $\overline x_i$, adding a bit or a carry bit (indicated by reading $b_i$ resp.\ $c_i$) flips the counter, so $x_i$ is reached.
 In $\overline x_i$, reading $c_{i+1}$ leads to $\hat{q_1}$ with weight one, because adding something to the $i$th bit of the counter while it is zero cannot set the $i+1$th carry bit to one.
 In \cref{fig:automaton-overview} this transition (among others) is symbolized by the transition with label \emph{counting error}.
 Taking this transition ensures that the value of the run can not go below one anymore.
 In $x_i$, adding a bit or a carry bit (indicated by reading $b_i$ resp.\ $c_i$) flips the counter, however, $c_{i+1}$ has to be read eventually, because this operation sets the $i+1$th carry bit to one.
 In \cref{fig:counter} the unnamed states represent states where the counter waists to see $c_{i+1}$, if as input some vertex $v \in V$ is seen, we know the addition of a duration to the counter is finished, because input has chosen a successor vertex.
 Thus, the necessary carry bit was not seen and an error was detected, so seeing $v \in V$ leads to $\hat{q_1}$.
 If the first waiting state was reached with $c_i$ the input $b_i$ might be seen before $c_{i+1}$, that is why there is the second waiting state.
 Reading $c_{i+1}$ directly from $x_i$ leads to $q_1$ with weight one, because this would indicate that the $i+1$th carry bit was set without adding something to the $i$th bit of the counter which is an error.
 
 Regarding the $n$th bit, from $x_n$ everything goes to $\hat{q_1}$ with weight one, because this means the counter has surpassed the value $N$, see \cref{fig:automaton-overview}.

 As mentioned above, in the game part, output can trigger to stop the game by the output symbol $\blacktriangleright$, then input has to answer with $\blacktriangleright$ as input; seeing the input $\blacktriangleright$ in the counter gadgets triggers to exit the counter gadget, see \cref{fig:automaton-overview}.
 The intention is that exiting the counter gadget with $\blacktriangleright$ should only be beneficial, i.e., lead to a run with value less than one, if the counter value represents exactly $N$.
 This is achieved as follows, see also \cref{fig:automaton-overview}.
 Let $y_0\dots y_{n-1}$ be the little-endian bit encoding of $N$.
 If $y_i$ is zero resp.\ one, then from $\overline{x_i}$ resp.\ $x_i$ a transitions with $\blacktriangleright$ leads to $\hat{q_1}$ with weight \emph{zero} for each $i$, ensuring that the value of the run will be less than one.
 For all other states (that expect input), reading $\blacktriangleright$ leads to $\hat{q_1}$ with weight one.

 In the bit gadgets, for all inputs that are not $\square$ and were not described together with some output symbol lead back to the same state, e.g., reading $b_2\outp$ from $x_i$ (with $i \neq 1$) goes back to $x_i$. 
 From every state (that expects input), reading $\square$ leads to $q_{1,\square}$ with weight \emph{zero}.
 This ensures that the value of the run is less than one, because it does not indicate a counter value that has surpassed $N$, nor an error in updating the counter.

 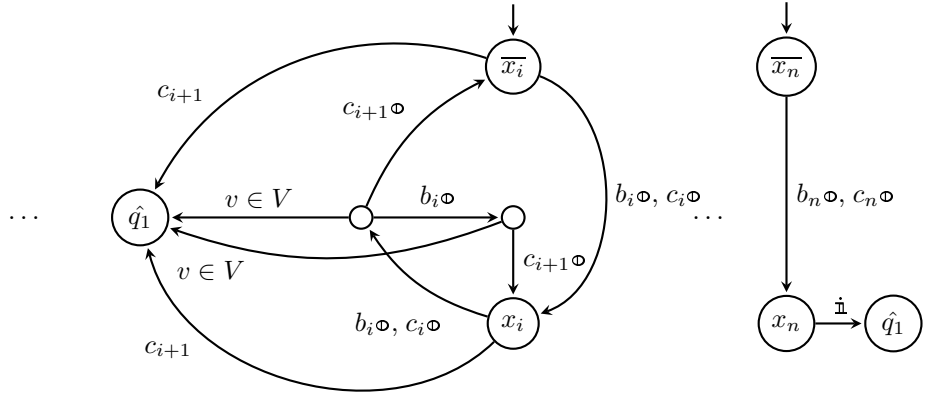
\begin{figure}[th!]
   \centering
   \begin{tikzpicture}[thick]
      \tikzstyle{every state}+=[inner sep=3pt, minimum size=3pt];
      \node[state, initial above]             (nxi) {$\overline{x_i}$};
      \node[state, below of=nxi, yshift=-2cm]   (xi) {$x_i$};
      \node[state, initial above, left of=nxi, xshift=5cm]   (nxn) {$\overline{x_n}$};

      \node[state, above of=xi]   (m2) {};

      \node[state, below of=nxn, yshift=-2cm]   (xn) {$x_n$};
      \node[state, above of=xi, xshift=-2cm]  (m) {};
      \node[state, left of=m2, xshift=-3.5cm]  (bot) {$\hat{q_1}$};
      \node[state, right of=xn]  (bot2) {$\hat{q_1}$};

      \node[state, left of=m, draw=none, xshift=-3cm]  (n1) {$\dots$};
      \node[state, left of=m, draw=none, xshift=6cm]  (n2) {$\dots$};


     \draw[->] (nxi)  edge[bend right=40,swap] node[near end] {$c_{i+1}$}   (bot);
     \draw[->] (nxi)  edge[bend left=70] node {$b_i\outp$, $c_i\outp$}   (xi);
     \draw[->] (xi) edge[bend left=20] node[near start] {$b_i\outp$, $c_i\outp$}   (m);
     \draw[->] (xi) edge[bend left=60] node[near end] {$c_{i+1}$}   (bot);
     \draw[->] (m) edge[bend left=20] node {$c_{i+1}\outp$}   (nxi);
     \draw[->] (m) edge node[swap] {$v\in V$}  (bot);

     \draw[->] (xn)  edge node {$\inp$}   (bot2);
     \draw[->] (nxn)  edge node {$b_n\outp$, $c_n\outp$}   (xn);

     \draw[->] (m) edge[] node {$b_i\outp$}   (m2);
     \draw[->] (m2) edge[] node {$c_{i+1}\outp$}   (xi);
     \draw[->] (m2) edge[bend left=20] node[near end] {$v \in V$}   (bot);

     \end{tikzpicture}
  \caption{Counter gadgets. One gadget for each bit.}
  \label{fig:counter}
 \end{figure}

 We describe the game part.
 Each transition with source and target in this part has weight zero.
 The idea behind the weights in this part is that each run should have value zero, unless the input stops to faithfully simulate a play in the counter game, in which case it should have a value greater than zero.

 From a state $v \in V$ an output symbol must be read.
 It is possible to go to a state $(v,d)$ with output $d$ if there exists a transition of the form $(v,d,v') \in T$.
 Furthermore, the output $\blacktriangleright$ is possible leading to the state $q_\blacktriangleright$ with weight zero, see \cref{fig:automaton-overview}.
 The intention is to use this output if the counter part of the automaton represents the value $N$.
 In $q_\blacktriangleright$ the input $\blacktriangleright$ leads to $q_0$ with weight zero, all other inputs (except $\square$) lead to $q_0$ with weight \emph{one}, $\square$ leads to $q_{0,\square}$ with weight \emph{one}.
 These transitions are not shown in \cref{fig:automaton-overview}, to not clutter the figure further, if inserted this would also be visualized by a blue transition labeled with \emph{input error}.
 As explained before, the automaton is designed such that reading the input $\blacktriangleright$ in the counter part leads to a value less than one if the counter value is indeed $N$.

 From a state $(v,d)$ the next symbols must ensure that the duration $d$ is added to the counter, i.e., the sequence read from here in the game part must -- when read in the counter part -- update the automaton representation of the bit counter.
 We show how this is done for a concrete example, see \cref{fig:adder-gaget}.
 We assume that a $4$-bit counter is used, that we are in a state $(s,13)$ and that the possible successors of $s$ with duration $13$ in the counter game are $s_1$ and $s_2$.
 The number $13$ is $1011$ in little-endian notation, so the next inputs that describe the duration are $b_0$, $b_2$ and $b_3$.
 When adding $1011$ to the $4$-bit counter, some carry bits might become one.
 The outputs $y$(es) and $n$(o) are used to indicate which ones, the input $c_i$ indicates that the $i$th carry bit becomes one, the input $\diamond$ is used for the carry bits that remain zero.
 Whether the chosen carry bits are correct is checked in the counter part of the automaton.
 If it is incorrect, the run in the counter part will have value one and the run in this adder gadget will have value zero.
 Eventually, the next input symbol indicates which successor is picked.

 From a state $(v,d)$, the input $\square$ leads to $q_{0,\square}$ with weight zero.
 From every other state (that expects input), the input $\square$ leads to $q_{0,\square}$ with weight \emph{one} (not pictured in \cref{fig:automaton-overview} to save space).
 The target of a transition for any state (that expects input) and an input symbol for which we did not describe a transition leads to $\hat{q_0}$ with weight \emph{one}.
 These transition are symbolized by the transition labeled with \emph{input error} in \cref{fig:automaton-overview}.
 The first input possibility marks a valid point to stop simulating the game.
 The latter two input possibilities mark invalid points to stop simulating the game, so it is ensured that the value of run is not zero.

 \begin{figure}[th!]
   \centering
   \begin{tikzpicture}[thick]
      \tikzstyle{every state}+=[inner sep=3pt, minimum size=3pt];
      \node[state, initial right, inner sep=0,outer sep=0]    (b0) {\scriptsize $(s,13)$};
    
      \node[state, left of=b0] (c1) {};
      \node[state, left of=c1, xshift=-.75cm] (c2) {};
      \node[state, left of=c2, xshift=-.75cm]  (b3) {};
      \node[state, left of=b3]  (ob3) {};
      \node[state, left of=ob3, yshift=-3cm] (c3) {};

      \node[state, right of=c3, xshift=.75cm]  (b2) {};
      \node[state, right of=b2]  (ob2) {};
      \node[state, right of=ob2] (c4) {};
      \node[state, right of=c4, xshift=.75cm] (d) {};
      \node[state, right of=d]  (e) {};
      \node[state, above right of=e] (s1) {$s_1$};
      \node[state, below right of=e] (s2) {$s_2$};

      \node[state, above left of=c1] (ac1) {};
      \node[state, above left of=c2] (ac2) {};
      \node[state, above right of=c3] (ac3) {};
      \node[state, above right of=c4] (ac4) {};

      \node[state, below left of=c1] (bc1) {};
      \node[state, below left of=c2] (bc2) {};
      \node[state, below right of=c3] (bc3) {};
      \node[state, below right of=c4] (bc4) {};

      \draw[->] (b0)  edge[swap] node {$b_0$}   (c1);
      \draw[->] (b3)  edge[swap] node {$\outp$}   (ob3);
      \draw[->] (ob3)  edge[swap,bend right=40] node {$b_2$}   (c3);
      \draw[->] (b2)  edge node {$\outp$}   (ob2);
      \draw[->] (ob2)  edge node {$b_3$}   (c4);
      \draw[->] (e)  edge[]          node        {$s_1$} (s1);
      \draw[->] (e)  edge[swap]          node        {$s_2$} (s2);

      \draw[->] (c1) edge[swap] node {$y$} (ac1);
      \draw[->] (c2) edge[swap] node {$y$} (ac2);
      \draw[->] (c3) edge node {$y$} (ac3);
      \draw[->] (c4) edge node {$y$} (ac4);

      \draw[->] (ac1) edge[swap] node {$c_1$} (c2);
      \draw[->] (ac2) edge[swap] node {$c_2$} (b3);
      \draw[->] (ac3) edge node {$c_3$} (b2);
      \draw[->] (ac4) edge node {$c_4$} (d);

      \draw[->] (c1) edge[] node {$n$} (bc1);
      \draw[->] (c2) edge[] node {$n$} (bc2);
      \draw[->] (c3) edge[swap] node {$n$} (bc3);
      \draw[->] (c4) edge[swap] node {$n$} (bc4);

      \draw[->] (bc1) edge[] node {$\diamond$} (c2);
      \draw[->] (bc2) edge[] node {$\diamond$} (b3);
      \draw[->] (bc3) edge[swap] node {$\diamond$} (b2);
      \draw[->] (bc4) edge[swap] node {$\diamond$} (d);

      \draw[->] (d) edge node {$\outp$} (e);
     \end{tikzpicture}
  \caption{Adder gadget. Example for the gadget corresponding to transitions $(s,13,s_1)$ and $(s,13,s_2)$ with $n = 4$. The duration $13$ is $1011$ in little-endian notation.}
  \label{fig:adder-gaget}
 \end{figure}
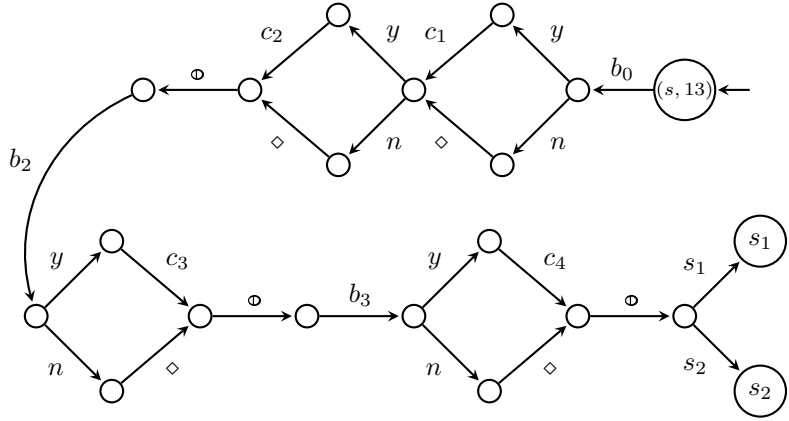

 We are ready to prove the correctness of our construction.
 We make three observations.
 First, it is easy to see that we have constructed a deterministic automaton with weights, thus, it defines a weighted specification $S$.
 Second, the domain of $S$ is $(\SigmaI\setminus\{\square\})^+\square$, furthermore, the domain of both parts of the automaton are the same.
 Third, we have $0 \leq S(u \otimes v) \leq 1$ for all $u \otimes v \in S$.

 We already established that a strictly $1$-approximate $S$-realizer must lie in the game part of the automaton if it exists.

 Assume that Eve has a winning strategy in the countdown game.
 Consider a transducer that behaves according to a winning strategy.
 
 We distinguish between input sequences that faithfully simulate a play in the game, and those that do not.
 
 In the latter case, the run in the game part has a strictly positive value, all values for alternative runs with same input sequence have a value of at most one (this property is also true for all other input sequences).
 So, the difference is less than one.
 
 In the former case, the output indicates to stop counting after the counter has reached the value $N$ by the output $\blacktriangleright$, the run in the game part has value zero.
 We take a look at all other output sequences that can be combined with the same input sequence.
 First, consider all alternative output sequences such that the run also lies in the game part.
 All runs in the game part have a value strictly less than one, so the difference is less than one.
 Secondly, consider all alternative output sequences such that the run lies in the counter part.
 The second output symbol fixes in which bit gadget the run lies.
 Note that after a bit gadget has been entered, the outputs do not influence the runs anymore, they depend only on the input.
 Thus, no matter which gadget is entered, when the input $\blacktriangleright$ is seen, the run is assigned a value less than one, because the bit value is the same as in $N$.
 So, the difference is less than one.

 What is left for this direction, is to consider the input sequences that faithfully simulate the game, but stop before the counter has reached $N$.
 The run in the game part induced by the output sequence provided by the transducer has value zero.
 All alternative output sequences that induce runs in the counter part have value less than one, because no counting error occurs and the $n$th bit does not flip to one.
 As argued before, all alternative output sequences that induce runs in the game part have a value less than one.
 
 Assume that Eve has no winning strategy.
 We consider input sequences that faithfully simulate the game.
 Consider the outputs of any transducer that is an $S$-realizer.
 The resulting run in the game part has value exactly zero, because the inputs faithfully simulate the game.

 We consider an input sequence such that the value of the counter eventually surpasses $N$ if the simulation of the game is faithful.

 If the provided output sequence is faithful, then the $n$th bit of the counter eventually flips to one.
 Considering the run for an alternative output sequence that lies in the gadget of the $n$th bit of the counter yields a value of exactly one.
 The difference is one, the transducer is not strictly $1$-approximate.

 The other possibility is that the output provided by the transducer contain errors when updating the carry bits in order to maintain a counter value below $N$.
 Since such an error is detected in the corresponding counter bit, again there exists an alternative output sequence such that the corresponding run in the counter part witnesses the error and is assigned value exactly one.
 The difference between the run in the game part and the alternative is one, thus, the transducer is not strictly $1$-approximate.

 We turn to the case of a non-strict threshold.
 We can easily change the described specification such that there exists a non-strict $1$-approximate realizer if, and only if, Eve has a winning strategy in the counter game.
 For this we change the weight of two transitions.
 First, the weight of the transition that goes from $q_y$ to $q_{0,\square}$ is changed from $4$ to $5$, to ensure that a non-strict $1$-approximate realizer has to still lie in the game part.
 Secondly, the weight of outgoing transitions from $q_n$ for some input that is not $\square$ is changed from $1$ to $2$.
 This ensures that the runs in the counter part have a value of slightly above one unless something beneficial for output happens, then the values will be decreased to exactly one.
 In the game part the values will be as before, i.e., zero, unless there is an input error, then the values will be slightly above zero.
 These changes ensure that the difference between the runs will always be exactly one if, and only if, Eve has a winning strategy in the counter game.
\end{proof}

\subsubsection{Proof of \cref{thm:approximate}: Discounted-sum}

\begin{theorem}\label{thm:approximate-dsum}
 The approximate synthesis problem for a \DSUM-specification with a discount factor $\lambda$ of the form $\frac{1}{n}$ with $n\in \mathbbm N$ and a strict threshold is decidable in \textsc{NEXPtime} and a non-strict threshold in \textsc{EXPtime}.
\end{theorem}

\begin{proof}
 We reduce the problems from the statement to threshold synthesis problems for \DSUM-specifications.

 In \cite{DBLP:journals/corr/BokerH14}, it is shown that \DSUM-automata with a discount factor $\lambda$ of the form $\frac{1}{\ell}$ with $\ell \in \mathbbm N$ are determinizable with the same discount factor.

 The idea is to (disregard the labels of output transitions which yields a nondeterministic specification automaton and) construct the product automaton of the specification automaton (without output labels) and a deterministic version of it.
 The weights of a transition in the product is the weight of the transition in the deterministic version subtracted from the weight of the transition in the specification automaton (without output labels).

 We show that there is a transducer that implements an approximate realization of the specification (without output labels) for a threshold $r$ if, and only if, there is a transducer that implements a realization of the specification (without output labels) defined by the product automaton that ensures a value of $-r$ for each pair. 
 
 Let $\mathcal A = (Q_\mathcal A,\SigmaIO,q_i^\mathcal A,\Delta_\mathcal A,F_\mathcal A,\gamma_\mathcal A)$ be a complete \DSUM-automaton with a discount factor $\lambda$ of the form $\frac{1}{\ell}$ with $\ell \in \mathbbm N$ defining a \DSUM-specification $S$.
 Let $\mathcal B$ be the nondeterministic \DSUM-automaton that is obtained from $\mathcal A$ by replacing all transition labels that are output symbols by the same arbitrary output symbol, let $o_\mathrm{fix}$ denote this output symbol.
 Let $\mathcal D = (Q_\mathcal D,\SigmaIO,q_i^\mathcal D,\Delta_\mathcal D,F_\mathcal D,\gamma_\mathcal D)$ be a trim version of $\mathcal B$.

 Formally, the \DSUM-automaton $\mathcal C = (Q_\mathcal C,\SigmaIO,q_i^\mathcal C,\Delta_\mathcal C,F_\mathcal C,\gamma_\mathcal C)$ is defined as follows.
 The state set is $Q^\mathcal A \times Q^\mathcal D$, the initial state is $(q_i^\mathcal A,q_i^\mathcal D)$, the final state set is $F_\mathcal A \times F_\mathcal D$, and $\Delta_\mathcal C$ contains the following transitions
\begin{itemize}
   \item $(p,q)$ to $(p',q')$ if $(p,\sigma,p') \in \Delta_\mathcal A$ and $(q,\sigma,q')\in\Delta_\mathcal D$ for all $(p,q) \in Q^\mathcal A_\inp \times Q^\mathcal D_\inp$, the weight of the transition is $\gamma_\mathcal A((p,\sigma,p'))-\gamma_\mathcal D((q,\sigma,q'))$, 
   \item $(p,q)$ to $(p',q')$ if $(p,\sigma,p') \in \Delta_\mathcal A$ and $(q,o_\mathrm{fix},q')\in\Delta_\mathcal D$ for all $(p,q) \in Q^\mathcal A_\outp \times Q^\mathcal D_\outp$, the weight of the transition is $\gamma_\mathcal A((p,\sigma,p'))-\gamma_\mathcal D((q,o_\mathrm{fix},q'))$. 
\end{itemize}

Given a threshold $r \in \mathbbm Q$ and $\triangleleft \in \{<,\leq\}$, it is easy to see that there is a transducer that implements a $r_\triangleleft$-approximate $S$-realization if, and only if, there is a transducer that implements a $S_\mathcal C$-realization that ensures a value $\triangleright -r$ for each pair, where $S_\mathcal C$ is the \DSUM-specification defined by $\mathcal C$, and $\triangleright = >$ if $\triangleleft = <$ and $\triangleright = \geq$ otherwise.

It is left to show the claimed complexity bounds.
We assume that the weights are given in binary and the discount factor $\lambda$ is given as a pair of binary numbers.
Following the construction presented in \cite{DBLP:journals/corr/BokerH14}, $\mathcal D$ is of size $m^n=2^{n \cdot \mathrm{log}\ m}$, where $n$ is the number of states of $\mathcal A$ and $\mathrm{log}\ m$ is a number in the size of the representation of its weights.
Thus, the number of states of $\mathcal D$ is exponential in the number of states of $\mathcal A$ and polynomial in the weights of $\mathcal A$.
The weights in $\mathcal D$ are polynomial (the representation of the weights grows by one bit at most) in the weights of $\mathcal A$.
Hence, the size of the game arena is exponential in the number of states and polynomial in the weights of $\mathcal A$, and its weights are polynomial in the weights of $\mathcal A$.

For a strict threshold, \cref{thm:deterministic-threshold} yields that solving critical prefix threshold games is in \textsc{NP} for discounted-sum.
Since the constructed arena is of exponential size, the claimed complexity bound follows.

For a non-strict threshold, \cref{thm:deterministic-threshold} yields that solving critical prefix threshold games in $\textsc{NP} \cap \textsc{coNP}$ for discounted-sum.
To obtain our desired complexity bound, we make a more precise analysis.
Threshold synthesis problems for discounted-sum are reduced in polynomial time to critical prefix discounted-sum threshold games which are reduced in polynomial time to discounted-sum games, see the proof of \cref{thm:threshold-games-dsum}.
Using the value iteration algorithm from \cite{zwick1996complexity} to solve discounted-sum games yields the claimed complexity bound, because it runs in polynomial time in the size of the arena, logarithmic in the absolute maximal weight of the arena, and exponential in the representation of the discount factor, i.e., polynomial in the discount factor.
An analysis of the run time of the value iteration algorithm is given in \cite{haddad2015value}.
\end{proof}

\subsection{Infinite words and Church synthesis}

\subsubsection{Some observation on the finite-memory requirement for
  synthesis problems}\label{rem:memory}\label{app:memory}

In the definition of synthesis problems of \cref{subsec:synthdef}, we ask for the realizability of the
    specification by a (finite-state) transducer. This
    requirement can be relaxed to asking whether the specification is
    realizable by some sequential function $f$, i.e. a function of type $\SigmaI^*\rightarrow
    \SigmaO$ prescribing which output symbol should be produced by
    Eve, depending on the sequence of input symbols Adam has provided so
    far. In other words, we could relax the requirement that $f$ is
    computable by a (finite-state) transducer. First, we impose this
    requirement because we are interested in functions which can
    be finitely represented and computed. Second, we claim that realizability in the
    relaxed sense entails realizability by transducers. To see that, consider first the threshold synthesis
    problems for all measures $V \in \{ \SUM,\AVG,\DSUM\}$. They are
    all solved by reduction to corresponding critical prefix games, with
    back-and-forth translations from finite-memory strategies to transducers. For those games, it has been
    shown in \cref{sec:games} that finite-memory suffice for
    Eve to win. It entails our claim for threshold synthesis
    problems.

    The claim also holds for (strict and non-strict) approximate
    synthesis (and therefore for best-value synthesis by taking
    non-strict threshold $0$). Indeed, for \SUM-specification $S$, the 
    approximate synthesis algorithm for $S$ works by reduction
    to the regret determinization problem for a nondeterministic
    \SUM-automaton (\cref{thm:approximate-sum}). The algorithmic solution to the latter problem itself relies on solving
    a game whose finite-memory strategies can be back-translated into
    (finite-state) regret determinizers~\cite{DBLP:conf/lics/FiliotJLPR17}, which in turn are
    back-translated into transducers realizing $S$ for the threshold synthesis problem as
    shown in \cref{lem:approx-to-regret}. In~\cite{DBLP:conf/lics/FiliotJLPR17}, it is shown
    that  finite-memory strategies are sufficient to win, entailing
    our claim for \SUM-specifications. For \AVG-specifications
    (\cref{thm:approximate-avg}), the
    approximate synthesis algorithms work by reduction to some critical prefix
    energy games of imperfect information with fixed initial credit
    for which finite-memory are sufficient to win by
    \cref{cor:fin-mem}. Likewise for \DSUM-specifications and
    discount factor $1/n$, the approximate synthesis problems
    (\cref{thm:approximate-dsum}) reduce
    to critical prefix threshold games for which finite-memory are
    known to suffice (\cref{thm:threshold-games-dsum-positional}).


\subsubsection{Proof of \cref{thm:church}}

\thmchurch*

These synthesis problems can be directly reduced to the
corresponding synthesis problems on finite words. E.g., synthesis
problem for $\Thres^{> t}(W)$ reduces to the strict threshold
synthesis problem for $W$ (on finite words). The main difference
is that our quantitative synthesis problems on finite words require
realizability by a transducer. This is however
not a problem. Indeed, we could have relaxed this condition to
asking for realizability by a possibly infinite-state transducer or equivalently, a function of type
$\SigmaI^*\rightarrow \SigmaO$. However, the algorithms we provide
to solve quantitative synthesis problems over finite words
guarantee that (finite-state) transducers are sufficient. So we do
not need infinite memory.

\begin{proof}
    We consider the case of threshold specifications. For best-value
    and approximate synthesis, the proof is the same. The proof
    is also parametric in $V$. Given a deterministic $V$-automaton defining
    a weighted specification $W$ and a threshold $t$, we reduce the
    synthesis problem for $\Thres^{\triangleright t}(W)$ to the
    threshold synthesis problem (on finite words) for $W$, which is decidable by \cref{thm:deterministic-threshold}. Let us show the correctness of the
    reduction.

    Assume that $\Thres^{\triangleright t}(W)$ is realizable by a (total) function $\lambda :
    \SigmaI^*\rightarrow \SigmaO$. We construct a function $f \colon
    \dom(W)\rightarrow \SigmaO^*$ and prove that it realizes
    $W$ (over finite words for the threshold problem). To define
    $f$, let $u = i_1\dots i_k\in \dom(W)$. Then, let $f(u) =
    \lambda(i_1)\lambda(i_1i_2)\dots \lambda(i_1\dots i_k)$. By
    definition of Church synthesis,
    since $\lambda$ realizes $\Thres^{\triangleright t}(W)$, for any
    infinite continuation $i_{k+1}\dots\in\SigmaI^\omega$, we have for
    all $j\geq 0$, if $i_1\dots i_j\in \dom(W)$, then
    $W(i_1\lambda(i_1)i_2\lambda(i_1i_2)\dots i_j\lambda(i_1\dots
    i_j))\triangleright t$. It is true in particular for $j = k$ and therefore, 
    $u \otimes f(u) \in W$ and $W(u \otimes f(u))\triangleright t$.

    Conversely, assume that $W$ is realizable (for threshold $t$) by a function $f :
    \dom(W)\rightarrow \SigmaO^*$, which can be assumed to be computable by
    a transducer thanks to the observation of \cref{rem:memory}. We construct a function
    $\lambda : \SigmaI^*\rightarrow \SigmaO$ and prove that $(i)$ it
    realizes $\Thres^{\triangleright t}(W)$ and $(ii)$ it is computable
    by a Mealy machine. The function $\lambda$ has to be defined by
    all finite words over $\SigmaI$. Let $u = i_1\dots i_k$ be such a
    word and let $o_0\in \SigmaO$ be some arbitrarily chosen output symbol. We consider two cases: 
    \begin{enumerate}
      \item there exists $v\in\SigmaI^*$ such that $uv\in
        \dom(W)$. Then, we let $\lambda(u)$ be the $k$th letter of
        $f(uv)$ (since $f$ is computable by a transducer, this definition is independent from the choice of $v$)
      \item there is no such $v$. Then, we let $\lambda(u) = o_0$.
    \end{enumerate}

    $(i)$ Let us show that $\lambda$ realizes $\Thres^{\triangleright t}(W)$. Let $i_1i_2\dots \in \SigmaI^\omega$ be some input
    $\omega$-word and for all $j$, let $o_j=\lambda(i_1\dots i_j)$. 
    Assume that $i_1\dots i_k\in
    \dom(W)$. Then, by the choice of $o_1,\dots,o_k$ (in particular
    by taking $v = \epsilon$ in the definition of $\lambda$ above), we
    have $f(i_1\dots i_k) = o_1\dots o_k)$ and since $f$ realizes $W$,
    we get $W(i_1o_1\dots i_ko_k)$. It is true for all $k$, therefore 
    $i_1o_1i_2o_2\dots \in \Thres^{\triangleright t}(W)$. 

    $(ii)$ Let $T$ be a transducer computing
    $f$. We construct a Mealy machine for $\lambda$. Since $T$ is
    defined only over $\dom(W)$, we have to extend its domain to
    any finite word. To do so, we just complete $T$ whenever there is
    a missing transition towards a sink state which outputs $o_0$
    whenever it reads an input letter.

    The reduction (and proof of its correctness) 
    is exactly the same for best-value and approximate
    synthesis. The only difference is the theorems we rely on to get
    decidability. For best-value and \SUM, decidability is given by
    \cref{thm:best-value}, for best-value and \AVG, by 
    \cref{thm:best-value-avg}, for best-value and \DSUM, by 
    \cref{thm:best-value-dsum}. For the approximate synthesis
    and \SUM it is given by \cref{thm:approximate-sum}, for
    \AVG by \cref{thm:approximate-avg}, and for \DSUM by
    \cref{thm:approximate-dsum}. 
\end{proof}


\end{document}